\documentclass[aps,prd,twocolumn,superscriptaddress,nofootinbib]{revtex4-1}

\usepackage{xspace}
\usepackage{xcolor}
\usepackage{graphicx}
\usepackage{amssymb}
\usepackage{amsmath}
\usepackage{bm}% bold math
\usepackage{epsfig}
\usepackage{import}
\usepackage{epstopdf}
\usepackage[section]{placeins}
\usepackage[export]{adjustbox}
\usepackage[utf8]{inputenc}
\usepackage[english]{babel}
\usepackage{blindtext}
\usepackage{enumitem}
\usepackage{algpseudocode}
\usepackage{verbatim}
\usepackage{subcaption}
\usepackage{hyperref}
\usepackage{caption}
\newcommand{\tsc}[1]{\textsc{#1}}
\newcommand{\Py}{\tsc{Pythia}}

\begin{document}

% Use the \preprint command to place your local institutional report
% number in the upper righthand corner of the title page in preprint mode.
% Multiple \preprint commands are allowed.
% Use the 'preprintnumbers' class option to override journal defaults
% to display numbers if necessary
%\preprint{}

%Title of paper
\title{Probabilistic, Structure-Intrinsic Clustering with a Hierarchical Embedding (PSICHE) in Time and Space Applied In Particle Physics Jet Reconstruction}

\author{Margaret Lazarovits}
\email{mlazarovits@ku.edu}
\affiliation{University of Kansas, Department of Physics \& Astronomy, Lawrence, Kansas, USA 66045}
\author{Christopher Rogan}
\email{crogan@ku.edu}
\affiliation{University of Kansas, Department of Physics \& Astronomy,
  Lawrence, Kansas, USA 66045}

%\date{\today}

\begin{abstract}
Jet reconstruction is an active and open area of particle physics research, with  questions related to jet size, multiplicity, substructure, and experimental performance in the presence of pileup and noise. Analogous challenges are ubiquitous throughout other applications of unsupervised learning in the context of clustering and association. This paper introduces a new algorithm for clustering in the context of jet reconstruction: Probabilistic, Structure-Intrinsic Clustering with a Hierarchical Embedding (PSICHE). PSICHE addresses several challenges and shortcomings of existing jet clustering methods with novel features, including but not limited to: (i) dynamically-learned and variable jet (cluster) sizes, (ii) unsupervised, multi-scale learning of emergent features, like jet substructure and multiplicity, (iii) clustering in space and time, and (iv) incorporation of domain-specific experimental uncertainties. All of these properties are achieved in a self-consistent, probabilistic framework that is computationally tractable. Furthermore, demonstrations of these features are presented for a range of jet clustering examples in LHC scenarios for various physics phenomena, clustering inputs, pileup conditions, and detector performance parameters.
\end{abstract}
%\tableofcontents{}

% insert suggested PACS numbers in braces on next line
%\pacs{Pacs numbers}
% insert suggested keywords - APS authors don't need to do this
%\keywords{}

\maketitle

\section{Introduction}
The desire to learn associations between elements of a dataset -- clustering -- in an unsupervised manner is a common and compelling process, tool, and/or primary goal of data analysis. Jet reconstruction in particle physics is one such clustering application. There are a number of approaches to clustering \cite{lloyd1982least, goos_comparing_2003, mclachlan_finite_2019, heller_bayesian_2005, Stolcke_1994, Banfield1993ModelbasedGA, vaithyanathan2013modelbasedhierarchicalclustering, rasmussen_nite_nodate, chen_variational_2016}, including both heuristic and probabilistic methods, with many featuring the same impositions and limitations, frequently related to determination of the number and scale of clusters, and further latent structure to be discovered. These same elements appear throughout the topic of jet clustering \cite{grigoriev_optimal_2003, Li_2025, Chekanov_2006, De_Simone_2019, brehmer2020hierarchicalclusteringparticlephysics, cms_collaboration_cambridge-aachen_2009, ju_supervised_2020, larkoski_jet_2023, ellis_qjets_2012, mackey_fuzzy_2016, cacciari_anti-k_t_2008, ellis_successive_1993, cacciari_dispelling_2006, Huth:1990pfa, Krohn_2009, Lapsien_2016, Sirunyan_2020} in resolving jet multiplicity, size, and associated internal substructure. We introduce PSICHE (Probabilistic, Structure-Intrinsic Clustering with a Hierarchical Embedding), a general and modular algorithm that addresses many of these challenges in the domain of jet clustering and beyond.

In high-energy, subatomic particle collisions, jets are defined as associations between stable, relatively light constituents, in the form of calorimeter depositions, ionization signals, or even reconstructed particle candidates. Jets typically correspond to strongly-interacting, colored quarks and gluons \cite{gell-mann_schematic_1964, Zweig:570209, GellMann:1962xb} which, through the processes of hadronization and fragmentation, manifest as collimated showers of these elements. Reconstructing jets, or learning these associations, is not monolithic; one can cluster groups of jets into larger structures \cite{Lee:2023tfx, Scott:2024txs, Jackson_2017a, Jackson_2017b}, study the latent structure within jets, mitigate pileup and noise contaminating jets \cite{larkoski_soft_2014, Cacciari_2008, Krohn_2010, Berta_2014, cacciari_softkiller_2015, Krohn_2014, bertolini_pileup_2014}, analyze correlations between jets, or use jets' four-vectors in calculations of physically-meaningful quantities. While these tasks are typically factorized, PSICHE attempts to perform them simultaneously. This is achieved through a probabilistic, variational algorithm \cite{attias1999inferring, blei_variational_2017, blei_variational_2004, chen_variational_2016, guan_variational_nodate}, where each of these elements can inform the learning of others.

In contemporary jet clustering algorithms \cite{cacciari_anti-k_t_2008, cacciari_fastjet_2012, ellis_successive_1993, cacciari_dispelling_2006, grigoriev_optimal_2003, Ellis_2010, ju_supervised_2020, larkoski_jet_2023, ellis_qjets_2012, mackey_fuzzy_2016, chekanov2002jetalgorithmsminireview}, dependencies on prescribed jet sizes, energy scales, or multiplicities are frequent characteristics, and challenges. PSICHE inherently accommodates jets of varying sizes and types, without having to \textit{a priori} specify any of these algorithmic inputs. Larger jets can often contain rich substructure, the study of which is a broad and ongoing topic of investigation \cite{Ellis_2010, komiske_energy_2018, Thaler_2011, chien_telescoping_2020, marzani_looking_2019, dillon_uncovering_2019, klimek_time_2021, dasgupta_towards_2013, feige_precision_2012, butterworth_jet_2008, delafuente2026simplexdemixingdisentanglingmultiple,bogatskiy_explainable_2024}. In PSICHE, the emergent, internal structure of jets is considered alongside the definitions of the jets themselves, simultaneously modeling clusters and subclusters according to the most jointly-probable configuration. While jets are often restricted to a projective, spatial domain of observations, current and next-generation particle detectors are capable of much more, including precision timing measurements of emissions and depositions \cite{cms_collaboration_mip_2019, 2091129}. The probabilistic framework of PSICHE natively incorporates these experimental resolutions as precisions among the clustering-domain dimensions, while also including parameters to encode the intensities and measurement precisions of clustered observations.

Overall, PSICHE is comprised of nested, modular probabilistic models, whose details are provided in Section \ref{sec:algo_details}. A variational mixture model is embedded into a tree-based, hierarchical model, knit together with general inductive biases to symbiotically learn cluster multiplicity, size, and structure. The application of PSICHE to jet clustering is introduced in Section \ref{sec:nom_jet_clustering}, where optimized posteriors in a probabilistic model are physically translated into jets and their associated properties, facilitated by domain-specific inductive priors. Section \ref{sec:novel_features} explores the utility of PSICHE's hallmark features -- dynamic sizing, multi-scale learning, spatiotemporal clustering, and inclusion of observational uncertainties -- in several scenarios of jet production at a hadron collider, with a demonstration of potential strategies for using PSICHE jet and subcluster observables in pileup mitigation. Finally, directions for future development and studies in both the PSICHE algorithm and jet clustering application are summarized in Section~\ref{sec:future}.

\section{Algorithm Details} \label{sec:algo_details}
PSICHE is a probabilistic clustering algorithm that assimilates the recursive tree structure of Bayesian Hierarchical Clustering \cite{heller_bayesian_2005} with an embedded, variational mixture model \cite{chen_variational_2016}. By integrating these elements in a single, probabilistic framework, PSICHE learns high-level associations through the overarching tree structure while symbiotically resolving lower-level substructure with the mixture model. Sections \ref{sec:vGMM} and \ref{sec:bhc_model} respectively describe the variational mixture and tree-consistent partitioning modules of the algorithm. Beyond combining these elements, there are several other, notable algorithmic additions featured in PSICHE, namely, the inclusion of observational data precisions and weights in the model likelihood and a history-informed mixture model initialization procedure that mirrors the merges of the overarching tree level within the internal substructure, as described in Section \ref{sec:unique_algo_features}.

\subsection{Variational Gaussian Mixture Model} \label{sec:vGMM}
Variational Gaussian mixture models (vGMMs) are a class of well-studied methods for clustering locational data in continuous and infinite spaces \cite{attias1999inferring, chen_variational_2016, bishop_pattern_2006, Banfield1993ModelbasedGA, rasmussen_nite_nodate}. In PSICHE, this model captures local information as clusters within a self-contained group of observations. Multiple vGMMs are embedded into a higher-level, tree-consistent model for multi-scale learning, with more information about this nested design given in the following sections. This section introduces the notation that describes the form, optimization, and probabilistic observables of a PSICHE vGMM.

\begin{figure}
\centering
\includegraphics[width=\linewidth]{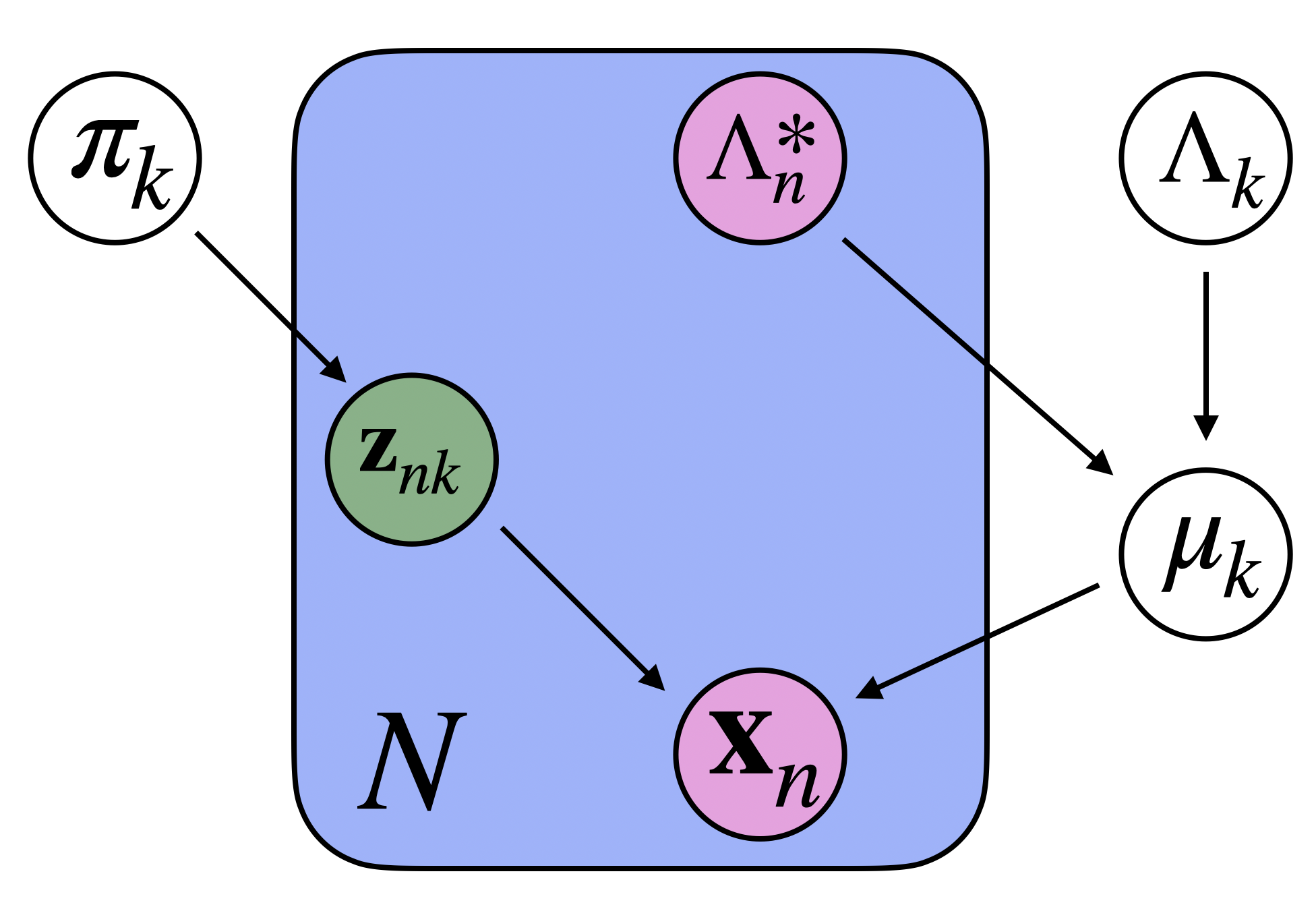}
\caption{\small Plate notation of the likelihood $p(\boldsymbol\chi | \mathbf{Z}, \boldsymbol\mu, \mathbf{\Lambda})$ in Equation \ref{eq:vGMM_LH}. Variables in the box correspond to each of the $N$ observations, as denoted by subscripts, with parameters appearing outside the box corresponding to each of the $K$ subclusters of the model. The pink nodes denote the observations from data while the green node denotes the latent, random variables governing cluster assignment. The mixture model random variables $\boldsymbol\zeta \equiv \{\boldsymbol\pi, \boldsymbol\mu, \boldsymbol\Lambda\}$ are uncolored.}
\label{fig:plate_notation}
\end{figure}

These mixture models employ a group of multivariate Gaussians to describe clusters of observations. After maximizing the joint probability of data and model parameters, the outputs are posterior distributions on the random variables for cluster description (location, shape, and multiplicity), as well as for the latent variables dictating assignments of observations to clusters. To tractably infer these posteriors, the vGMM leverages exponential-family distributions and exploits a factorized approximation of the ``true" joint distribution, variationally updating cluster description posteriors independently from those of assignment probabilities \cite{dempster_maximum_1977, attias1999inferring}. Therefore, inference amounts to iteratively and alternatingly updating the sufficient statistics of these posterior distributions, while maximizing a variational lower bound to the log model evidence. Notation for each factor of the approximated joint distribution, including variables and parameters for the likelihood and priors, is introduced below; this factorized distribution is later expanded in Equation \ref{eq:vGMM_LH} and graphically depicted in plate notation in Figure \ref{fig:plate_notation}. 

In the model likelihood, a cluster indexed by $k \in \{1, \ldots, K\}$ is represented by a Gaussian probability distribution $\mathcal{N}^d(\mathbf{X} | \boldsymbol\mu_k, \Lambda_k^{-1})$ over a set of observations  $\mathbf{X} \equiv \{ \mathbf{x}_n \}$ ($n \in \{1, \ldots, N\}$) for instance $\mathbf{x}_n \in \mathbb{R}^d$. The mean ($\boldsymbol\mu_k \in \mathbb{R}^d$) and precision ($\Lambda_k \in \mathbb{R}^{d \times d}$) parameters are random variables in the variational approach, with the posterior probabilities of the locational cluster parameters ($\boldsymbol\mu \equiv  \{\boldsymbol\mu_k \}$, $\boldsymbol\Lambda \equiv  \{\Lambda_k \}$) to be inferred by the algorithm. An additional, observational precision, $\Lambda^*_n$, is associated with each data point and included in PSICHE's vGMM likelihood, with its motivation and implementation detailed in Section \ref{sec:meas_errs}.

For each observation $\mathbf{x}_n$, the model introduces a latent random variable $\mathbf{z}_n$, which indicates a hard assignment to one of the $K$ clusters. It is represented as a discrete, $K$-dimensional, ``one-hot" vector with $z_{nk} = \delta_{k,k'}$ for a single $k$. The set $\mathbf{Z} \equiv \{ \mathbf{z}_n \}$ is modeled with categorial probability distributions $p(\mathbf{z}_n | \boldsymbol\pi) = \prod_k \pi_k^{z_{nk}}$, where $\boldsymbol\pi \in \mathbb{R}^{k,+}$ is a random variable, with $\sum_k \pi_k = 1$. This vector $\boldsymbol\pi$ can be interpreted as the ``mixing coefficients" of clusters, or the relative intensity of each cluster in the overall model. The posterior probability distribution of these random variables is inferred from data, such that the hard assignments in $\mathbf{z}_n$ are effectively replaced by continuous expectations $\mathbb{E} [ \mathbf{z}_n ]$, resulting in probabilistic assignments of \textit{responsibility} for each data point $\mathbf{x}_n$ over the $K$ clusters. The responsibility $\mathbb{E} [z_{nk} ]$ for a specific cluster $k$ can be interpreted as how much of $\mathbf{x}_n$ is explained by cluster $k$, and will correspondingly inform how a given $\mathbf{x}_n$ influences this cluster's parameters.

The data $\mathbf{X}$, latent assignment variables $\mathbf{Z}$, and collection of cluster sufficient statistics $\boldsymbol\zeta \equiv \{\boldsymbol\pi, \boldsymbol\mu, \boldsymbol\Lambda \}$ are combined in a factorized, variational distribution approximating the joint distribution $p(\mathbf{X},\mathbf{Z}, \boldsymbol\zeta)$ as $\tilde{p} = p(\mathbf{Z} | \mathbf{X})p(\boldsymbol\zeta | \mathbf{X})$. In this way, the joint probability factorizes in a hierarchical manner that facilitates inference of posterior probability distributions through a variational Bayes fixed-point algorithm. These distributions exhibit the same factorization and functional forms as their respective priors.

For example, the prior probability distribution for $\boldsymbol\pi$ -- and due to likelihood-conjugacy, the posterior -- is a Dirichlet, with $p(\boldsymbol\pi | \boldsymbol\alpha) = \text{Dir}(\boldsymbol\pi | \boldsymbol\alpha)$, where $\boldsymbol\alpha$ is a $K$-dimensional vector. For these prior distributions, $\boldsymbol\alpha_0$ are set in advance as hyperparameters, whereas the $\boldsymbol\alpha_\text{posterior}$ are inferred from data \cite{raiffa_applied_1961, 80cb41ee-7d47-3d90-ae24-2998213a9236}. Similarly, the prior and posterior distributions for the cluster location parameters have Normal-Inverse-Wishart forms, with $p(\boldsymbol\mu_k, \Lambda_k) = \mathcal{N}(\boldsymbol\mu_k | \mathbf{m}, (\beta \Lambda_k)^{-1})\mathcal{W}(\Lambda_k | W, \nu)$. These distributions factorize such that $p(\boldsymbol\mu_k)$ depends on mean parameter $\mathbf{m} \in \mathbb{R}^d$, the random precision matrix $\Lambda_k$, and precision parameter $\beta \in \mathbb{R}^+$, while $p(\Lambda_k)$ depends on scale matrix $W \in \mathbb{R}^{d \times d}$ and precision parameter $\nu \in \mathbb{R}^+$. In PSICHE, the initial values of the sufficient statistics of the prior distributions, $\boldsymbol\Psi_0 \equiv \{ \boldsymbol\alpha_0, \mathbf{m}_0, \beta_0, W_0, \nu_0 \}$, are hyperparameters, with the posterior values learned through optimization.

We augment the space of location observations in a typical vGMM with additional information: a measurement precision, $\Lambda^*_n \in \mathbb{R}^{d \times d}$ ($\boldsymbol\Lambda^* \equiv \{\Lambda^*_n \}$) and weight $\omega_n \in \mathbb{R}^+$ ($\boldsymbol\Omega \equiv \{\omega_n\}$) associated with each $\mathbf{x}_n$. The measurement precision (and its inverse, covariance) terms encode uncertainties that are potentially nonuniform among the clustering dimensions. Whereas, weights incorporate information about prior probabilities, concentration, or effective importance of individual points. The data that appears in the PSICHE vGMM likelihood is, then, an augmented set of fixed observations $\boldsymbol\chi \equiv \{ \mathbf{X}, \boldsymbol\Lambda^*, \boldsymbol\Omega \}$ with the details of how they are incorporated into the likelihood described in Section \ref{sec:unique_algo_features}.

The factorized, joint distribution of the model, including data observations, $\boldsymbol\chi$, latent random variables $\mathbf{Z}, \boldsymbol\zeta$, and hyperparameters $\boldsymbol\Psi_0$, is summarized in Equation \ref{eq:vGMM_LH}. 
\onecolumngrid

\begin{equation} \label{eq:vGMM_LH}
p(\boldsymbol\chi, \mathbf{Z}, \boldsymbol\zeta | \boldsymbol\Psi_0) = p(\boldsymbol\chi | \mathbf{Z}, \boldsymbol\mu, \mathbf{\Lambda}) \tilde{p}(\mathbf{Z} | \boldsymbol\pi) \tilde{p}(\boldsymbol\pi | \boldsymbol\alpha_0) \tilde{p}(\boldsymbol\mu | \mathbf{\Lambda}, \mathbf{m}_0, \beta_0) \tilde{p}(\mathbf{\Lambda} | W_0, \nu_0) \,.
\end{equation}
\twocolumngrid

The variational Bayes algorithm effectively minimizes the KL-divergence between the variational approximation $p(\boldsymbol\chi, \mathbf{Z}, \boldsymbol\zeta | \boldsymbol\Psi_0)$ and $\tilde{p}(\mathbf{Z}, \boldsymbol\zeta)$ by maximizing a lower bound (ELBO, Equation \ref{eq:elbo}) to the ``true" model evidence $p(\boldsymbol\chi)$ given by,
\begin{align}
\ln p(\boldsymbol\chi) &\equiv \ln \sum_\mathbf{Z} \int p(\boldsymbol\chi, \mathbf{Z}, \boldsymbol\zeta) d\boldsymbol\zeta \\
\label{eq:elbo}
&\geq \sum_\mathbf{Z}  \int \tilde{p}(\mathbf{Z}, \boldsymbol\zeta) \ln \left\{ \frac{p(\boldsymbol\chi, \mathbf{Z}, \boldsymbol\zeta )}{\tilde{p}(\mathbf{Z}, \boldsymbol\zeta)}\right\} d\boldsymbol\zeta  \,,
\end{align} 
with the factorization exhibited in Equation \ref{eq:vGMM_LH} facilitating an analytical evaluation of Equation \ref{eq:elbo}. The optimization of this model can be visualized as a variational message passing algorithm over the directed graph in Figure \ref{fig:plate_notation}. While directional arrows indicate conditional dependencies in the hierarchical graph, information during optimization travels in both directions. The initial hyperparameters, $\boldsymbol\Psi_0$, corresponding to priors of latent random variables $\boldsymbol\zeta$, set the starting sufficient statistics and expectations for the model. The information contained in the data observations, $\boldsymbol\chi$, and in the empirical sufficient statistics thereof, then percolates upwards through the graph to inform optimized posteriors summarized by the final parameters, $\boldsymbol\Psi$. Therefore, each vGMM depends only on the input data $\boldsymbol\chi$ and hyperparameters $\boldsymbol\Psi_0$. With multiple vGMM instances simultaneously incorporated into a larger probabilistic model, as in PSICHE, their initializations will occur recursively throughout the full algorithm. As opposed to treating the initializations as an amorphous challenge, requiring a consistent, arbitrary solution, we will exploit the vGMM's dependency on initial hyperparameters to both provide additional regulation during optimization and to aid in the learning of structure in the larger model, as described later in Section \ref{sec:unique_algo_features}.

\begin{figure}[h]
\centering
\includegraphics[width=\linewidth]{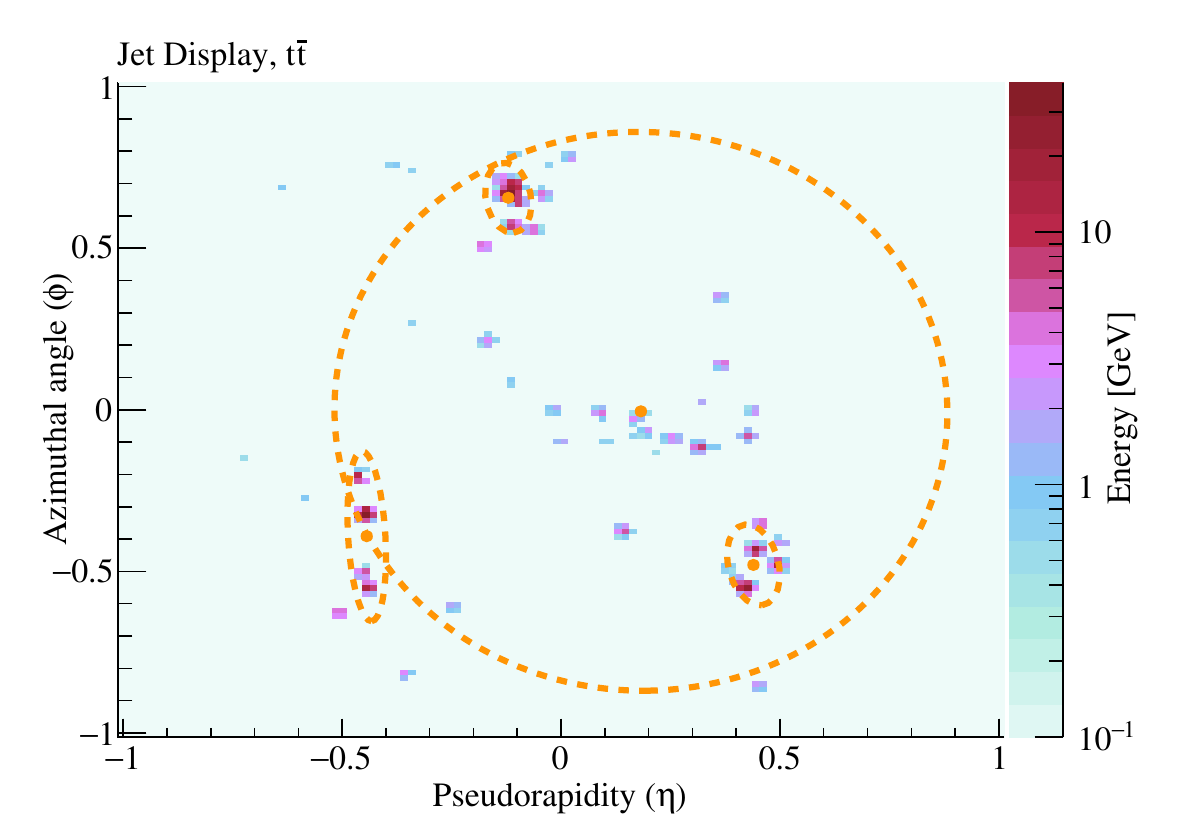}
\caption{\small Illustration of observational inputs and model output for a multivariate vGMM applied to calorimeter cell data, projected into two spatial dimensions. Each cell corresponds to a measured energy (with magnitude indicated by the $z$ color scale) with the collection clustered into four Gaussian components represented as orange ellipses.}
\label{fig:vgmm_ex}
\end{figure}

Anticipating the use of PSICHE for particle physics jet clustering, discussed throughout the remainder of this paper, the application of a vGMM to the energy depositions of a jet, reconstructed by PSICHE from particle detector energy depositions, is shown in Figure \ref{fig:vgmm_ex}. This jet corresponds to the decay of a top quark with its three products (two light quarks from an intermediate $W$ boson and a $b$ quark) successfully discovered by the algorithm as three, separate, localized clusters. Unsupervised, PSICHE also reconstructs an additional, dispersive component, with dramatically different properties than the other three clusters, that we will see later corresponds to radiation and other soft emissions in the jet environment. The learning of this specific, physically-meaningful structure is not achieved by a vGMM in a vacuum, but rather, is facilitated by PSICHE's full, probabilistic model (described in the following section) in which the vGMM is embedded.

\subsection{Bayesian Hierarchical Clustering of Probabilistic Mixtures} \label{sec:bhc_model}
While the vGMM is a probabilistic clustering algorithm in its own right, in PSICHE multiple vGMM instances are themselves incorporated into a larger, probabilistic framework through Bayesian Hierarchical Clustering (BHC) \cite{heller_bayesian_2005}. In the BHC of probabilistic mixture models, the vGMMs describe local associations while the higher-level clustering dictates which observations are included in the respective vGMMs. This nested design allows for structure at multiple scales to be simultaneously learned in a self-consistent framework.

\begin{figure}
\centering
\includegraphics[width=\linewidth]{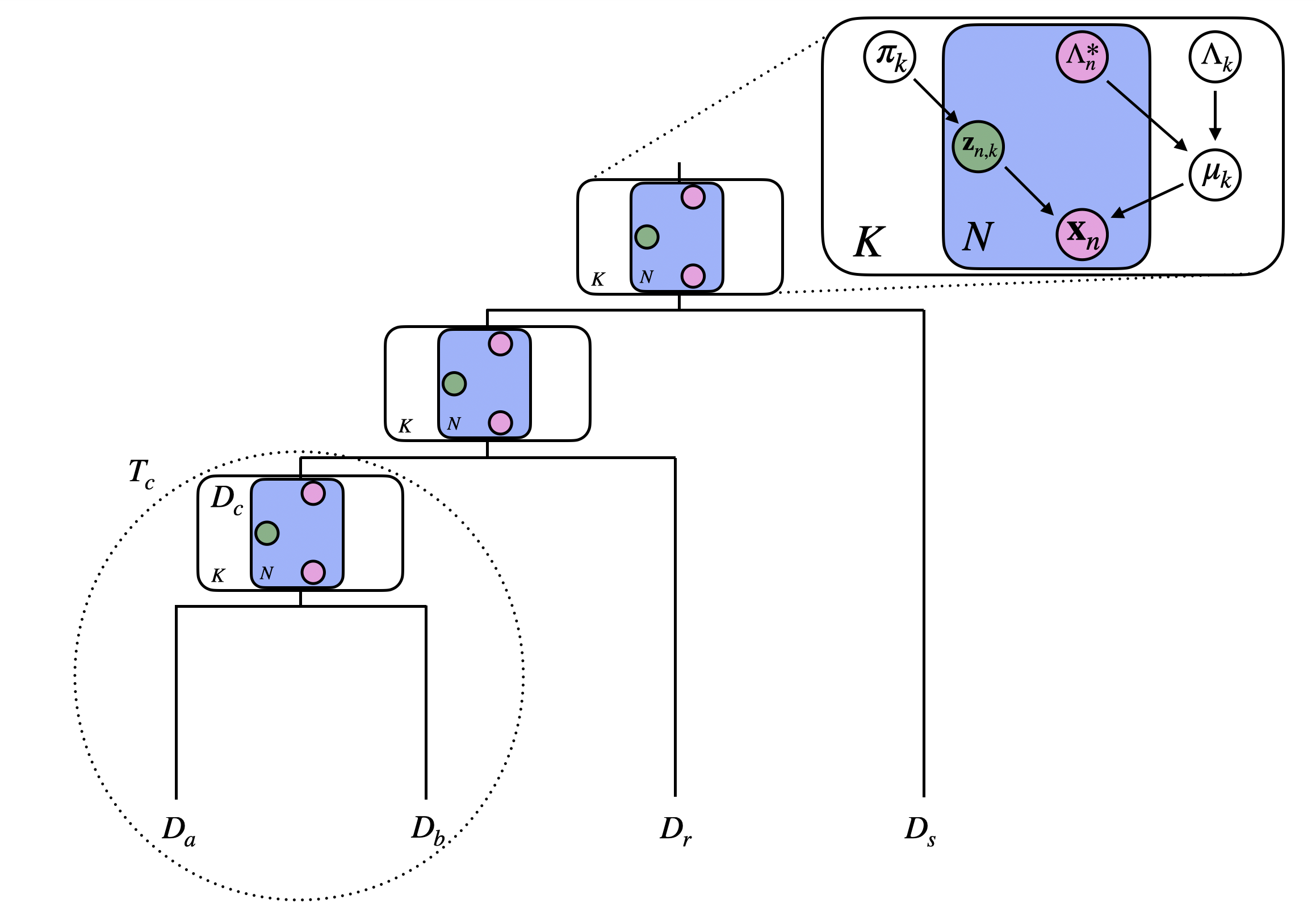}
\caption{\small A cartoon representation of PSICHE's nested design, embedding the vGMM plate notation within each higher-level, tree-consistent merge $T_i$ of datasets $D_i$. Here, the merge under consideration is between trees $T_a$ and $T_b$ (defined by datasets $\mathcal{D}_a$ and $\mathcal{D}_b$, respectively) to potentially create $T_c$ (dataset $\mathcal{D}_c$) to be used as a branch in future merges.} \label{fig:bhctree}
\end{figure}

The BHC of vGMMs is depicted in Figure \ref{fig:bhctree}. The general BHC algorithm clusters elements by recursively considering mergers of increasingly large partitions (agglomerative approach). In this scheme, the probability of merging two partitions of observations, $\mathcal{D}_a$ and $\mathcal{D}_b$, into a single dataset, $\mathcal{D}_c$ (hypothesis $\mathcal{H}_1$), is compared against the alternative of them remaining distinct ($\mathcal{H}_2$). In PSICHE, these probabilities correspond to vGMMs evaluated on partitions of observations. The tree-like structure of these merges in Figure \ref{fig:bhctree} follows from the restriction of only considering tree-consistent partitions.

For an unordered collection of data observations, the number of ways to partition them is combinatorially intractable and beyond explicit evaluation. Many of these possibilities, though, are self-evidently of low probability as they include numerous, disjointed and unintuitive associations. By restricting potential recursive merges to only consider tree-consistent partitions, the BHC algorithm implicitly ignores any assignments whose history cannot be represented in this manner, aligned with expectations of continuity associated with most clustering domains. The practical effect is that these evaluations become asymptotically tractable with a growing dataset. Depending on the clustering application, additional merge restrictions can also be incorporated. For example, in the application of jet clustering discussed throughout this paper, only merges between spatial nearest neighbors in a Voronoi tessellation are considered (Section \ref{sec:voronoi}). 

In these tree-consistent partitions, we are able to tractably define and evaluate the probabilities of the two hypotheses in a Bayesian comparison. Associated with the merge hypothesis ($\mathcal{H}_1$) is the likelihood of data $\mathcal{D}_c \equiv \boldsymbol\chi_c$ under a common model, defined as,
\begin{equation} \label{eq:bhc_H1}
p(\mathcal{D}_c | \mathcal{H}_1) = \text{exp} \left [ \sum_\mathbf{Z}  \int \tilde{p}(\mathbf{Z}, \boldsymbol\zeta) \ln \left\{ \frac{p(\boldsymbol\chi_c, \mathbf{Z}, \boldsymbol\zeta )}{\tilde{p}(\mathbf{Z}, \boldsymbol\zeta)}\right\} d\boldsymbol\zeta \right ] \, .
\end{equation}
The set of parameters and probabilities appearing in the right hand side of Equation \ref{eq:bhc_H1} corresponds to the evidence of the vGMM optimized to data $\mathcal{D}_c$, as described in Section \ref{sec:vGMM}. As such, the probabilities of merging vGMMs in the BHC are approximated by a lower bound (ELBO), with mixtures embedded into each potential merge step, as illustrated by the placement of plate notations in Figure \ref{fig:bhctree}.

The probability of the alternative hypothesis ($\mathcal{H}_2$), where the trees $T_a$ and $T_b$ remain separate, can be evaluated in the factorized form $p(\mathcal{D}_c | \mathcal{H}_2) = p(\mathcal{D}_a | T_a)p(\mathcal{D}_b | T_b)$ due to tree-consistent partitionings. While the greedy BHC algorithm will ultimately select either this hypothesis or $\mathcal{H}_1$ when constructing merges, the two coexist probabilistically in that the probabilities can be marginalized recursively over this choice. For example, the probability of a observing a dataset given a tree ($p(\mathcal{D}_c | T_c)$) can be expressed as a convex sum of hypothesis probabilities,
\begin{equation} \label{eq:bhc_H2}
p(\mathcal{D}_c | T_c) = \rho_c p(\mathcal{D}_c | \mathcal{H}_1) + (1 - \rho_c)p(\mathcal{D}_a | T_a)p(\mathcal{D}_b | T_b) \, ,
\end{equation}
where $\rho_c$ is the prior $p(\mathcal{H}_1)$. Mergers are evaluated recursively until all data observations are included in a single partition. While all the considered alternatives coexist probabilistically in the model, a specific partition is realized by ``cutting" the tree at the step when a particular merge probability becomes less favorable than not merging. The ratio of probabilities ($r^c_{1,2}$) between two hypotheses at a step $c$ can be written as,
\begin{equation}
r^c_{1,2} = \frac{ \rho_c p(\mathcal{D}_c | \mathcal{H}_1)}{  (1 - \rho_c)p(\mathcal{D}_a | T_a)p(\mathcal{D}_b | T_b)} \,,
\end{equation}
such that the tree is cut, and an optimal partitioning chosen, when $r^c_{1,2} < 1$.

The factor $\rho_c$ follows from the marginalization of a Dirichlet process prior, which effectively accounts for a potentially infinite-component mixture model \cite{heller_bayesian_2005}. Like the Dirichlet distribution in the vGMM, the Dirichlet process prior depends on its own concentration hyperparameter that dictates the probabilities of new cluster creation and the addition of points to existing clusters.

\begin{figure}
\centering
\includegraphics[width=\linewidth]{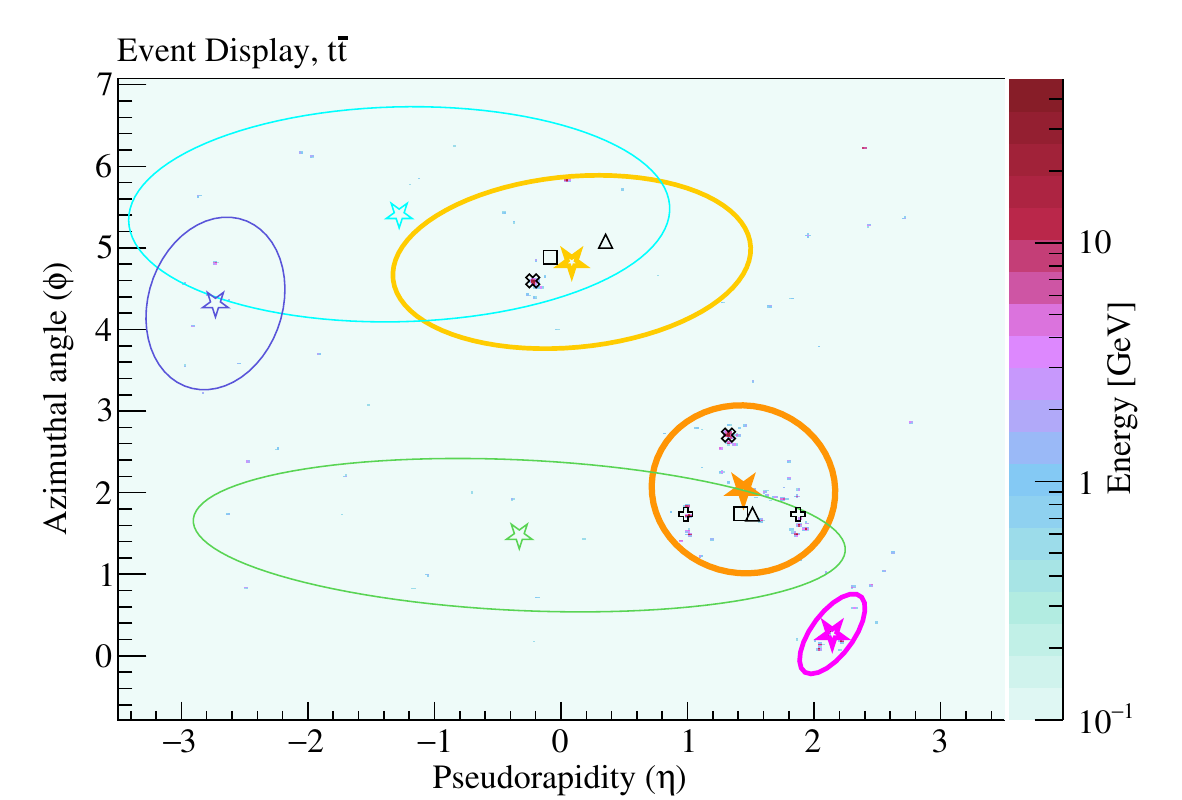}
\caption{\small PSICHE clusters (ellipses centered on correspondingly colored stars) projected in the 2D spatial plane and overlaid onto the input calorimeter energy deposits (red-to-blue colored cells with energy on the $z$ color axis). Local subclusters for each cluster are not shown for clarity and emphasis on the global structure. Black markers denote true particles of interest in this simulated decay of two top quarks (black squares); each decays to one $b$ quark (black x's) and a $W$ boson (black triangles), further decaying to either two leptons (yellow cluster, leptons not denoted) or two light quarks (orange cluster, black crosses). Further interpretations of these clusters, and their nested subclusters, in a physics context will be discussed in Section \ref{sec:nom_jet_clustering}.} 
\label{fig:bhc_ex}
\end{figure}

To continue the example of high-energy particle jet clustering, representing jets as probabilistically partitioned vGMMs allows the PSICHE BHC model to describe events with multiple jets, each with their own subclusters and properties. Figure \ref{fig:bhc_ex} illustrates the output of the full PSICHE clustering algorithm on a collection of particle emissions, where the subclusters shown earlier in Figure \ref{fig:vgmm_ex} now correspond to the substructure within a single (orange) cluster among others. PSICHE is able to identify and model these clusters with a diversity of scales and minimal prior specification as to what structure or features should emerge. Combining the BHC and vGMM components probabilistically is perhaps a necessary but insufficient condition to achieving this multi-scale learning; other important elements, like further regulation and hyperparameter decisions, are discussed in the following section.

\subsection{The Addition of General Inductive Priors} \label{sec:unique_algo_features}
Even though the building blocks of PSICHE are constructed from established models, there are several new features of this algorithm that contribute to the regularization necessary to learn meaningful structure at multiple scales. One set of these inductive priors, or biases, is obtained from data, relying on measurements to provide a gauge of importance for inputs. The other set is algorithmic, introducing mechanisms for parsimoniously constraining degrees of freedom within hierarchical, probabilistic models.

\subsubsection{Observational Measurement Precisions and Weights} \label{sec:meas_errs}
In order to accommodate experimental variations among observations, the vGMM likelihood, introduced in Equation \ref{eq:vGMM_LH}, is augmented with terms that dictate both the overall importance of an observation within the model, as well as its importance relative to others. The concept of observational importance is incorporated in PSICHE by associating data weights $\boldsymbol\Omega \equiv \{\omega_n\}$ and measurement precisions (inverse covariances) $\boldsymbol\Lambda^* \equiv \{\Lambda^*_n\}$ with all observations $\mathbf{X} \equiv {\mathbf{x}_n}$. These terms appear in the model likelihood as,
\begin{equation} \label{eq:aug_vgmm_lh}
p(\boldsymbol\chi | \mathbf{Z}, \boldsymbol\mu, \mathbf{\Lambda}) = \prod_N \left [ \prod_K \mathcal{N}(\mathbf{x}_n | \boldsymbol\mu_k, (\Lambda_k + \Lambda^*_n)^{-1})^{z_{nk}} \right ]^{\omega_n} \, .
\end{equation}

The data weights $\boldsymbol\Omega$ effectively scale the contribution from $\mathbb{E}[z_{nk}]$ when accumulating totals, expectations, and precisions in posterior parameter updates. For example, the posterior mean parameter ($\mathbf{m}_k$) of a Gaussian mixture component $k$ (when the prior precision parameter $\beta_0$ is arbitrarily small) is given by a weighted, empirical mean,
\begin{equation}\label{eq:mean_update}
\mathbf{m}_k = \bar{\mathbf{x}}_k \equiv \frac{  \sum_n  \omega_n\mathbb{E}[z_{nk}]\mathbf{x}_n }{  \sum_n \omega_n\mathbb{E}[z_{nk}]} \, .
\end{equation}
These weights also scale the effective number of data observations, with posterior precision parameters updating as, for example, $\nu_k = \nu_0 + \sum_n \omega_n \mathbb{E}[z_{nk}]$ for the inverse-Wishart component covariance. Hence, these weights provide a mechanism for the incorporation of supplementary information -- such as additional probabilities, intensities, or other relevant features -- when dictating the relative and absolute contributions of each datum. 

When observations, $\mathbf{x}_n$, correspond to many and different types of multidimensional measurements, a single, scalar weight $\omega_n$ may not be sufficient for optimally weighing relative contributions to posterior updates. For example, in the jet clustering application of PSICHE (Section \ref{sec:nom_jet_clustering}), observations are in both spatial and temporal dimensions, which can have dramatically different measurement uncertainties in both magnitude and uniformity. The inclusion of the set of measurement precisions $\boldsymbol\Lambda^* \equiv \{\Lambda^*_n\}$ into the model likelihood allows for precisely this flexibility in relative contributions of observations. 

With the addition of these factors to the model likelihood in Equation \ref{eq:aug_vgmm_lh}, the component posterior mean update of Equation \ref{eq:mean_update} becomes,
\begin{equation} \label{eq:meas_errs}
\mathbf{m}^*_k = \mathbf{m}_k + \frac{ \sum_n \omega_n  \mathbb{E} [z_{nk} ] \Lambda^*_{n} \mathbf{x}_n }{ \sum_n \omega_n \mathbb{E}[z_{nk}] (\mathbb{E}[\Lambda_k] + \Lambda^*_n) } \, .
\end{equation}
where the second term appears as a dimension- and observation-dependent weighted sum. Unfortunately, the addition of these observational precisions to the vGMM is not seamless, as the likelihood in Equation \ref{eq:aug_vgmm_lh} is no longer conjugate to the inverse-Wishart of the component posterior precisions $\Lambda_k$. As such, $\boldsymbol\Lambda^*$ is not directly propagated to this posterior. Rather, the $\boldsymbol\Lambda^*$ are only included in refining the responsibilities and relative importance of information for the inverse-Wishart update via relative weights in the empirical sufficient statistics in the manner of Equation \ref{eq:meas_errs}.

The result of this compromise is that the inverse-Wishart posterior of the random precisions $\Lambda_k$ is not made to be overly precise in that $\Lambda^*_n$ are not counted as effective observations in its update. Similarly, these adoptions are accounted for within the higher-level probabilistic model through the ELBO's dependence on these posterior parameters, with this approximate model still providing a consistent lower bound.

\subsubsection{Subcluster Initialization and Regulation \\ via Hierarchical Inheritance} \label{sec:mm_init}
A recurring question in statistical model building is that of initialization. Variational algorithms, like the vGMM used in PSICHE, are guaranteed to converge to a local optimum, but can still be sensitive to the starting position of the parameters governing cluster multiplicity, location, and shape. One route for the initialization of location and shape parameters is to employ a general seeding algorithm, like K-means clustering \cite{lloyd1982least}, or a domain-specific one, as in Ref.~\cite{mackey_fuzzy_2016}. However, these pre-processing algorithms can also be sensitive to their own initialization, with questions related to both multiplicities and locations still unresolved. For hyperparameters appearing in the exponential-family distributions of mixture components, it is not unreasonable to set them to their least informative values, such as setting the variance of a Gaussian prior to effectively infinity. Unfortunately, an analogously uninformative approach is poorly defined in the case of structural unknowns, like component multiplicity, potentially introducing sensitivity of what can be learned to these initializations.

\begin{figure}
\centering
\includegraphics[width=\linewidth]{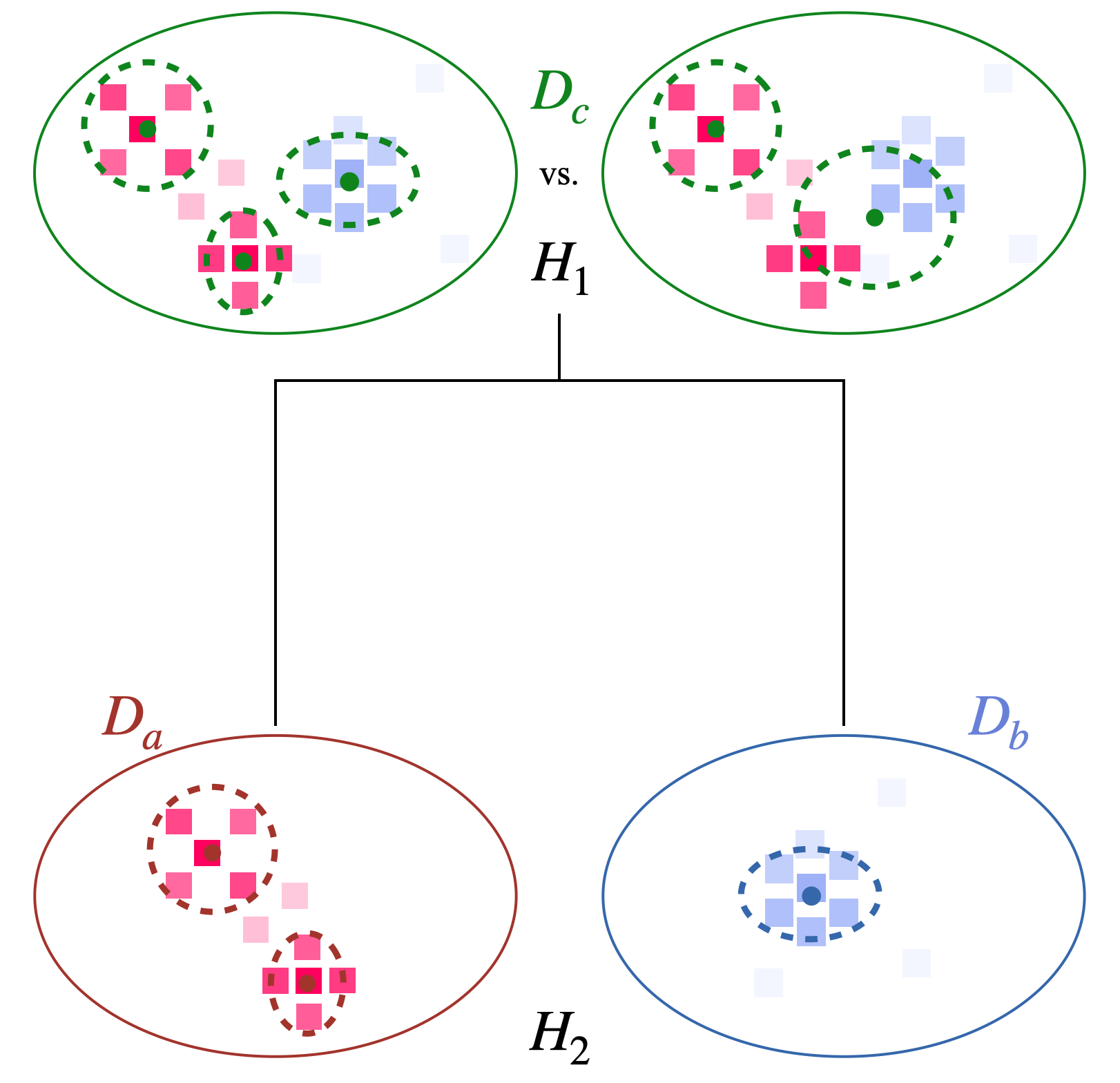}
\caption{\small An instance of a tree-consistent partition between parent clusters $\mathcal{D}_a$ (red) and $\mathcal{D}_b$ (blue) that considers the merge ($\mathcal{H}_1$, top) and separate ($\mathcal{H}_2$, bottom) hypotheses. The merged, child cluster $\mathcal{D}_c = \mathcal{D}_a + \mathcal{D}_b$ (green) inherits vGMM components from its parents and also considers comparisons of its own subclusters. Higher-level clusters are denoted by solid ellipses, vGMM subclusters are dotted ellipses, and the input observations $\mathbf{x}_n$ are squares with opacity proportional to their data weight $\omega_n$ and color matching their parent cluster.} \label{fig:subcl_merging_cartoon}
\end{figure}

With PSICHE, we break the degeneracy of this chicken-and-egg problem by using vGMMs from previous hierarchical steps within a growing BHC tree history to inform the initial conditions of subsequent mixture models. Specifically, the number of subclusters and their starting exponential-family distributions are set by the posteriors from the parent vGMMs considered in a merge. A representation of this inheritance scheme is illustrated in Figure \ref{fig:subcl_merging_cartoon}. In this considered merge, the posterior parameters of mixtures from datasets $\mathcal{D}_a$ (red) and $\mathcal{D}_b$ (blue) become the initial values for the vGMM describing dataset $\mathcal{D}_c$. Subsequently, the multiplicity of subclusters in the child vGMM under merge consideration is set initially to the sum of multiplicities of its parents. This initialization scheme directly regulates the proliferation of subcluster multiplicity throughout recursive BHC merges and introduces a restriction corresponding to a particular inductive bias: that, when merging datasets and their vGMMs, a subcluster multiplicity greater than the sum of its parts will never be probabilistically favorable. Practically, this scheme, which evolves with the PSICHE algorithm, allows mixture models to begin from a more informed position in their parameter spaces, yielding faster convergence and more intuitive and stable minima. Whether the assumption on subcluster multiplicity is sensible in all cases and domains is debatable, but here it can be desirable to temper unnecessary complexity in unsupervised learning by ensuring that the inductive, tree-consistent structure of the BHC is consistently communicated to vGMMs. 

\subsubsection{Subcluster Regulation via Merged Model Comparisons} \label{sec:subcl_parsimony}
Even with this inheritance scheme for vGMM initializations, it is not necessarily the case that mixture models will converge to global optima, particularly regarding the subcluster multiplicity degree of freedom. For example, as illustrated in Figure \ref{fig:subcl_merging_cartoon}, there may be a mixture component that is effectively redundant between two models being merged, in that the highest probability configuration of the merged vGMM may have fewer components than the sum of its parents. Unfortunately, the merged vGMM without these redundancies removed is singular, and it is not guaranteed (maybe even unlikely) that the optimization will discover, unsupervised, the most parsimonious solution.

In Figure \ref{fig:subcl_merging_cartoon}, we consider the merge of two distinct datasets $\mathcal{D}_a$ and $\mathcal{D}_b$, each with unique mixture models, into a single vGMM for dataset $\mathcal{D}_c$. Two alternative, merged vGMMs are considered at the top of the figure: in one, the multiplicity and starting conditions of the subclusters are summed from the two parent vGMMs (top left), in the other a pair of subclusters between the parents has been merged (top right). Before selecting either model as the representative vGMM of $\mathcal{D}_c$, PSICHE performs an explicit, probabilistic comparison of the two. Enumerating all potential pairwise subcluster merges, and which to include in a given vGMM, is itself combinatorially intractable for large multiplicities. PSICHE mitigates this challenge by severely restricting potential subcluster merges to only those heuristically evaluated to be favorable and feasible.

In order to quantify which subcluster pairs from parent vGMMs may have higher probabilities as a single subcluster, we define a measure of ``overlap" between subclusters $k$ and $k'$ according to the L2 inner product of the respective subcluster Gaussians, $\int \mathcal{N}_k(\boldsymbol\chi | \boldsymbol\mu_k, \Lambda_k) \mathcal{N}_{k'}(\boldsymbol\chi | \boldsymbol\mu_{k'}, \Lambda_{k'}) d\boldsymbol\chi$. All pairs are enumerated and ranked such that those with more overlap are considered first, as they are more likely to be redundant in the mixture model. 

For the highest-ranking potential subcluster mergers, the original subcluster components are removed and replaced with a single, combined component in the probabilistic comparison. Rather than re-fitting the entire vGMM with this new combined component, for all possibilities, the probabilistic responsibilities associated with each of the parent vGMMs are used to project out the contributions from any subclusters that are not currently under consideration. This means that decisions for which merges to entertain between subclusters can be made independently for a given pair. Simplified, surrogate mixtures with either one (merge) or two (not merge) components are optimized on this responsibility-projected data, and the likelihoods (or more accurately, ELBOs) are compared to choose the most probable alternative.

The fully-merged model then sets starting parameters from the posteriors of each selected surrogate, pairwise model. Returning to the question of the potential overall merger of the vGMMs of $\mathcal{D}_a$ and $\mathcal{D}_b$ into $\mathcal{D}_c$, the vGMM ELBO is again used to greedily select the most probable model. Similar to the tree-consistent partition in the BHC module of the algorithm, this subcluster merger scheme allows the overall optimization to be exposed to a combinatorially large set of configurations without actually considering the ones deemed to be unintuitive or improbable in advance. This schema further couples the cluster merge history and evolution with that of the subclusters, mirroring the same probabilistic comparisons, now unified between scales. \\

The general, inductive biases outlined in this section symbiotically coalesce in PSICHE's structure-intrinsic design. Observational precisions and weights introduce an indication of relative and absolute importance of data points, informing their vGMM component contributions. Subcluster multiplicity is regulated in preparation for optimization by inheriting substructure from parent merges, and is regulated in further optimization via a probabilistic subcluster model comparison, maintaining a pervasively parsimonious substructure model of meaningful features. In the following sections, we explore the application of PSICHE in the task of jet clustering in high energy physics. This will introduce physical meaning to both the BHC and embedded mixture models with domain-specific choices and adaptations. An instance of clustering in this context is illustrated in Figure \ref{fig:jet_3dview}, with the shape and multiplicity of clusters (jets) interacting non-trivially with mixtures of subclusters (subjets) over both spatial and temporal observations. As we will see, this learned model has real, physically-interpretable correspondences with subatomic particles and their interactions.

\section{Probabilistic Jet Modeling and Reconstruction with PSICHE} \label{sec:nom_jet_clustering}
All of the aforementioned probability models and general inductive biases of PSICHE come together to create an algorithm whose design is resonantly-matched to the task of jet clustering in the field of high energy physics. Jet clustering, or jet reconstruction, aims to group collimated showers of particles produced in high energy collisions, interpreting those clusters as self-contained objects with emergent substructure.

\begin{figure}[h]
\centering
\includegraphics[width=0.87\linewidth]{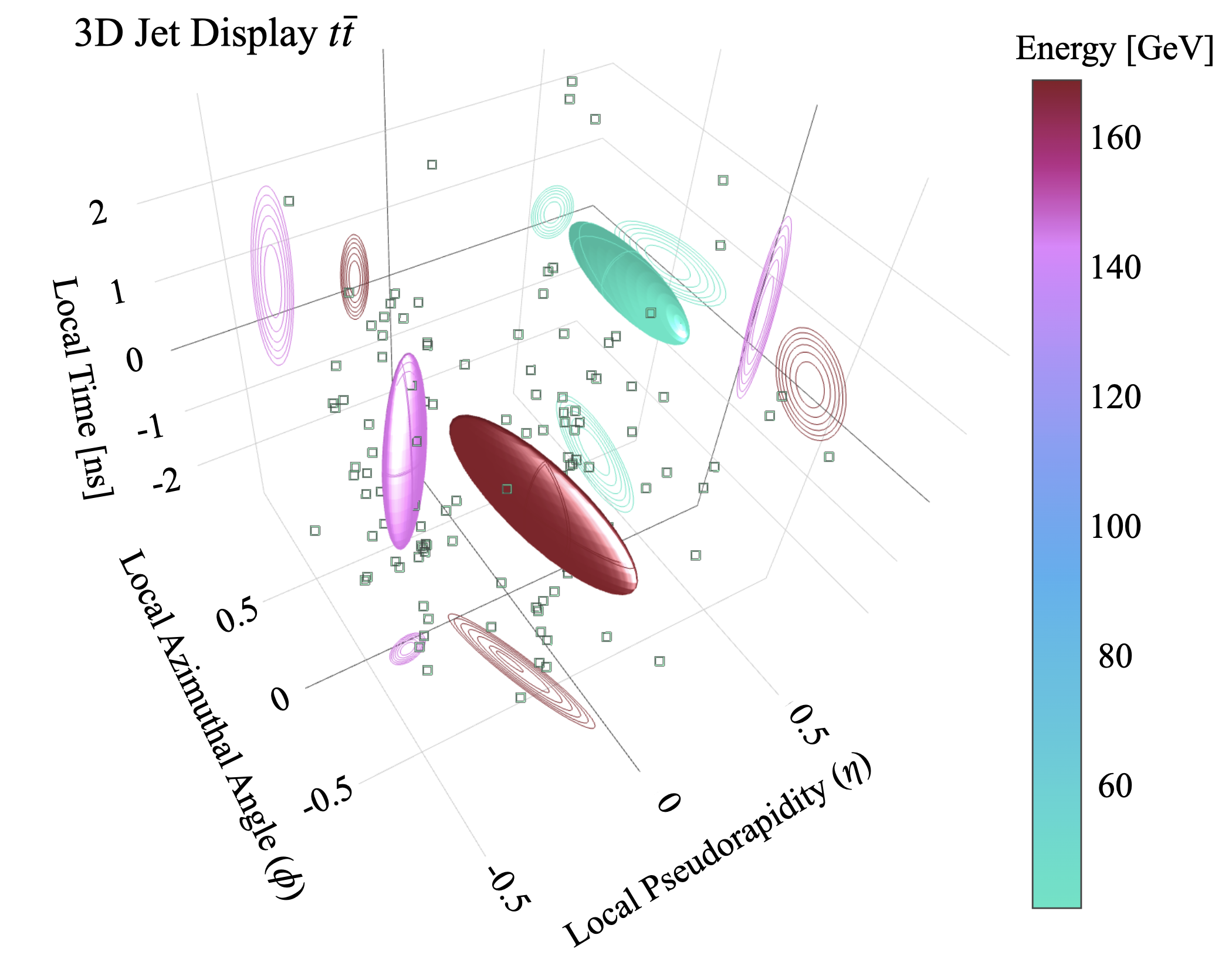}
\caption{\small Example of clustered PSICHE jet and corresponding substructure in its full dimensionality. Calorimeter cells (green squares) represent the energy deposits of particles in space and time and colored ellipsoids correspond to vGMM components, with the color scale denoting the subcluster energy. Every component is projected onto each 2D plane in this 3D clustering space for multidimensional views of subcluster diversity of shape.}\label{fig:jet_3dview}
\end{figure}

An example result of PSICHE's jet clustering in its full dimensionality is depicted in Figure \ref{fig:jet_3dview}. Relevant jet scales are learned simultaneously via PSICHE's hierarchical embedding, with a conversion of probabilistic models to physical jets outlined in Section \ref{sec:jet_features}. 

While the underlying principles of this multi-scale algorithm hold generally, several features of PSICHE were developed with this domain-specific application in mind. These features are enumerated in Section \ref{sec:domain_specific}, with an emphasis on PSICHE's structure-intrinsic design. Finally, PSICHE jets reconstructed from simulated datasets are presented in Section \ref{sec:simreco}, while a demonstration of PSICHE's novel features in a particle physics context is presented in Section \ref{sec:novel_features}.

\onecolumngrid

\twocolumngrid

\subsection{From Probabilities to Jets} \label{sec:jet_features}

We apply PSICHE's probabilistic clustering model, detailed in Section \ref{sec:algo_details}, to the problem of jet clustering in high-energy particle physics. The BHC clusters now correspond to physically-meaningful groupings of sub-atomic particle emissions and depositions -- jets -- following from the collimated decay of a single particle whose identity and properties may be of interest. Within this shower there may be further jet substructure, with intermediate partonic states producing local modes of intensity, correlations between emissions from QCD color connections, and patterns in particle multiplicity and type. We hope to capture this information using the embedded vGMM subclustering inside a jet, effectively modeling the probability density of jets themselves. Beyond the spatial coordinates in which jet clustering is typically performed, we will also augment the space of experimental observations with a new dimension -- time -- and incorporate experimental estimates of measurement precision into the framework so that clustering can be performed uniformly over these multiple dimensions. This requires a translation between the probabilities, components, and parameters of PSICHE and the physics of jets. 

Jet clustering, and the particle detectors which facilitate it, conventionally operates in a hermetic, cylindrical coordinate system, with the $z$-axis defined along colliding particle beams. The three spatial dimensions are condensed into two, $\eta$ and $\phi$, where the former is the pseudorapidity (relativistic rapidity in the massless limit) and uniform along the $z$-axis. The latter is an azimuthal angle indicating the coordinate in the plane transverse to $z$. While detector elements, and hence measurements, are distributed in 3D space throughout the detector, by working in $\eta$ and $\phi$ we are effectively considering a projective geometry of particle momenta, corresponding to the interaction point (or vertex) where emissions of interest are produced. When adding timing information to this space, a similar projective convention is used, whereby times are projected to this vertex location (now in 4D) for comparison and evaluation.

Within this observation space, what is being clustered into jets could include calorimeter cell energy depositions, ionization signals of charged particles, trajectories, or full particle candidates already reconstructed from lower-level inputs. We assume each input element to be clustered will have four features: i) location in pseudorapidity ($\eta_n$), ii) location in azimuthal angle ($\phi_n$), iii) location in time ($t_n$), and iv) recorded energy ($E_n$). PSICHE will cluster these inputs into jets corresponding to exclusive, or hard, assignments of the BHC partitions, with further jet subclustering performed by the embedded vGMM, allowing for probabilistic, or soft, assignments between subclusters. An instance of this three-dimensional, multi-scale clustering was illustrated in Figure \ref{fig:jet_3dview} for a single jet with a three-prong substructure.

Each clustering input is defined by its spatiotemporal location in the detector $\mathbf{x}_n = (\eta_n, \phi_n, t_n)$ and its energy $E_n$ such that (with the convention that these emissions are massless) its corresponding four-vector is $\mathbf{p}_n \equiv \frac{E_n}{\cosh(\eta_n)}(\cosh(\eta_n), \cos(\phi_n), \sin(\phi_n), \sinh(\eta_n))$. Non-trivial element masses can be incorporated into these four-vectors without loss of generality. From these inputs, jet and subcluster four-vectors are defined by sums over their associated emissions as,
\begin{align}
\label{eq:jet_fourvec}
\mathbf{p}_\text{jet} &= \sum_{n \in jet} \mathbf{p}_n \\
\label{eq:subcl_fourvec}
\mathbf{p}_k &= \sum_{n \in jet} \mathbb{E} [z_{nk} ] \mathbf{p}_n \,,
\end{align}
where $\mathbb{E} [z_{nk} ]$ corresponds to a responsibility, or convex weight, of input $\mathbf{x}_n$'s association with subcluster $k$.

Perhaps the most important translation between the PSICHE probabilistic model and jet physics is the choice of weights, $\omega_{n}$, associated with each of the observational input elements $\mathbf{x}_n$. As described in Section~\ref{sec:meas_errs}, these weights represent intensity, concentration, or additional probabilistic information of inputs, providing a scale of relative and absolute importance. In this clustering application, we adopt the convention that these weights should be proportional to energy, with $\omega_n = \omega \times E_n$. The constant of proportionality, $\omega$, is a tunable hyperparameter within the PSICHE framework, with considerations and implications of its specification discussed in Section~\ref{sec:hyperparams}. Irrespective of the value of $\omega$, the choice of setting these weights proportional to energy is crucial to the jet clustering application. It reflects that electronic devices measure energy $E$ in counts both physically, counting secondaries from showers within particle detectors, and also electronically, measuring amplitudes with ADC counts. Linear energy-dependent scaling of responsibilities preserves the four-vector summation over emissions that allows jet- and subcluster-level kinematic observables to be calculated in Equations \ref{eq:jet_fourvec} and \ref{eq:subcl_fourvec}, respectively. This correspondence between energy and responsibility means that energy conservation in the decays and in the interactions of particles implies weight, or probability, conservation in both the physical diffusion of energy into different and smaller elements and in the algorithmic agglomeration during the clustering itself. 

This linear energy-weighting scheme also has desirable theoretical implications, in that it provides infrared (IR) and collinear (C) safety \cite{marzani_looking_2019} to clustered jets. The former tests clustering-insensitivity to the addition of infinitely soft emissions, while the latter probes for robustness against deviations due to collinear splittings of emissions. At the jet-level, the PSICHE BHC partitions are IRC safe in the same sense (and for similar reasons) to many conventional jet clustering algorithms. With partition merges first considering local nearest-neighbors, collinear emissions are immediately combined, while IR emissions do not affect clustering as they have zero probability, such that the model effectively disincludes them. At the subcluster level, the notion of IRC safety is more subtle, particularly in a probabilistic model. The addition or splitting of subclusters, no matter how unchanged the total weight is, changes the mixture model in terms of subcluster multiplicity, and hence model probability (via entropy contributions to the vGMM ELBO). Due to probabilities being proportional to energies in PSICHE, the optimal model configuration remains the same, but convergence to that optimum can depend on initialization, and potentially multiplicities of even IR emissions. We observe that PSICHE's dedicated subcluster merging scheme, mirroring BHC jet merges, provides explicit IRC protection at the subcluster level by immediately and consistently combining collinear inputs and efficiently removing redundant mixture components. This was tested through the artificial addition of multiple ``ghost" subclusters, each with a numerically small contribution to their vGMM, in the initialization of every mixture model at each hierarchical merge step, where no ghost subclusters were observed to remain.

\subsection{ Inductive and Bayesian Priors for \\ Jet Clustering} \label{sec:domain_specific}
PSICHE's structure-intrinsic, probabilistically-based design provides avenues for domain adaptation, for example through the observational weights $\omega_n$ and measurement precisions $\Lambda^*_n$. Careful definition of these elements from first principles naturally leads to desirable algorithmic traits, like IRC safety in a high energy physics context, rooted in probabilistic interpretations. These choices act as qualitative priors, or inductive biases, alongside the quantitative Bayesian priors of the variational model, relying on hyperparameters that can also be set with domain-specific choices. In PSICHE jet clustering, there are several important considerations in both selecting hyperparameter values and in the treatment of the observation space of measurements being clustered. 

\subsubsection{Local Jet Observable Transformations} \label{sec:telescoping}
Cluster locality -- defined as self-containment -- at any scale is an important consideration in many applications, including jet reconstruction. Algorithmically, this property follows from that of the explicit or implicit metric used to evaluate distances between elements in determining their association. As cluster assignments are built relationally, other questions related to jet locality can also appear. Should the definition of a jet, and what elements are assigned to it, exist independently of other jets, or only relationally? In PSICHE, elements are assigned to jets unambiguously and exclusively, meaning that a jet's substructure mixture model is defined independently of any other jets or their properties. This also means that observations, when associated with a particular jet, need not always be represented in the same basis or coordinate systems.

Strictly speaking, PSICHE jet clustering measures relational distances via relative probabilities, such that emissions are assigned to jets that can better describe them in their associated probabilistic mixture model. To further improve these assignments, we define a local jet reference frame and coordinate system, and project experimental observations into it when evaluating a given jet. Following from the nature of the projective geometry in which jet clustering is performed, we identify the jet origin as the empirical, energy-weighted mean of observations $\mathbf{\bar{x}}_{\rm jet}$ associated with the jet. Each of the observations is then adjusted to this origin before evaluating the vGMM. For the spatial coordinates $\eta$ and $\phi$, this means not only translation to that origin, but also a geometric projection that nonlinearly transforms locations to a new domain. The mean jet location defines a jet axis, $\hat{n}_{\rm jet} = \{ \bar{\eta}_{\rm jet}, \bar{\phi}_{\rm jet} \}$, and $\eta$ and $\phi$ observations (with the former transformed into a polar angle) are rotated into a local jet coordinate system $\eta \rightarrow \theta^{'}~, \phi \rightarrow \phi^{'}$. Here, $\theta^{'}$ and $\phi^{'}$ are perpendicular polar-angle axes with $\bar{\eta}_{\rm jet} \rightarrow 0, \bar{\phi}_{\rm jet} \rightarrow 0$.

There are several complications when modeling physical emissions captured by a hermetic, projective detector with multidimensional normal distributions, arising from periodic coordinates, limited instrumental acceptance, and physical intuition. As the associated distributions wrap around periodic dimensions, the resulting model will assign probability to observations multiple times from each cluster, while probability will also be assigned to space where no observation could ever be recorded. Physically, when clustering emissions in a projective momentum space, it is also undesirable for elements traveling in opposite directions to have a nonzero probabilistic overlap. We address these considerations by optimizing the vGMM in a domain representative of these constraints. To this end, observations' local jet coordinates $\theta^{'}$ and $\phi^{'}$ are further transformed by projecting them onto the 2D plane normal to $\hat{n}_{\rm jet}$ according to
\begin{align}
  \eta^{\rm jet} &= \tan \theta^{'} \\
  \phi^{\rm jet} &= \tan \phi^{'}~,
\end{align}
 where $\eta^{\rm jet}, \phi^{\rm jet} \in (-\infty, \infty)$ are no longer angles, but coordinates in a local Euclidean tangent space. Any observations in the direction opposite $\hat{n}_{\rm jet} $ ($|\theta^{'}|, |\phi^{'}| \geq \pi / 2$) are not mapped into this local coordinate system and cannot be assigned to the jet. 

The adoption of this local mapping enforces the physical, inductive bias of projective momentum-space clustering in PSICHE jets, with the effective distance from the central jet axis growing infinitely large as emissions become perpendicular to this axis. Even for a particle detector uniform in cylindrical or spherical coordinates, observations are warped non-uniformly in $(\eta^{\rm jet}, \phi^{\rm jet})$, effectively focused on the central jet axis. Practically, this reduces the (negligible) probabilistic overlap between the vGMMs applied to different data partitions and diminishes sensitivity to physical detector boundaries.

Most importantly, these transformations facilitate the multi-scale learning of PSICHE, in that they break probabilistic degeneracies between embedded mixture models. When the priors on location parameters in the vGMM are set weakly, a modestly isolated group of observations can have little probabilistic preference for which jet it is associated with, as it is equally well-described in either vGMM. Local jet transformations of coordinates break this degeneracy, with subclusters and their underlying emissions assigned to the jet with the best-suited perspective. As the jet axis is, in turn, defined by the emissions assigned to it, the PSICHE algorithm is variationally adapting these coordinates during each step of the clustering algorithm. 

This transformation can be interpreted as PSICHE jets observing emissions through a nonuniform aperture, with jets becoming increasingly ``blurry'' at their boundaries. The mixture model is applied on top of this underlying space, with subcluster evolution then coupled to the jet it is associated with. The result of incentivizing locality in this manner is interpretable and self-contained jets.

\subsubsection{Proximity-based Hierarchical Merges and \\ Computational Complexity} \label{sec:voronoi}
PSICHE jets are defined by the association between elements found with the Bayesian Hierarchical merges, whose enumerations are typically a computational bottleneck in algorithms, potentially requiring consideration of every pairing of elements. To consider these merges, a recursive, agglomerative tree must evaluate which potential merger to consider at each step, choosing the most probable.

The solution adopted here, as in standard jet clustering algorithms, is to limit element comparisons to geometric nearest neighbors. At each step in the BHC clustering a Delaunay triangulation \cite{Del34} of spatial observations is constructed in ($\eta, \phi$) coordinates, and is implemented with its dual graph, the Voronoi tessellation \cite{voronoi1908nouvelles}, in a subroutine based on that of \texttt{FastJet}~3.4 \cite{cacciari_dispelling_2006}. Hierarchical merges between partitions of observations are only considered among nearest-neighbors in this graph.

Practically, this heuristic reduces the asymptotic scaling of PSICHE from the BHC characteristic $N^2$ to order $N \mathrm{log} N$, with a milder quadratic dependence on the number of subclusters rather than total elements. Physically, this imposes an inductive bias on jets beyond probabilities of vGMMs and observable transformations. While the distance metric used in the Delaunay triangulation is not the same as that used in the PSICHE hierarchal mergers, the uniformity of spatial measurement precision largely mitigates the difference, as nearest-neighbors will be nearly identical in the different coordinate systems described previously. The time coordinate is absent from the nearest-neighbors evaluation, as its potentially nonuniform measurement precision breaks this correspondence. Ultimately, this spatial nearest-neighbor restriction is a refinement of the BHC tree-consistent partition, with jets forced to be spatially-continuous and distinct, while allowed to have variations and overlaps in time, if probabilistically favorable.

\subsubsection{Jet Interpretability with \\ Bayesian Prior Hyperparameters} \label{sec:hyperparams}
There are a variety of hyperparameters to set in both the PSICHE BHC and vGMM modules. Many general methods exist to find an ``optimal" set of hyperparameters for a given application, including Pareto optimization, likelihood ratios for model comparisons (as in general BHC~\cite{heller_bayesian_2005}), bootstrapping techniques, and brute force grid searches. With the frequent appearance of exponential family distributions throughout PSICHE, many of these hyperparameter decisions can be reasonably inferred from first principles, and used to enforce additional domain-specific inductive biases. These considerations not only motivate values for hyperparameters, but also imbue them with physical, as well as informational, interpretability. 

The value of any variance-related hyperparameter, like the Wishart prior scale matrix $\mathbf{W}_0$, is selected to match the expected spatial granularity and time resolution of the simulated detector. This parameter can be interpreted as the minimum resolution of each component in the mixture model, using this prior to avoid singular, point-like solutions. Similarly, the measurement precisions $\Lambda^*_n$ described in Section \ref{sec:meas_errs} are set to the spatial and temporal precisions of each simulated measurement to incorporate experimental uncertainties.

The hyperparameter for the location mean in subcluster local priors is chosen such that $\mathbf{m}_0 = \bar{\mathbf{x}}_{\rm jet} = \{0,0,0\}$, corresponding to the origin of the local jet coordinate system. This prior, transformed into the original observation space, is naturally interpreted as a jet's location. Beyond the local transformations of observations into this frame, $\mathbf{m}_0$ further enforces jet locality and self-consistency. 

Precision parameters $\boldsymbol\alpha_0, \beta_0, \nu_0$, as well as the concentration hyperparameter for the BHC Dirichlet process prior, correspond to multiplicity of observations and are set to relatively small, yet numerically stable values. This choice represents our lack of prior information, with the observed data predominantly dictating the inference of posterior values~\cite{patrick_breheny_wishart_2013}. These small priors correspond to no pseudo-observations of components prior to seeing data, which is the case when the clustering is initialized.

The transfer factor $\omega$ governing the translation of energies into observation weights is perhaps the most consequential, and domain-dependent, parameter in PSICHE. In general, it indirectly sets a minimum scale -- avoiding a hard cut-off as seen in deterministic methods -- by imbuing the concept of ``one" with meaningful units. Not only do these weights convey relative importance of {\it observations}, but the hyperparameter that defines them determines the relative importance of {\it scales} to resolve; with the intimate relationship between energy and probability, discussed in Section~\ref{sec:jet_features}, $\omega$ defines a probabilistic lower bound, encouraging PSICHE to resolve kinematic scales above it.

For example, if $\omega$ is larger-than or comparable to the masses of particle resonances we would like to study, the model will likely fail to resolve any partonic substructure, favoring a monolithic view of the jet.
% favoring the hard partonic substructure as will be explored in Section~\ref{sec:jet_substructure}.
If $\omega$ is set too small, the subclustering models may trivially identify the scale of inputs, individual particles, showers, or detector cells. For the examples described throughout this paper we adopt $\omega = 4~\text{GeV}$. This is sufficiently above the zero suppression threshold of the simulation (only reconstructing calorimeter cells with $E \geq 0.5$ GeV), avoiding subcluster proliferation well-below the mass and momentum scale of boosted $W$ bosons and top quarks. This particular value was selected from a heuristic scan and observed to be in a region of stable algorithmic behavior. Irrespective of domain, $\omega$ has a generic interpretation of $\omega \sim \frac{\mathrm{units}}{\text{\# of observations}}$. This hyperparameter is the bridge between the purely informational, mathematical interpretation of data and the observations' physical meaning.

While the specification of $\omega$ introduces a dependency on an external scale input to PSICHE jet clustering, it differs from cone size parameters in other clustering algorithms in that it only, softly, sets a lower bound on jet and subcluster scale. The jets, and their substructure, that emerge from the algorithm will typically be very different, allowing PSICHE to dynamically capture a wide range of jet physics.

\clearpage
\subsection{Jet Clustering in Time: \\ Simulation and Implementation} \label{sec:simreco}
Experimental realities in the collection of data are incorporated in PSICHE via mechanisms such as the observational weights and precision terms. When applying PSICHE to jet clustering in high energy physics, synthetic datasets were simulated, incorporating many realistic features of both particle detectors and quantum mechanical processes. Notably, timing information is also incorporated as an experimental observable in PSICHE, necessitating the further simulation of these measurements and their resolutions. Clusters are then three-dimensional probabilistic mixture models of jets and their substructure, with a colorful example of top-quark decay shown in Figure \ref{fig:jet_3dview}. PSICHE identifies stable, expressive, and interpretable jets within this probabilistic framework, as evidenced by basic kinematic and time-based observables discussed below and the more complex emergent behavior described in subsequent sections. 

Proton-proton collisions are simulated with the \Py~8.3 event generator \cite{bierlich_comprehensive_2022} at a center-of-mass energy of $\sqrt{s}  = 13$ TeV, corresponding to Run II LHC energy. Various phenomena arising from these high energy collisions, including dijet and electroweak boson production, are simulated at leading order with initial and final state radiation. Hadronization of color-charged particles leverages the widely-used Lund string model \cite{SJOSTRAND1984469, Andersson:1983jt} in \Py.

Particle interaction and reconstruction effects are represented in an idealized simulation of a discretized, cylindrical calorimeter, whose geometry resembles that of the CMS Electromagnetic Calorimeter \cite{acosta_cms_2006, cms_collaboration_cms_1997}. In the ``Calo Reco" reconstruction, showers are probabilistically simulated as normally distributed across the calorimeter cells with a constant Molier\`{e} radius of a half-cell width. For a ParticleFlow \cite{ParticleFlow_2017} (PF)-like reconstruction (``Particle Reco"), particle energy and momenta are convoluted with experimental resolutions, without explicit shower simulation. Regardless of the input, both have the energy and location information necessary for jet clustering with PSICHE. 

Similarly, synthetic timing measurements of emissions are simulated, including reconstruction and resolution effects. Variations in the time-of-arrival of particles can be caused by the intrinsic spread in the luminous region of interactions, additional interactions, differences in travel path lengths, and inherent material properties. The luminous region, or beam-spot, is simulated to have primary vertices (PVs) with normally distributed variations in $z$-location and time, with $\sigma_z = 30$~mm and $\sigma_t = 100$~ps, respectively. Additional interactions, called pileup (PU), are included in certain samples to represent other proton-proton collisions in the same bunch crossing. The number of PU interactions per event is sampled from a Poisson distribution, and the spatial and time coordinates follow the same beam-spot spread and are overlaid onto the hard scattering, or event of interest. 

All stable particles produced in interactions are propagated from their production vertex to the detector face. The path lengths of neutral particles are calculated with straight-line trajectories, whereas charged particles follow helical trajectories due to the assumed 4 T solenoidal magnetic field within the detector volume. In the examples presented in this paper, charged particles originating from pileup vertices are not reconstructed, reflecting our ability to filter these contributions when PU vertices are readily identifiable using tracking detector elements. Conversely, neutral particles from PU interactions are treated (incorrectly) as if they came from the event PV when reconstructing their time-of-flight, capturing a realistic experimental limitation.

Calorimeter energy and timing measurements, including experimental uncertainties, are simulated cell-by-cell. The resolutions for both energy $E$ and time $t$ can are modeled as $\sigma^2(x) = \left( \frac{ N_x  }{ E } \right) \oplus \left( \frac{ S_x }{ \sqrt{E} } \right) \oplus C_x$, where $\sigma(x)$ is the relative uncertainty for $x = E, t$. The coefficients appearing in these parameterizations are chosen to roughly correspond with existing detectors, distinctly for energy and time~\cite{cms_collaboration_time_2010, cms_collaboration_performance_2024}. These types of experimental uncertainties are not only included in simulation, but also incorporated into PSICHE as measurement precisions $\Lambda^*_n$ in its likelihoods, with values matching their corresponding resolutions. 

\begin{figure}[!t]
\centering
\includegraphics[width=\linewidth]{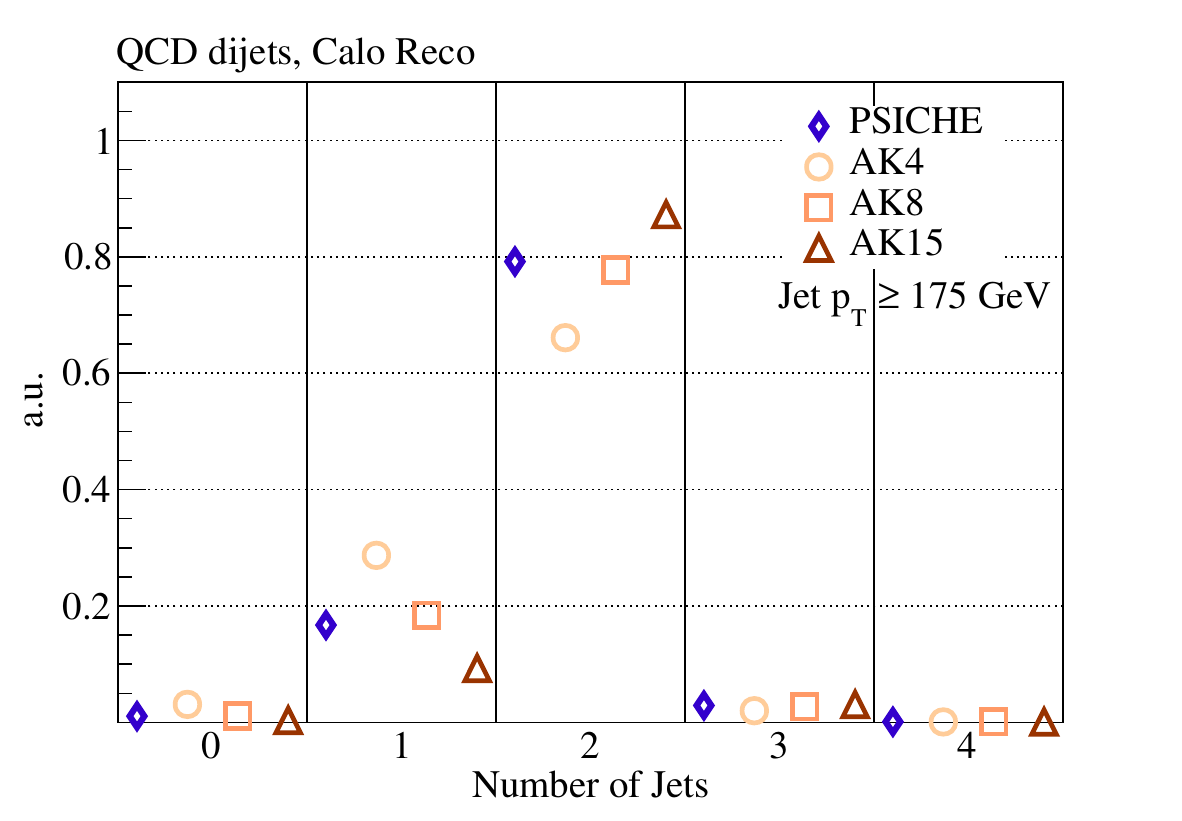}
\caption{\small Multiplicity of high-$p_T$ jets for simulated dijet events with a calorimeter cell reconstruction. Multiplicities of PSICHE jets (blue diamonds) are compared to those of AK4 (yellow circles), AK8 (orange squares), and AK15 (brown triangles) jets.}\label{fig:njets} 
\end{figure}

PSICHE's jet clustering is benchmarked against the commonly-used, inclusive anti-$k_T$ (AK) \cite{cacciari_anti-k_t_2008} algorithm, implemented in the \texttt{FastJet}~3.4 package \cite{cacciari_dispelling_2006, cacciari_fastjet_2012}. AK jets are reconstructed at various radii, or cone sizes, ($R = 0.4, 0.8, 1.5$) to capture different jet scales and phenomenology and to provide a familiar reference for comparison. A relatively simple dijet topology, characterized by two jets recoiling off of each other, is used to establish PSICHE's baseline performance in the absence of intricate jet substructure or kinematic features. Figure \ref{fig:njets} shows that PSICHE reconstructs the correct multiplicity of these dijet topologies as frequently as the AK jets.

\begin{figure}[!h]
    \centering
\includegraphics[width=0.40\linewidth]{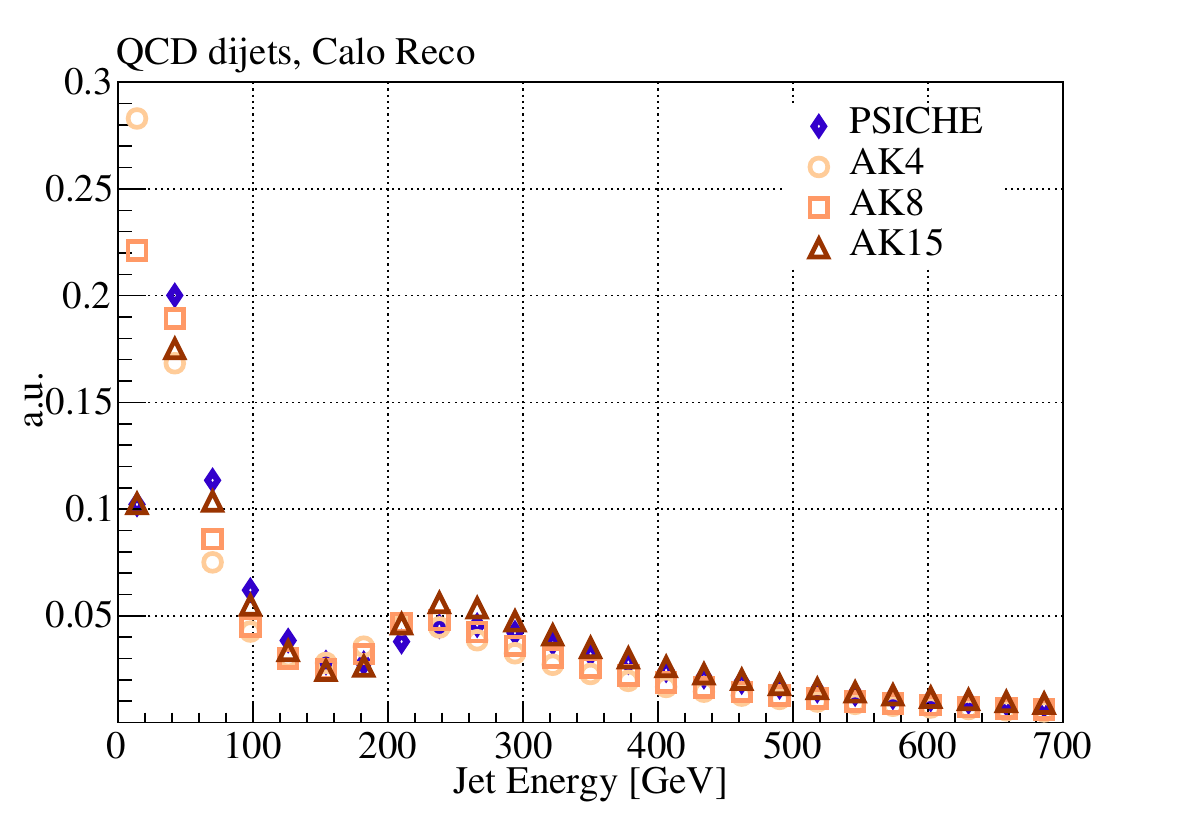}
   \hspace{1cm}
   \includegraphics[width=0.40\linewidth]{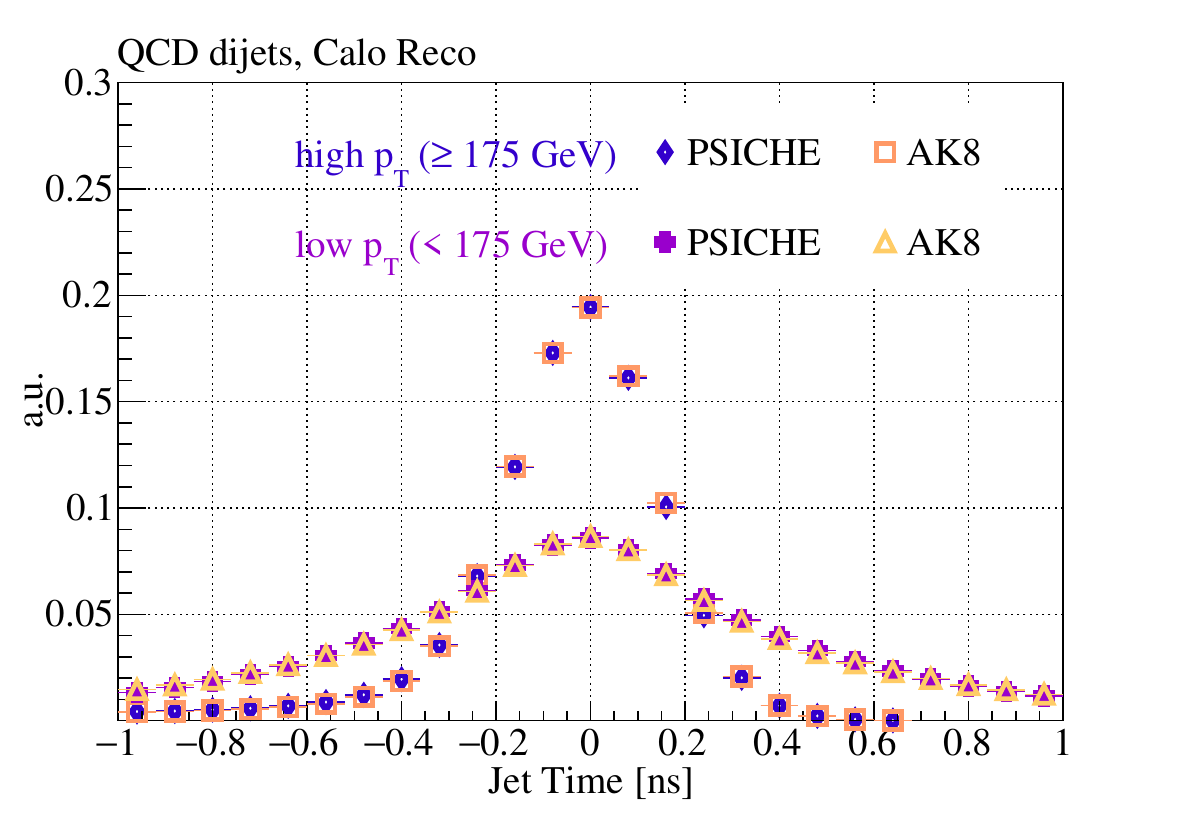}

\includegraphics[width=0.40\linewidth]{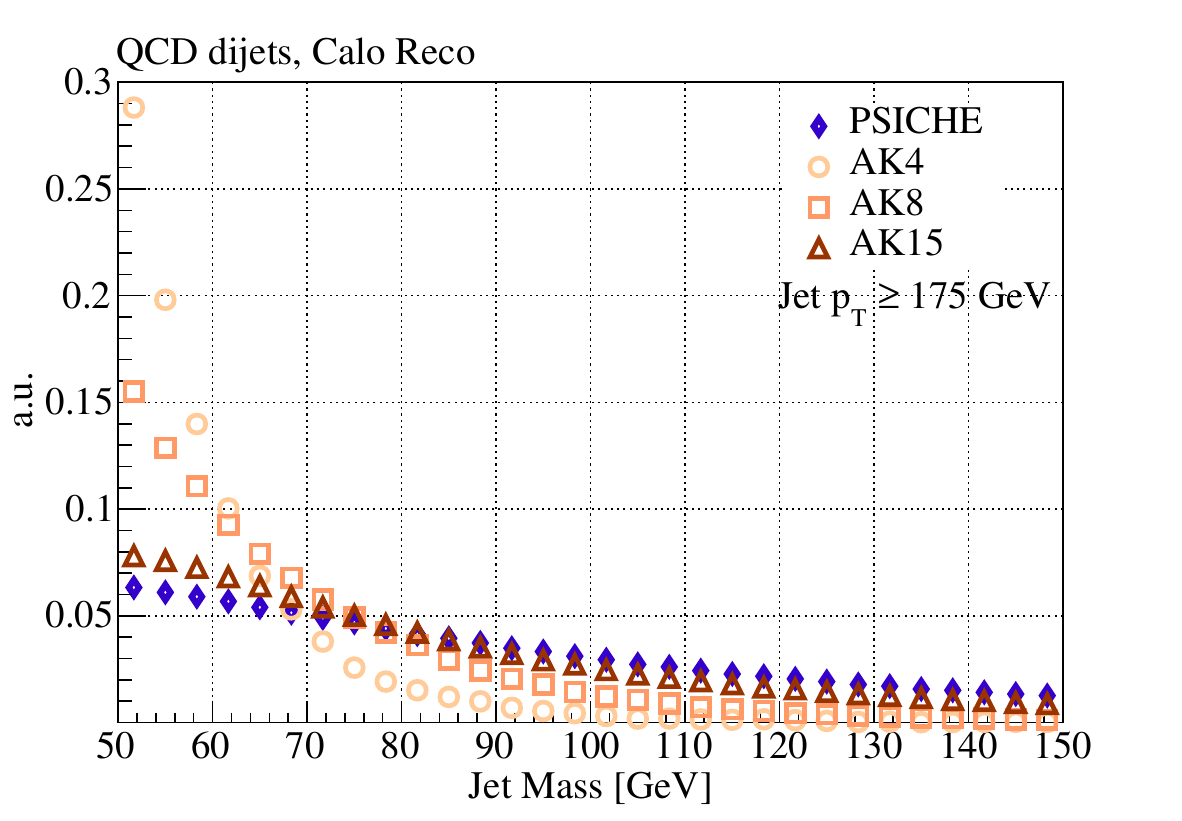}
\hspace{1cm}
\includegraphics[width=0.40\linewidth]{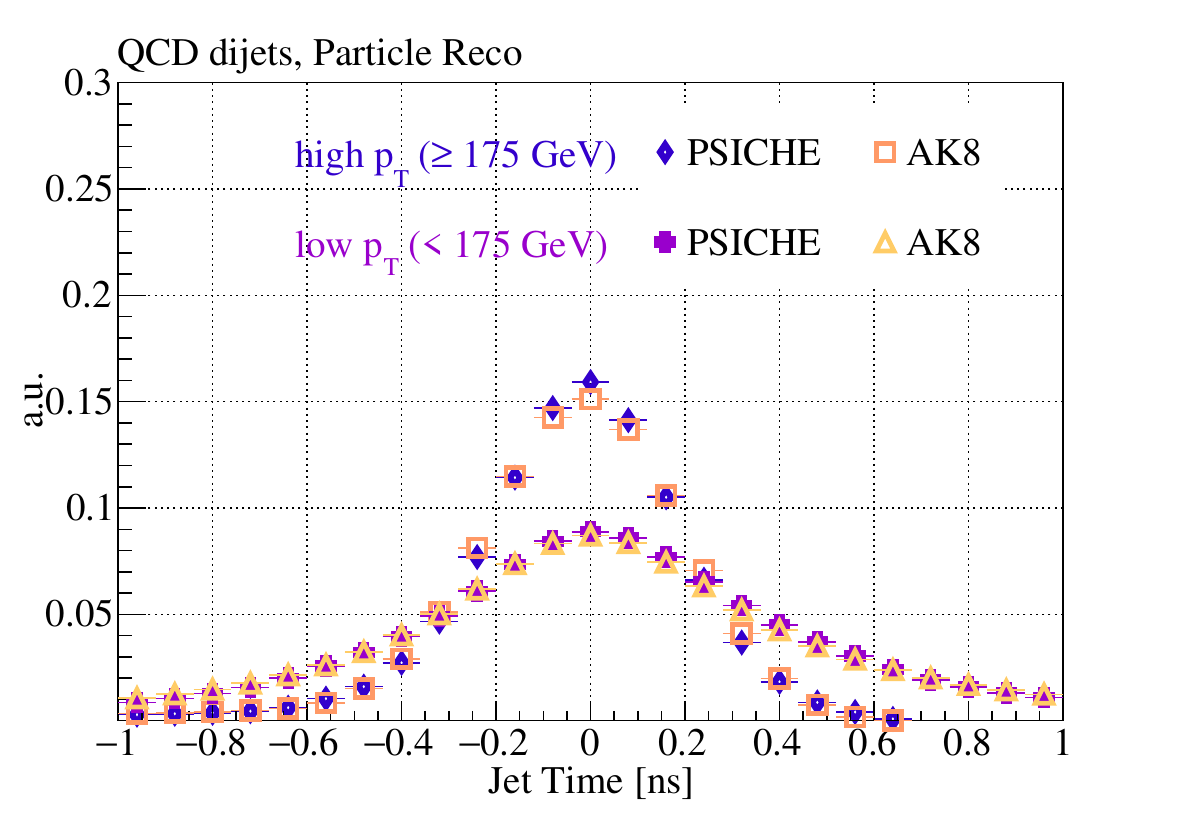}
    \caption{\small Distributions of different jet kinematic observables (left column) and jet time (right column) for two reconstruction methods. All distributions are from a dijet sample and compare PSICHE jets (blue diamonds) to AK jets at various cone sizes. Jet energy (top left) and jet mass (bottom left) demonstrate sensible agreement between PSICHE and AK jets. The jet time for calorimeter cell-based reconstruction (top right) and particle-based reconstruction (bottom right) methods both illustrate similar resolutions with the expected $p_T$ dependence.}\label{fig:jet_nom_obs}
  \end{figure}

As no resonances are simulated in this dijet sample, reconstructed jets should exhibit minimal substructure features. The kinematic properties of these different jet clustering algorithms are summarized in Figure \ref{fig:jet_nom_obs}. As expected, the mass distributions do not demonstrate any peaks denoting on-shell resonances. Across all jet types considered, the jet energy spectra show two types of jets: low-momentum jets ($E_\text{jet} < 200$) originating from radiative effects and high-momentum jets ($E_\text{jet} \gtrsim 200$) coming from the hard scattering. The momentum spectra of these harder jets is governed by the generator-level observable $\hat{p}_T^\text{min}$, which was set to 200 GeV for this sample. The jet time is shown for both types of considered reconstructions (Calo Reco and Particle Reco) in Figure \ref{fig:jet_nom_obs}, with each of the populations in the energy spectra separated. We observe a larger variance for lower momentum jets due to the energy dependence of resolutions and out-of-cone effects from the bending of low-momentum charged particles. The similarity between the jet time distributions for both methods illustrates PSICHE's generalizability, finding consistent, underlying probabilistic models instead of focusing on stochastic differences.

\onecolumngrid

\twocolumngrid

Figure \ref{fig:jet_nom_obs} confirms PSICHE's ability to reconstruct basic jets aligned with expectations from AK jets, in both time and kinematic observables. This correspondence is expected -- as no overlapping interactions nor jet substructure were simulated -- and purposeful. With more intricate topologies and collision conditions, the properties of PSICHE and AK jets begin to diverge. For example, emergent, parton-scale substructure is not inherently captured by the AK algorithm. Additionally, in high pileup conditions, where the inclusion of time becomes increasingly important, AK jets do not consider this dimension while reconstructing jets. While precision timing can be used to filter PU contributions from clustering inputs, non-trivially incorporating this information into jet clustering requires a metric for relatively weighing distances in space and time. PSICHE achieves this synthesis by uniformly translating all observations into the same probabilistic setting, and grading each with explicit experimental measurement precisions, $\Lambda^*_n$.

The incorporation of time opens possibilities of new, time-based substructure observables, including unique subcluster times and spatial-time covariance. These observables can be qualitatively assessed in Figure \ref{fig:jet_3dview}, depicting the mixture model components as ellipsoids in 3D space with slope and width along the time axis. %The timing-awareness of PSICHE works in concert with the other general and domain-specific inductive biases to dynamically learn jet size and jet substructure, simultaneously.

\section{Unsupervised Jet Clustering, Subclustering, and Kinematic Characterization with PSICHE} \label{sec:novel_features}
The act of jet clustering implicitly defines a jet, asking \textit{what is a jet}? 
Deterministic methods, like the AK algorithm, are able to learn overall jet features, like multiplicity, by specifying a fixed cone size, or scale, in advance. Conversely, algorithms like exclusive $k_T$ \cite{ellis_successive_1993}, capable of dynamic jet sizes, require the specification of multiplicity. Neither incorporates jet substructure information into clustering. Ideally, the main hallmarks of a jet -- a localized, interconnected collection of emissions with emergent substructure -- would be fully integrated and learned in a single jet clustering algorithm. PSICHE, by design, is well-matched to the question of defining a jet.

%With its structure-intrinsic design and probabilistic models, PSICHE achieves the assimilation of jet hallmarks in a comprehensive framework.

PSICHE jets are learned from data in an unsupervised way, grouping emissions with a high likelihood of association. Mirroring structure within jets, PSICHE's nested design addresses the challenge of simultaneously learning multiplicity and scale. The hierarchal model leverages a probability-based metric to determine associations among emissions, learning the jet scale, and is regulated by the complexity of its enveloped mixture models, controlling the jet multiplicity. The mixture model, in turn, whose multiplicity is regulated by both structural and analytic priors, provides learned features in the form of the subcluster posterior sufficient statistics. These subcluster posteriors can be recombined to describe the jet as a whole; this information forms the basis for a kinematic characterization across multiple jet scales, and can be used to, for example, mitigate pileup, an ever-present source of ambiguity at particle collisions.

\vspace{-0.3cm}
\subsection{Learning Dynamic Jet Sizes and Multiplicity} \label{sec:dynamic_jet_size}
Perhaps the most important property in interpreting jets is their multiplicity. In the zero-sum game of hard assignments, more jets means, on average, smaller jets. Furthermore, jet size is intimately related with energy, momenta, and mass. In PSICHE, these properties are learned with the tree-based model, and are balanced by the Dirichlet process prior and probabilities of the embedded mixture models. The learning of sensible jet multiplicities was demonstrated in Section \ref{sec:simreco}, in the simplest of event topologies. Moving towards phenomenologically richer cases, featuring jets with nontrivial masses, multiplicities and substructures, we will observe that PSICHE considers all of these elements simultaneously.

\onecolumngrid

\begin{figure}[!htb]
\centering
% \vspace{0.5cm}
\includegraphics[width=0.42\linewidth]{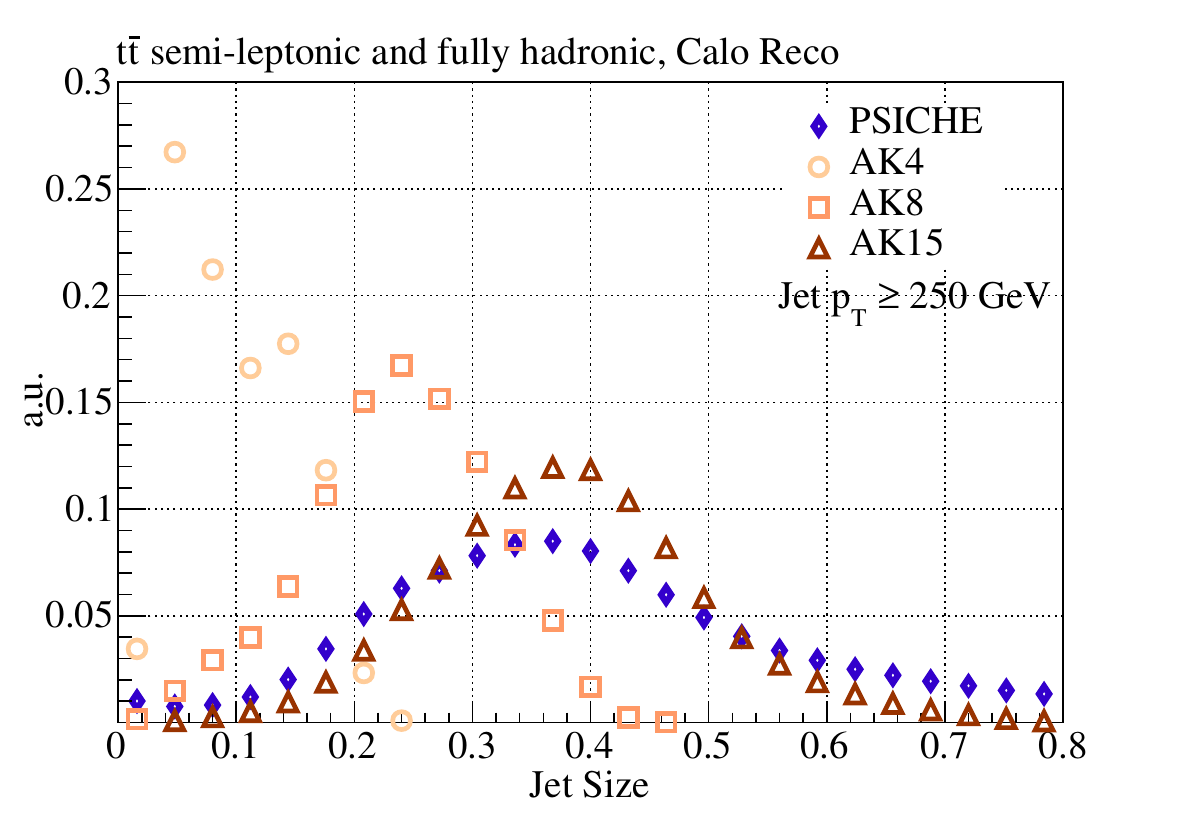}
       \hspace{1cm}
      \includegraphics[width=0.42\linewidth]{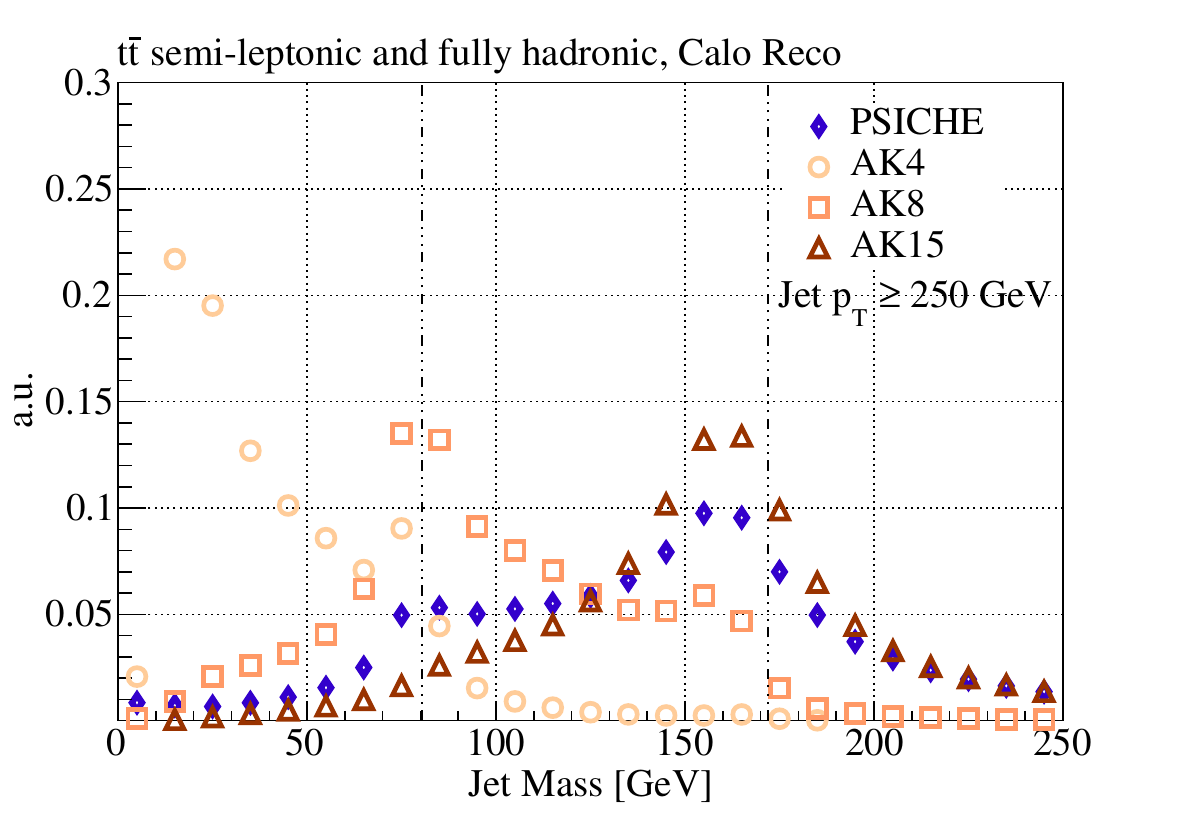}
  \includegraphics[width=0.42\linewidth]{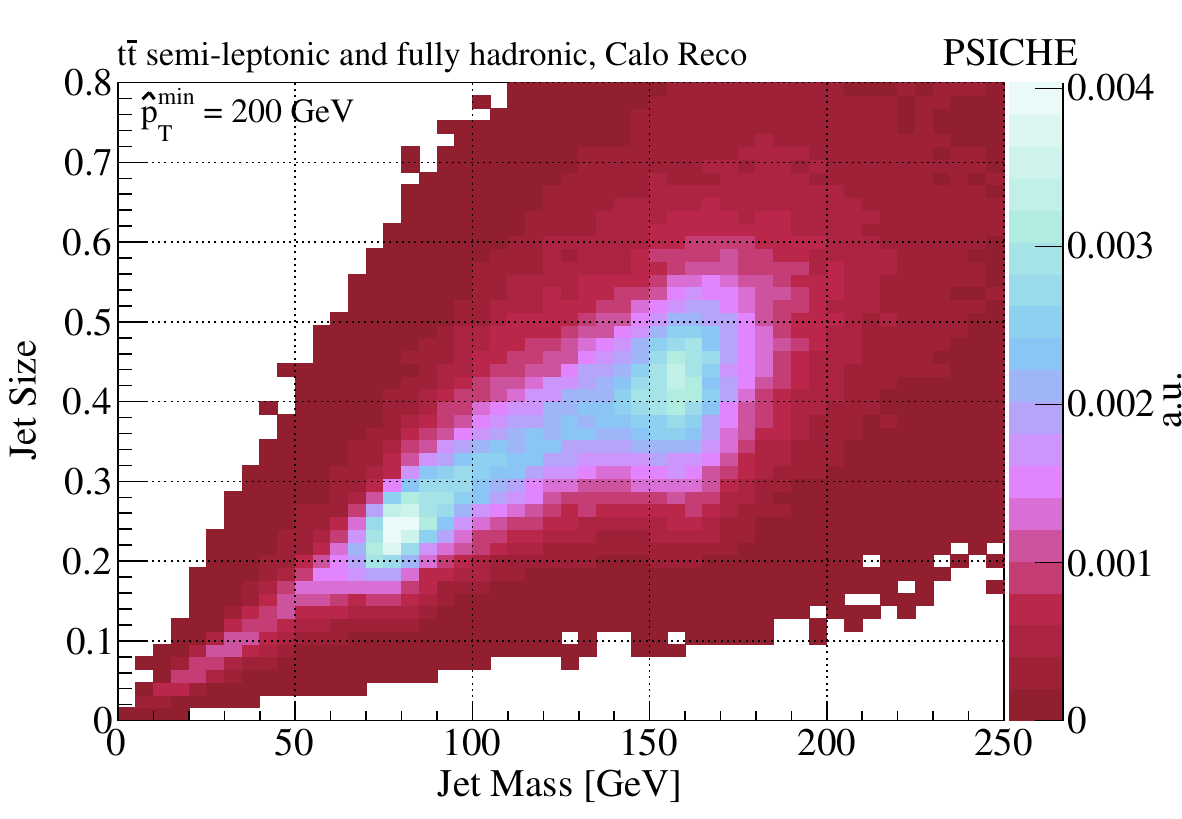}
   \hspace{1cm}
  \includegraphics[width=0.42\linewidth]{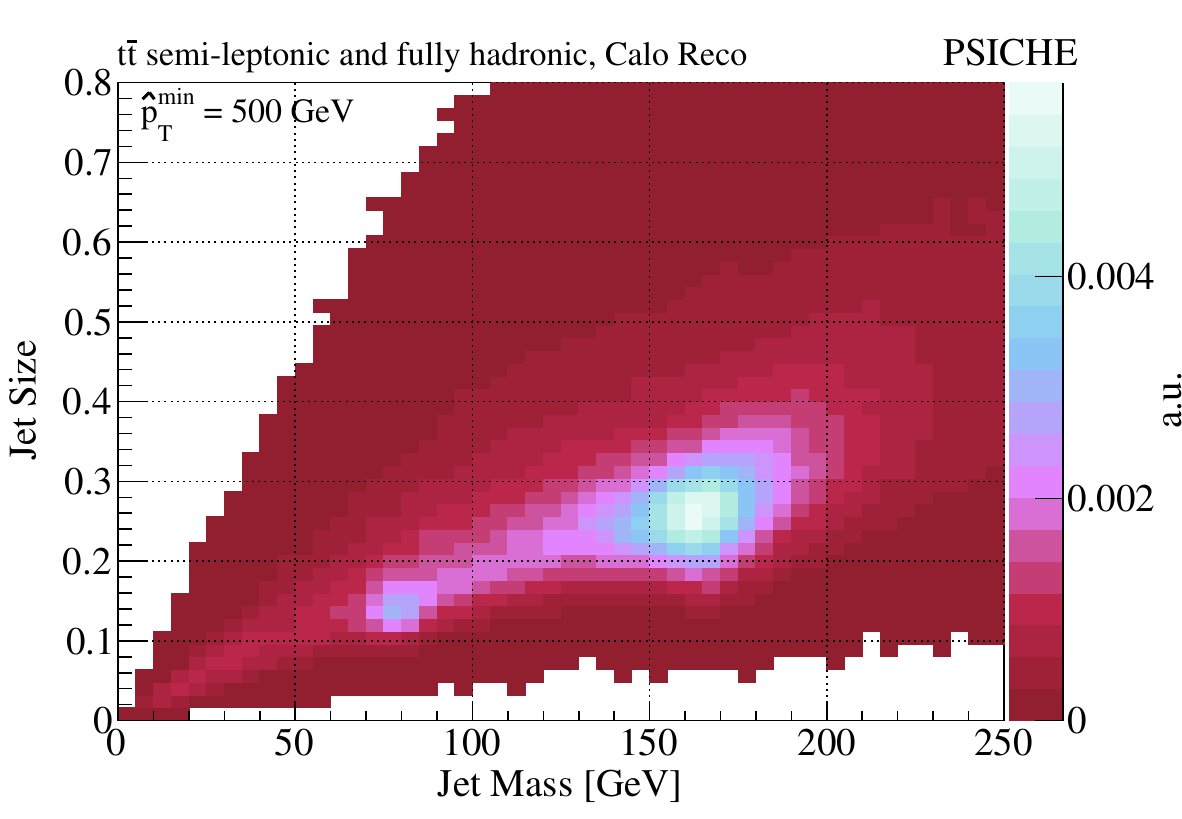}
    \caption{\small Distributions of jet size (top left) and mass (top right) for PSICHE jets (blue diamonds) and AK jets of various cone sizes, reconstructed in simulated $t\bar{t}$ events. Distributions in the 2D plane of jet size and mass (bottom row) illustrate their relationship, with PSICHE comprehensively reconstructing resolved  and boosted top quark jets. A descrease in jet size and increase in proportion of jets with mass consistent with a top quark (relative to $W$-boson) is observed when comparing distributions between samples with $\hat{p}_T^\text{min} = 200$ GeV (bottom left) and $\hat{p}_T^\text{min} = 500$ GeV (bottom right).}
    \label{fig:dyn_jet_size}
 \end{figure}
  
\twocolumngrid

We define a heuristic ``jet size" as the square-root of the lead eigenvalue of the jet $\eta-\phi$ covariance matrix. The distributions of this observable for different jet clustering methods are shown in Figure \ref{fig:dyn_jet_size} (top left) for a simulated di-top sample, including both $W$ ($m_W \approx 80$~GeV) and top ($m_t \approx 170$~GeV) decays. Across kinematic regimes, these resonances can produce a variety of jet types. AK jet sizes, and consequently jet masses (Figure \ref{fig:dyn_jet_size} top right), depend strongly on their cone parameters, with a hard cutoff at the prescribed scale. Since jet mass is a proxy for jet identity, AK jets do not just differ in size, but in what they fundamentally reconstruct. The AK4 radius is too small to fully capture any resonance, the AK15 radius is large enough to consistently capture the entire top decay, and the AK8 radius partially captures either resonance. PSICHE reconstructs an admixture of both resonances, dynamically adjusting jet size according to its mass and momentum

On average, a jet's size is related to its mass and momentum as $\text{jet size} \propto \frac{ m_\text{jet}}{ p^\text{jet}_T}$, with this relationship illustrated in the bottom row of Figure \ref{fig:dyn_jet_size}. We observe that each resonance in the PSICHE mass spectrum corresponds to a characteristic size, which is maintained and appropriately adjusted at higher momentum scales. Here, two different values of $\hat{p}^\text{min}_T$ are used in the event simulation, leading to two different spectra of top quark $p_T$ (left and right bottom Figure \ref{fig:dyn_jet_size}). As the top quarks are produced with higher momenta, their decay products become more collimated, with the size of this top-jet consequently shrinking. Correspondingly, the proportion of top-jets to $W$s grows larger. In the lower momentum regime, $W$-bosons in top decays are reconstructed on their own, as boosted jets, as well as resolved $W$s in the form of two, separate light quark jets. Both observed and hypothesized phenomena (for example, Refs.~\cite{cms_collaboration_search_2020, collaboration_measurement_2024}) exhibit different types of jets with their own characteristic masses and sizes, resulting in a panoply of jet types that PSICHE can comprehensively identify, dynamically evolving across kinematic regimes.

\subsection{Unsupervised Jet Subclustering \\ and Substructure} \label{sec:jet_substructure}

A quintessential trait of any type of jet is the presence of non-trivial, inherent substructure. PSICHE's variational mixture model forms the basis for learning this substructure in tandem with jet scale. In cases where jets represent decays of massive particles to multiple partons, the mass of the total jet and the number of effective ``prongs" in the decay will probabilistically inform what jet size, subcluster orientation, and kinematics are expected, as well as the interplay between these observables. The result is that different types of jets will have unique, characteristic features in the space of PSICHE observables. 

In the taxonomy of common jet decays of interest, hadronically-decaying $W$ bosons produce two light quarks, which may both be reconstructed as the same jet if the decay is sufficiently collimated. A characteristic signature, then, is a bimodal distribution of emissions, which could manifest as a two-subcluster mixture model, a readily identifiable but not overly-complicated pattern. 

 \begin{figure}[!t]
    \centering
         \includegraphics[width=\linewidth]{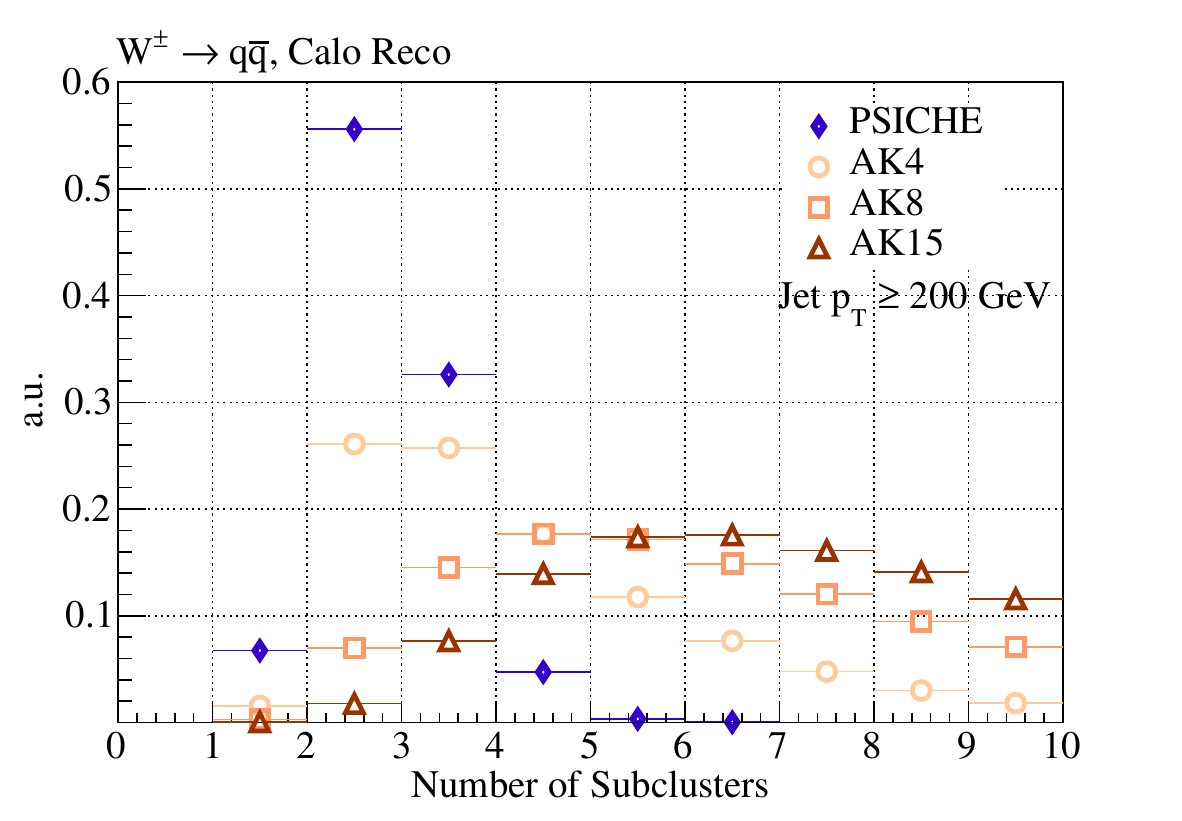}
\caption{\small Distribution of vGMM subcluster multiplicities for high-$p_T$ PSICHE (blue diamonds) and AK jets of various cone sizes for a single $W$ sample with a two-prong decay structure. Separate vGMMs were optimized for each pre-clustered AK jet, with the component multiplicity seeded to the number of calorimeter cells within the jet. PSICHE's subcluster merging scheme, facilitated by its embedded design, allows for successful regulation of subcluster multiplicity, leading to physically-interpretable substructure.}\label{fig:nsubcls_W}
\end{figure}

The distribution of subcluster multiplicities for different jet reconstruction methods is shown in Figure \ref{fig:nsubcls_W} for simulated single $W$ production. As the mixture model is not integrated with the clustering in the AK algorithm, a vGMM was run on the emissions assigned to these jets for comparison. PSICHE consistently reconstructs these boosted $W$s as two-subcluster jets, successfully resolving the parton-level substructure. In the absence of the regulation of subcluster initialization via hierarchal inheritance and recursive merging (described in Section \ref{sec:unique_algo_features}), the mixture model for AK jets is unable to identify this scale, instead finding significantly more subclusters and more closely resembling the scale of individual particle showers.

In PSICHE, multiplicity is qualitatively regulated by a Bayesian Occam's Razor mechanism, with entropy-based terms in the ELBO penalizing model complexity. These terms are included in the AK subcluster models, but what is absent are the inductive biases that work with the Bayesian priors to incentivize learning meaningful subcluster multiplicities. The regulation of subcluster merges, in collaboration with the hierarchical model for the overall jet scale, acts as a virtual annealing process, allowing PSICHE to find a more probable, or physically-meaningful, configuration with fewer subclusters. The inductive priors detailed in Sections \ref{sec:unique_algo_features} and \ref{sec:domain_specific} play a particularly important role in inference of subcluster multiplicity. The inheritance-based mixture model initialization directly sets the subcluster multiplicity starting conditions, while the merge model comparison further optimizes the multiplicity. Both contribute to the Ouroboros-style regulation of PSICHE, with the history of jet merges guiding the subcluster multiplicity while this multiplicity also informs the merges of jets.

Single $W$-bosons can even be subcomponents of more complicated jets, like the decays of top quarks, which PSICHE is able to generalize to. A simulated example of such an instance appears in Figure \ref{fig:jet_3dview}, where we observe a three-pronged structure, with modes in detector cell energies corresponding to individual subcluster contributions from the $W$ decay quarks and an additional $b$-quark jet. While identifying a particular partonic scale could typically require involved hyperparameter tuning, PSICHE is able to discover this structure unsupervised. Beyond multiplicity, Figure \ref{fig:jet_3dview} hints at interesting features in subcluster shapes, manifestations of the learned sufficient statistics of the mixture model, as described below.

\onecolumngrid

\begin{figure}[!h]
\centering
\includegraphics[width=0.48\linewidth]{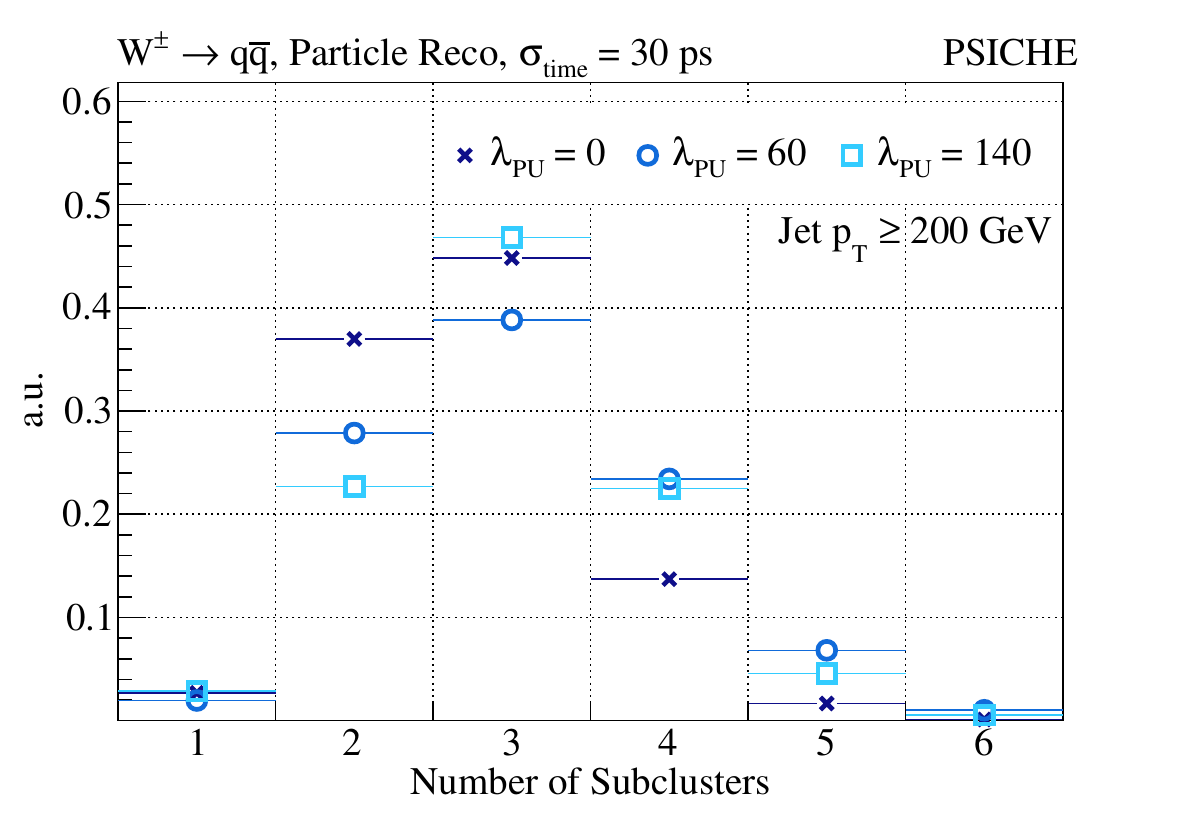} 
\hspace{0.3cm}
\includegraphics[width=0.48\linewidth]{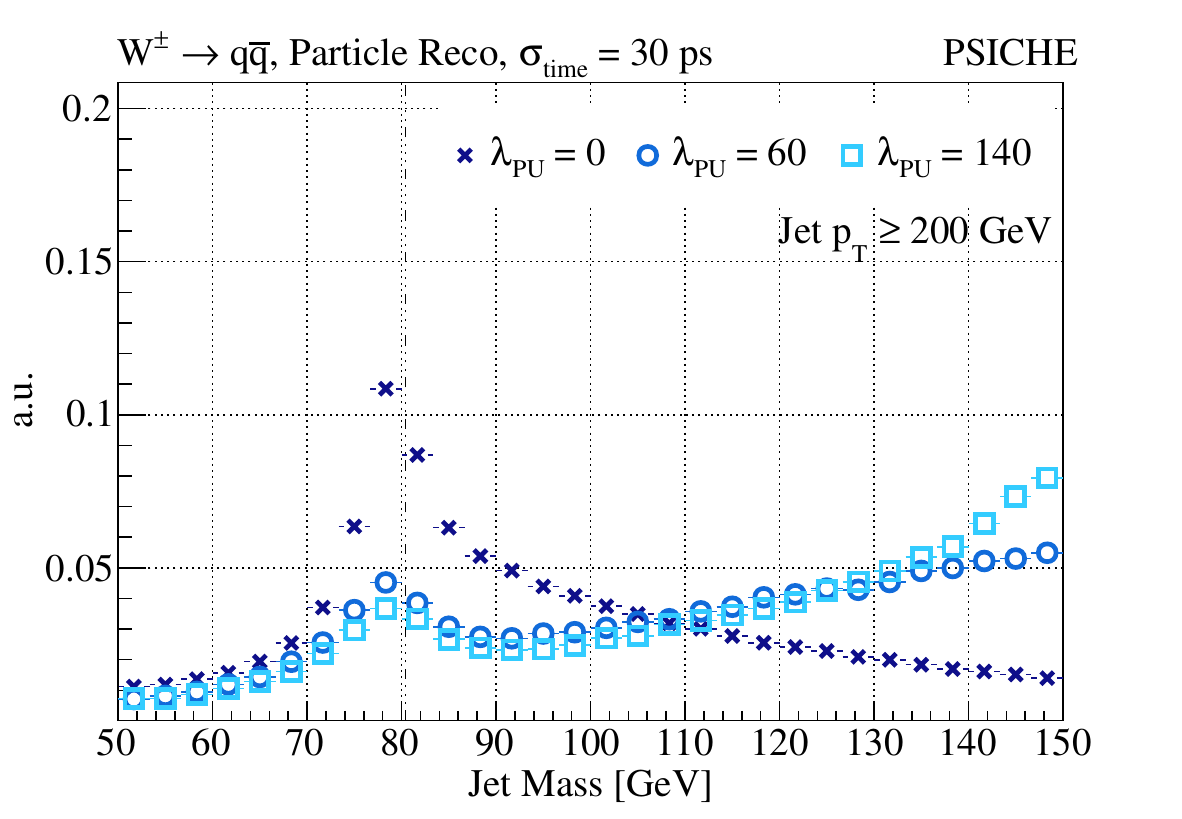}
\label{fig:puscenarios_nsubclusters}
\caption{\small Distributions of subcluster multiplicity (left, within a window of the $W$ mass) and mass (right) for PSICHE jets in single $W$ sample simulated with a particle-like reconstruction and increasing levels of pileup: no pileup (dark blue x's), moderate pileup (blue circles) and extreme pileup (light blue squares). Given that PSICHE is inherently timing-aware, we simulate precision, energy-independent timing effects on par with those expected for the HL-LHC, modeled after the hermetic CMS MIP Timing detector \cite{cms_collaboration_mip_2019}, with timing resolutions of $\sigma_\text{time} = 30$ ps, as denoted in the plots.}\label{fig:puscenarios_nsubclusters_mass} \vspace{0.1cm}
\end{figure}

\twocolumngrid
\subsection{Characterization and Application of Emergent Kinematics for Pileup Mitigation} \label{sec:pu_cleaning}
Beyond learning subcluster multiplicity, PSICHE can resolve details of the parton scale within jets via the optimized posterior parameters of the vGMM. These sufficient statistics -- location, covariance, and intensity -- encode a wealth of information on jets' substructure, which can be combined with subcluster multiplicity and other features for jet identification, or tagging \cite{Thaler_2011, chien_telescoping_2020, marzani_looking_2019, dillon_uncovering_2019, klimek_time_2021, dasgupta_towards_2013, feige_precision_2012, butterworth_jet_2008,bogatskiy_explainable_2024}. Here, we demonstrate the utility of this information by studying the deleterious effects of \textit{pileup contamination} in jets and illustrating potential mitigation strategies that leverage the kinematic characterization of jet subclusters with PSICHE observables.

Perhaps one of the biggest challenges in contemporary jet physics is dealing with emissions from pileup (PU) contaminating reconstructed jets. When hadrons collide at facilities like the Large Hadron Collider (LHC), it is not individual particles, but bunches of as many as hundreds of billions, that are passing through each other in the center of particle detectors. Tens to hundreds of additional, confounding interactions may be observed, piling up on top of any rare interactions of interest. Furthermore, underlying event (UE) emissions arise due to radiation from the ``spectator" partons not involved in the hard scattering \cite{marzani_looking_2019}; both PU and UE phenomena lead to relatively soft, dispersive emissions uniformly distributed over the face of the detector, potentially obfuscating the products of the hard scattering and infiltrating the reconstruction of jets. To study these effects, three representative PU scenarios are simulated -- none (notationally $\lambda_\text{PU} = 0$), moderate ($\lambda_\text{PU} = 60$), and extreme ($\lambda_\text{PU} = 140$) -- where the last two correspond to LHC Run II and projected HL-LHC conditions, respectively~\cite{CERN_HLLHC}.

The effects of PU contamination on jet mass and subcluster multiplicity can be seen in Figure \ref{fig:puscenarios_nsubclusters_mass} for a simulated sample of $W$ decays to two quarks. When these bosons have sufficient momentum, relative to their mass (are boosted), their decay products can be reconstructed in a single, collimated jet, resulting in a sharp peak in jet mass at the $W$ resonance ($m_W \approx 80$~GeV) and a two-prong substructure. Under more severe PU conditions, these properties of the $W$ resonance become obscured, as illustrated in the figure. The distributions of jet masses, sizes, and subcluster multiplicities become more dispersive and are biased to larger values as jets become increasingly addled with PU emissions. Furthermore, at extreme PU, jets form more readily, leading to jet and subjet-level ``saturation" due to the higher average energy density. In PSICHE, hyperparameters (particularly the energy-to-multiplicity transfer factor, $\omega$) could be adjusted to temper the effects of an increasing PU scale, although doing so may, in turn, discard information about the actual decays of interest. Fortunately, we will see that there are much stronger mitigation strategies beyond simple hyperparameter tuning that leverage the complementary information learned by PSICHE.

Pileup mitigation and its interaction with jet substructure is a well-studied subfield in its own right \cite{larkoski_soft_2014, Cacciari_2008, Krohn_2010, Berta_2014, cacciari_softkiller_2015, Krohn_2014, bertolini_pileup_2014, butterworth_jet_2008, dasgupta_towards_2013}. Some of these strategies are already present in PSICHE through the parameters and mechanisms of the algorithm. Other existing techniques can be quantitatively or qualitatively adapted to PSICHE through the calculation of derived observables from the posterior statistics of jets and subclusters. For example, the PSICHE model inherently incorporates emission-by-emission probabilistic weights, which are used to assign responsibilities of emissions to each subcluster within a jet. This weights-based approach is similar to algorithms like PUPPI~\cite{bertolini_pileup_2014}, with PSICHE using an augmented space of information in these probabilistic assignments beyond just local, relative distances between and momenta of emissions.

In the spirit of learning jets and their qualities in an unsupervised way, rather than introducing by hand additional scales when analyzing these subclusters we will frequently instead look at ratios, particularly those between subclusters properties, such masses or variances, and their jet analogues. Similar approaches include mass and soft drop algorithms \cite{butterworth_jet_2008, dasgupta_towards_2013, larkoski_soft_2014}, which leverage ratios of kinematic jet observables and externally-imposed relative scales to separate ambient emissions from the hard processes of interest. 

\onecolumngrid

\begin{figure}[!h]
    \centering
        \includegraphics[width=0.85\linewidth]{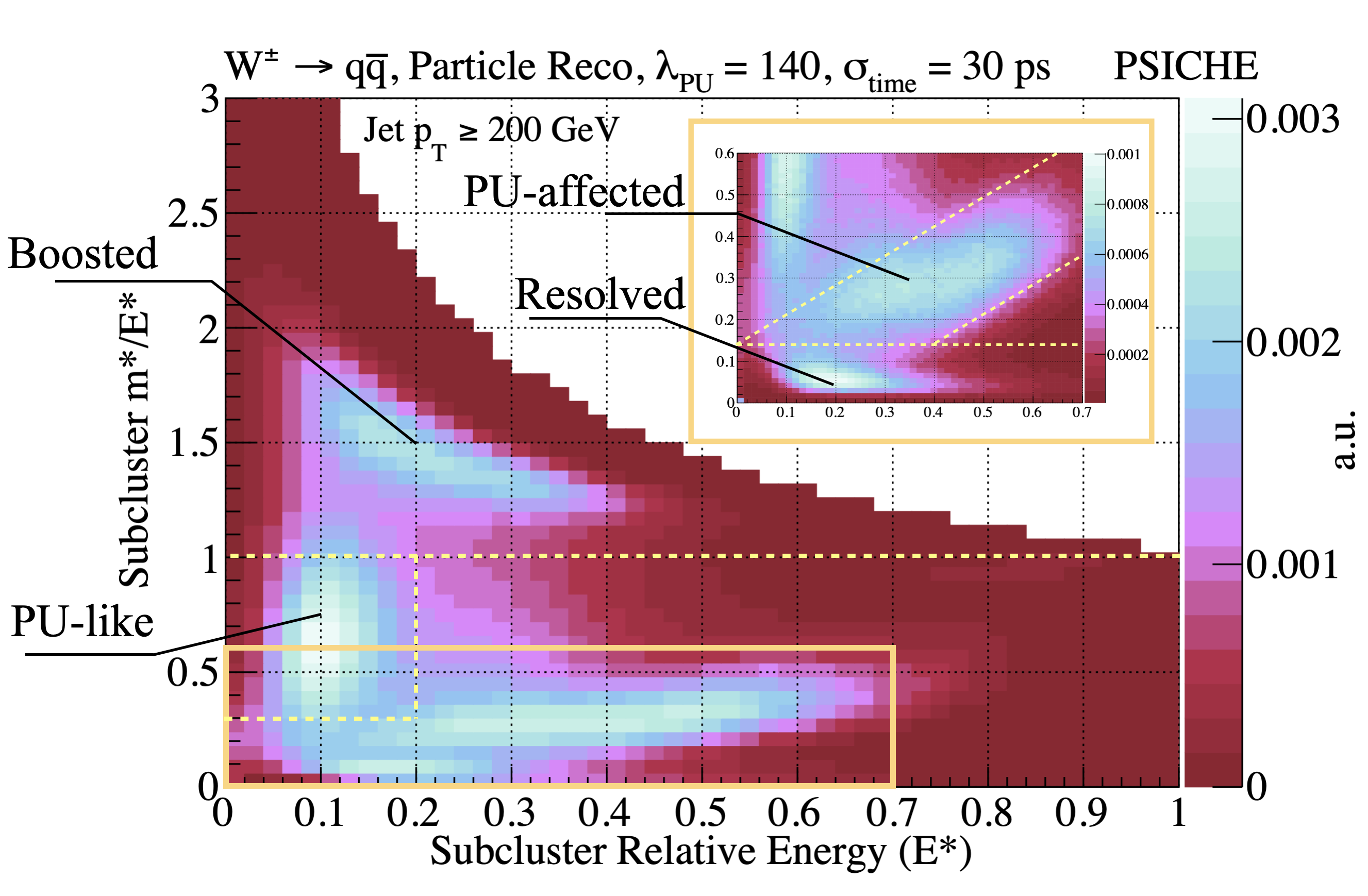}
\caption{\small A 2D plane of PSICHE relative subclusters observables, with each observable $q$ defined relative to the jet such that $q^* = q_\text{subcluster} / q_\text{jet}$. The $x$-axis corresponds to relative subcluster energy ($E^*$) and the $y$-axis is relative subcluster mass divided by relative energy ($m^*/E^*$). These jets were reconstructed from a simulated single $W$ sample with an extreme level of pileup ($\lambda_\text{PU} = 140$). Different subcluster topologies can be identified and are labeled accordingly, with dotted yellow lines denoting the selections in these observables used to isolate these populations. For clarity, a finer perspective of the area demarcated by solid yellow lines is included as an inlaid plot (upper right).}\label{fig:doubleratio_rele}
\end{figure}

\twocolumngrid

An example distribution of relative subcluster observables derived from the PSICHE jet model is illustrated in the 2D plane of Figure~\ref{fig:doubleratio_rele} for simulated decays of boosted $W$-bosons, under extreme PU conditions. Here, the relative subcluster energy, $E^* \equiv E_\text{subcluster}/E_\text{jet}$, is compared with the double ratio of relative mass  ($m^{*} \equiv m_\text{subcluster} / m_\text{jet}$) divided by relative energy, $m^* / E^*$. These constructions have strong similarities to the observables used in the Lund jet plane analysis of emissions within jets~\cite{dreyer_lund_2018} and, correspondingly, several distinct populations of subclusters can be identified in this distribution. In order to characterize these emergent subcluster types in the following discussion, and to later demonstrate the utility of subcluster classification for PU mitigation,  we define simple selection regions in the 2D plane of Figure~\ref{fig:doubleratio_rele}, indicated by the dotted yellow lines. We will refer to both these populations and their associated classification cuts according to interpretive labels: {\it PU-like}, {\it PU-affected}, {\it Boosted}, and {\it Resolved}. While a further refined approach would incorporate more subcluster information beyond these two observables and use a more expressive model for classification and selection, this simple characterization is able to illustrate a diverse subcluster phenomenology. 

While the kinematic double ratio on the $y$-axis of Figure~\ref{fig:doubleratio_rele} is a measure of relative spread of a subcluster, the jet-normalized energy, $E^*$ on the $x$-axis gives an indication of the relative importance of that subcluster in determining the jet's kinematic properties. The population appearing at lower values, labeled ``PU-like,'' denotes soft subclusters/emissions that represent a small fraction of the jet's activity. Correspondingly, these subclusters have negligible masses, and their ratio $m^* / E^*$ is spread somewhat uniformly between 0.5 and 1. Notably, the distributions of these two observables is largely uncorrelated for these subclusters, indicating that they are determined through independent stochastic variations. These properties are consistent with low-energy, dispersive emissions from PU and UE.

With larger relative importance and $E^*$, the other subcluster populations appearing in Figure~\ref{fig:doubleratio_rele} all contain some portion of the emissions following from the quarks appearing in the decay of the $W$-boson. The ``Resolved'' class appears at modest $E^*$ but small $m^* / E^*$, and corresponds to individual quarks reconstructed in separate subclusters within the same jet. At large $m^* / E^*$, ``Boosted'' subclusters appear in jets where the momentum of the $W$-boson is especially high, such that all its decay products are associated predominantly with a single subcluster. When both ambient PU emissions and products of $W$-decays are assigned to a subcluster, the relative energy and $m^* / E^*$ ratio will increase proportionally with the energy of PU contributions, due to the fixed mass scale of the $W$-boson decay and non-trivial opening angles between these elements. These erroneous assignments introduce a conspicuous correlation in the 2D plane of Figure~\ref{fig:doubleratio_rele} for ``PU-affected'' subclusters. 

While there are several complementary properties of subclusters, beyond the observables of Figure~\ref{fig:doubleratio_rele}, that can be used to characterize and distinguish these qualitative classes, perhaps what is the most striking is how these subclusters {\it look}. In fact, several of these subcluster archetypes are present in the example jet clustering illustrations of the earlier Figures~\ref{fig:vgmm_ex},~\ref{fig:bhc_ex}, and~\ref{fig:jet_3dview}. Each of these were snapshots of a simulated $t\bar{t}$ pair-production interaction, with an image of the vGMM mixture model and subclusters of one of the resulting top quark jets in Figure~\ref{fig:vgmm_ex}. The three-prong structure of the top quark decay is observed in three ``small'' (low variance in $\eta$ and $\phi$), but pronounced (significant energy) subclusters with similar size and shape. These are examples of Resolved subclusters, each with a strong one-to-one correspondence to the partons of the top decay. However, a fourth subcluster, much ``wider'' in space and of modest total energy, also appears in the PSICHE jet model. This PU-like subcluster acts an effective ``halo'' component of the jet, taking responsibility for dispersive PU and UE emissions within the jet (as opposed to the resolved partonic subclusters). We note that there is typically, at most, only a single PU-like component in each PSICHE jet, as the necessarily large spatial overlap of multiple such subclusters is probabilistically disfavored. This emergent phenomena is responsible for the subcluster multiplicity distribution mode at `three,' observed in Figure~\ref{fig:puscenarios_nsubclusters_mass} for two-prong $W$-boson decays, and means that natively appearing in many PSICHE jets is a built-in ``hypothesis test'' between PU emissions and those from resonance decays. The responsibilities, or weights of association, between a clustered element and each subcluster is based on the ratio of probabilities between alternative assignments, such that the derived kinematics of Resolved subclusters of interest are implicitly weighted by the per-element probabilities of not being from PU. 

\begin{figure}[!b]
    \centering \includegraphics[width=0.98\linewidth]{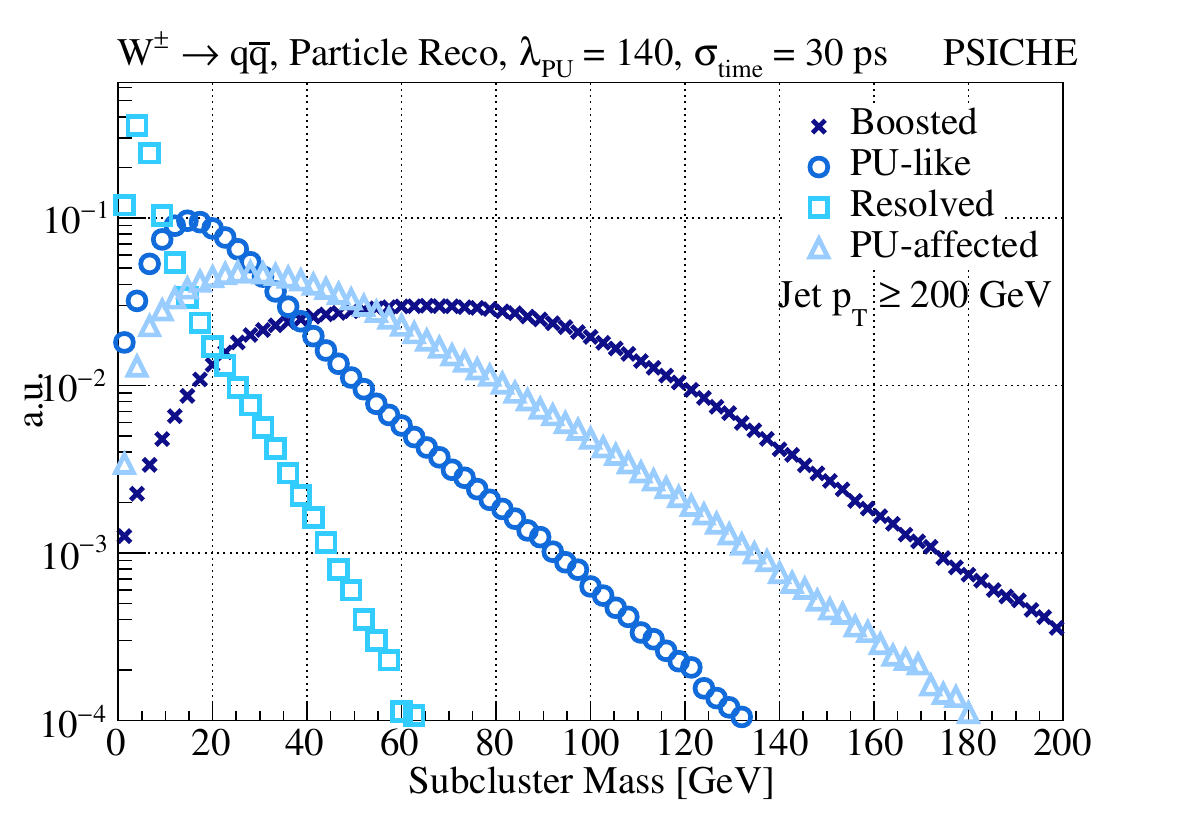}

  \caption{\small Distribution of subcluster mass for populations selected from Figure \ref{fig:doubleratio_rele} in a single $W$ sample with extreme pileup. Boosted (dark blue x's), PU-like (blue circles), Resolved (light blue squares), and PU-affected (lighter-blue triangles) subclusters each have distinct mass spectra.}\label{fig:sub_mass}
\end{figure}

\begin{figure}[!t]
    \centering

\includegraphics[width=0.98\linewidth]{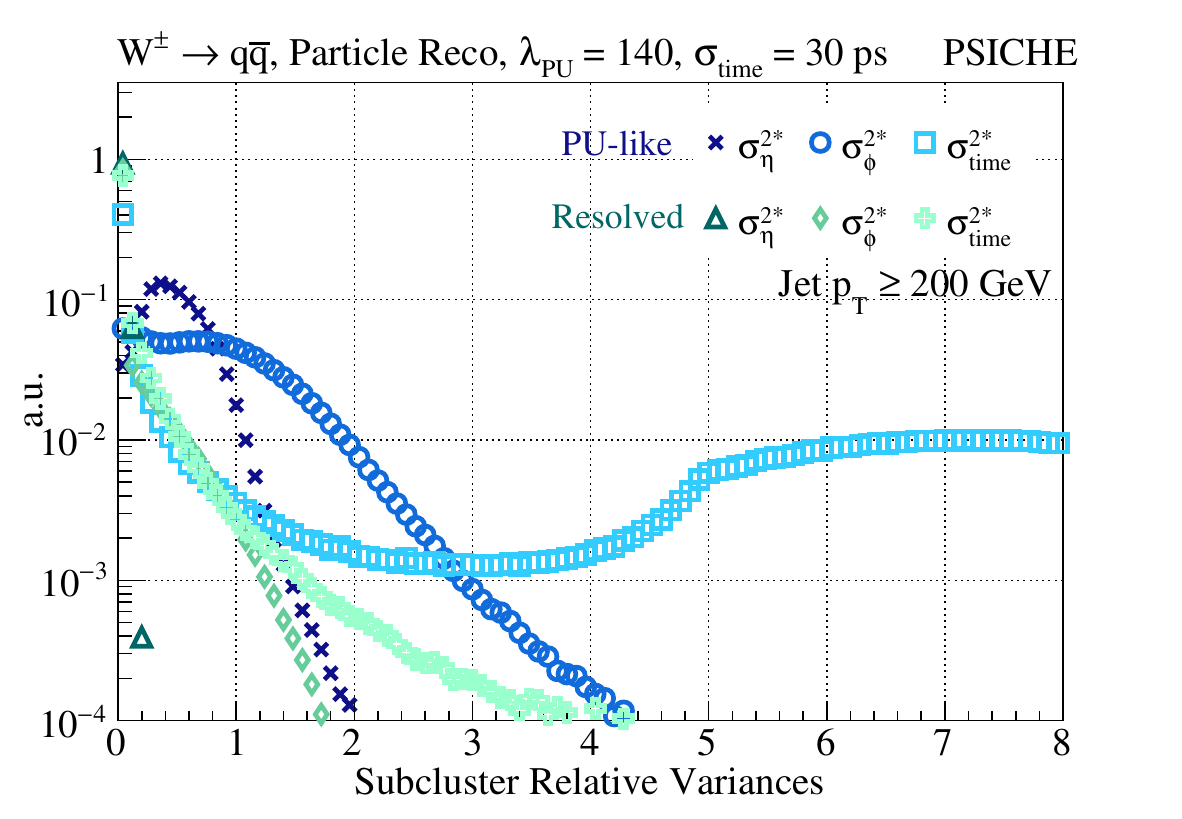}
        \caption{\small Distributions of relative variances for PU-like (blue) and Resolved (green) subclusters in jets from single $W$-boson decays. Relative subcluster variances are normalized by the corresponding variance of the entire jet for three different spatiotemporal dimensions: $\eta$ (x's, triangles), $\phi$ (circles, diamonds), and time (squares, crosses).}\label{fig:sub_relvar}
\end{figure}

The distribution of subcluster masses in $W$-boson jets for different classes, shown in Figure~\ref{fig:sub_mass}, further supports their interpretation. With the largest masses at the scale of the $W$-boson pole mass, Boosted subclusters are consistent with the majority of the $W$ decay products being assigned to a single subcluster (along with some contamination from PU emissions). At the other extreme are Resolved subclusters, with a highly suppressed mass distribution and mode near the lowest resolvable scale of light quark masses and selected vGMM energy-to-counts transfer factor, $\omega$. While the PU-like subclusters emerging as probabilistic halos shadowing jets tend to have large spatiotemporal variances, as seen in Figure \ref{fig:sub_relvar}, their masses in absolute terms are modest. Falling in between these extremes, PU-affected subclusters have masses consistent with the inclusion of some of the $W$-boson decay products and additional PU emissions. 

The qualitative characterization of subclusters' size can be quantified through the evaluation of relative spatiotemporal variances, normalized by the corresponding jet quantities, as was done for energy and mass. Figure~\ref{fig:sub_relvar} compares the distributions of these relative variances, $\sigma^{2*}_q \equiv \sigma_{q, \text{subcluster}}^2 / \sigma_{q, \text{jet}}^2$, between Resolved and PU-like subclusters, including all three of the coordinates ($\eta$, $\phi$, time) describing the clustering domain. Consistent with high energies and low relative masses, Resolved subclusters exhibit small relative variances in each of the spatiotemporal dimensions. Conversely, PU-like subclusters have expansive relative variances, frequently with values $> 1$, indicating that they are, in fact, ``wider'' than the jet itself (which is possible due to the probabilistic sharing of elements between subclusters). While the qualitative behavior of relative time variance, $\sigma^{2*}_{\rm time}$, matches that of $\eta$ and $\phi$, it is clear from the distribution of PU-like subclusters in Figure~\ref{fig:sub_relvar} that a larger range of element times can be assigned to a single jet. This is due to the explicit omission of time in the evaluation of nearest neighbors in the recursive merger scheme of hierarchical jet clustering, described in Section~\ref{sec:bhc_model}. The inclusion of time in the jet vGMM domain incorporates this information in the determination of subcluster assignments, and implicitly in the jet probabilities that inform the clustering scheme. But elements overlapping in space, even with dramatically different times, are unlikely to be assigned to different jets. This asymmetric treatment of time is both computationally convenient (as explained in Section~\ref{sec:voronoi}), and important for enforcing a self-consistent definition of jets that corresponds to our intuitive understanding. Practically, this treatment incentivizes the emergence of dispersive, PU-like subclusters associated with jets, as it is probabilistically favorable to parsimoniously isolate soft PU and UE emissions in a single subcluster component distinct from better resolved ones. It also allows for a large dynamic range of expressive time-related jet phenomenology, where latent structure and correlations can be resolved.

\onecolumngrid

\begin{figure}[!h]
    \centering
        \includegraphics[width=0.48\linewidth]{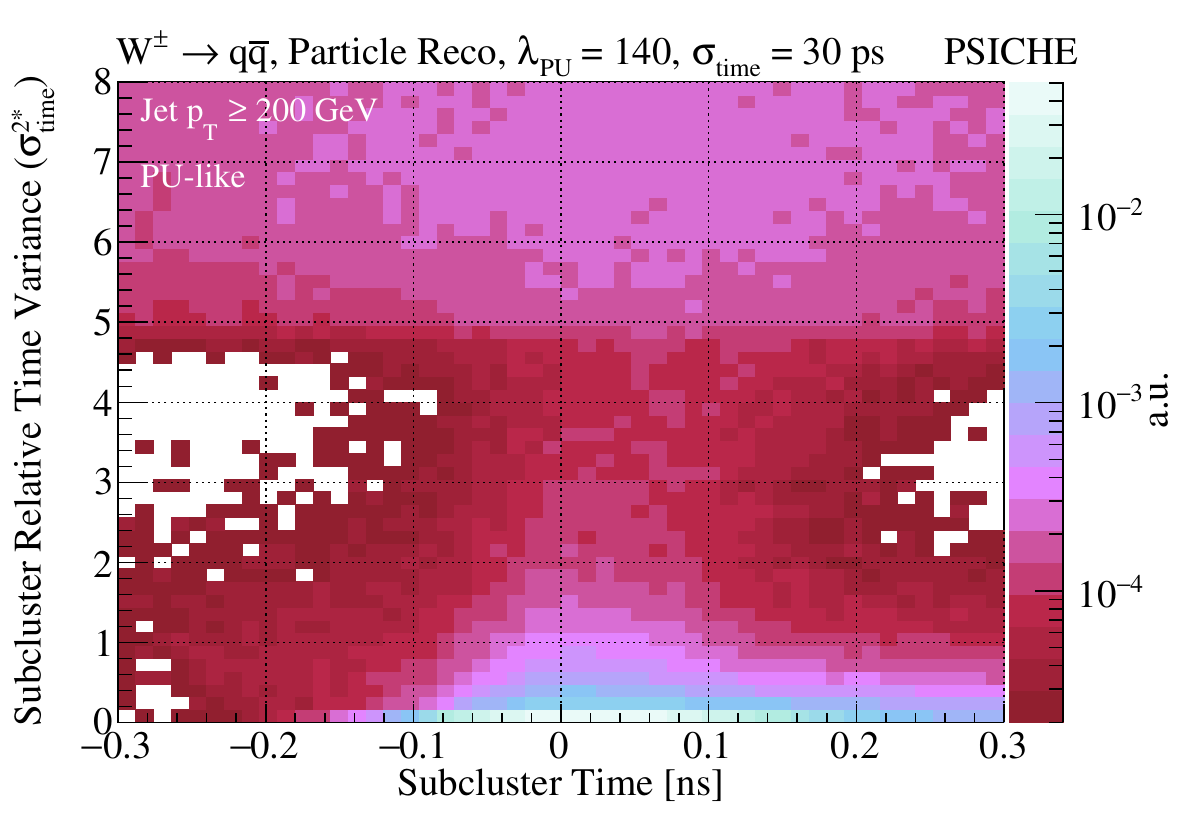}
             \hspace{0.3cm}  \includegraphics[width=0.48\linewidth]{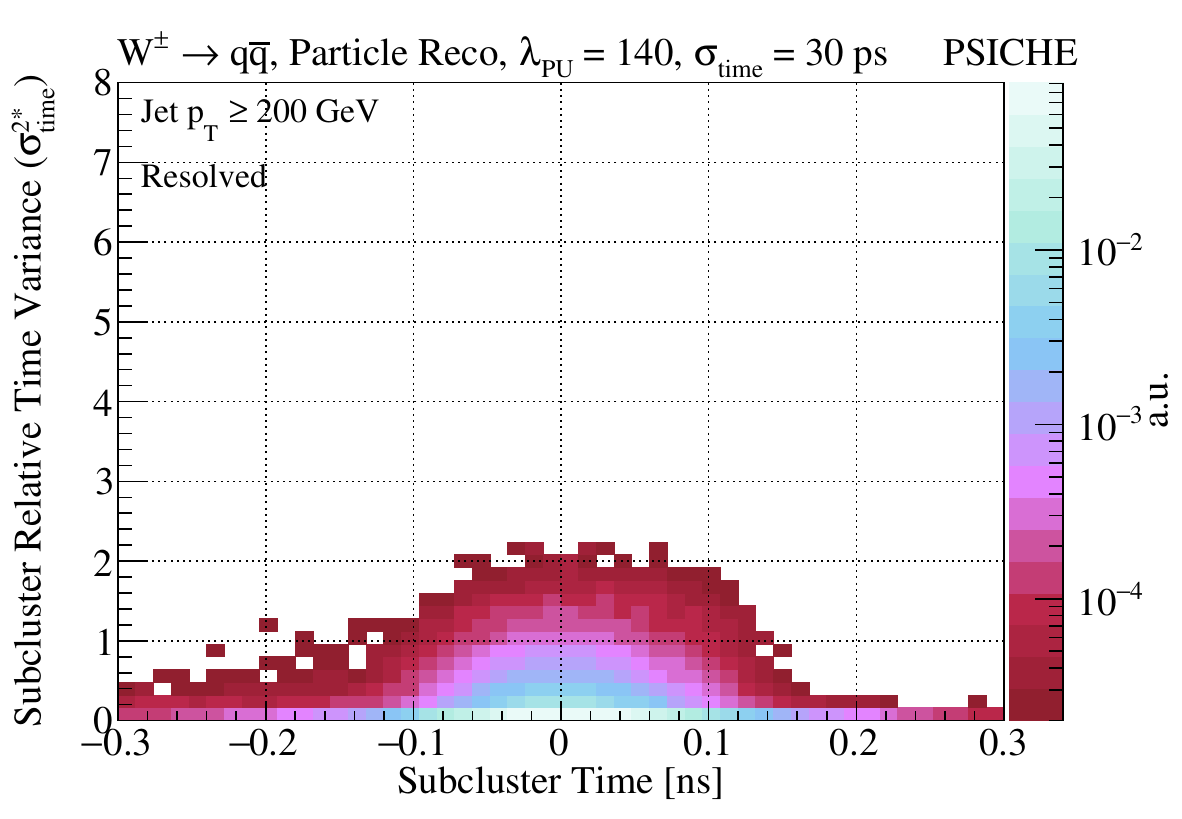}
\caption{\small Subcluster distributions in the 2D plane defined by relative time variance ($y$-axis) and absolute subcluster time ($x$-axis) for PU-like (left) and Resolved (right) subclusters. Jets are reconstructed from to simulated decays of single $W$-bosons with extreme PU.}\label{fig:rel_var_time}
\end{figure}

\twocolumngrid

In addition to relative time variances, the subcluster times themselves (with origin at the generator-level PV time) are useful for characterization. The 2D distributions of $\sigma^{2*}_{\rm time}$ and time are compared in Figure~\ref{fig:rel_var_time} for PU-like and Resolved subclusters. Both classes contain a population centered at zero with small relative variance, consistent with subclusters contains only elements from a single interaction, and a consensus among the most energetic subclusters in the jet. There is also evidence of low-variance, PU-like subclusters with times displaced from zero at later values, which correspond to soft, resolved PU and UE emissions, potentially co-appearing with a larger, PU-like halo subcluster. The presence of the latter is conspicuous in the distributions for PU-like subclusters in Figure~\ref{fig:rel_var_time}, appearing at large $\sigma^{2*}_{\rm time}$ over a wide range of indeterminate times. While the $W$-boson decay, PU, and UE processes simulated here exhibit relatively simple behavior in these observables, more exotic signatures could exhibit more expressive structure in time, time variance, and even spatiotemporal covariances within and between subclusters. 

The space of jet and subcluster observables derived from PSICHE can be modeled, explored, or exploited in applications like PU mitigation in a variety of ways. The distributions of emergent populations in relative energy, mass, and spatiotemporal variance, illustrated in Figures~\ref{fig:doubleratio_rele},~\ref{fig:sub_mass}, and~\ref{fig:sub_relvar}, could themselves be learned and modeled, in an unsupervised manner. Subcluster or jet observables could be used as inputs to a neural network classifier or regressor, with outputs used to probabilistically weigh or filter subclusters or emissions. In the case of PU-affected subclusters, contaminating emissions obfuscate elements of interest in their PSICHE assignments. These subclusters could be ``cleaned'' of such PU contributions by using the posterior jet and subcluster model to re-examine the initial clustering inputs, and potentially re-cluster events with down-weighted or removed elements. Such embellishments are beyond the scope of this work and left for future exploration. Here, we will instead adopt a simple approach as a proof-of-concept, performing PU mitigation for $W$-boson jets by removing subclusters that are undesirable or ambiguous. 

We refine the inclusive collection of reconstructed $W$-boson jets, whose subcluster multiplicities and masses in Figure~\ref{fig:puscenarios_nsubclusters_mass} were symptomatic of significant PU contamination, by removing subclusters inconsistent with the expected resolved, two-pronged, time-coherent structure. Retained subclusters are restricted to those that are classified as Resolved (according to the criteria illustrated in Figure~\ref{fig:doubleratio_rele}) and have self-consistent subcluster times ($|t_\text{subcluster}| < 0.1$ ns) and relative time variances ($\sigma^{2*}_\text{time} \leq 1$).

PSICHE naturally presents a mechanism through which to down-weight or remove the contributions of undesireable subclusters by using their associated probabilistic responsibilities for elements, $\mathbb{E}[z_{nk}]$, as evaluated from the vGMM jet posterior model (described in Section~\ref{sec:vGMM}). Subclusters $k$ that do not satisfy the above requirements have their respective responsibilities removed from the jet's four-vector calculation by removing a fraction of energy $\mathbb{E}[z_{nk}]$ from each input element $n$, equivalently subtracting subclusters' four-vectors (Equation~\ref{eq:subcl_fourvec}) from the jet total (Equation~\ref{eq:jet_fourvec}).

\onecolumngrid

\begin{figure}[!h]
    \centering	\includegraphics[width=0.48\linewidth]{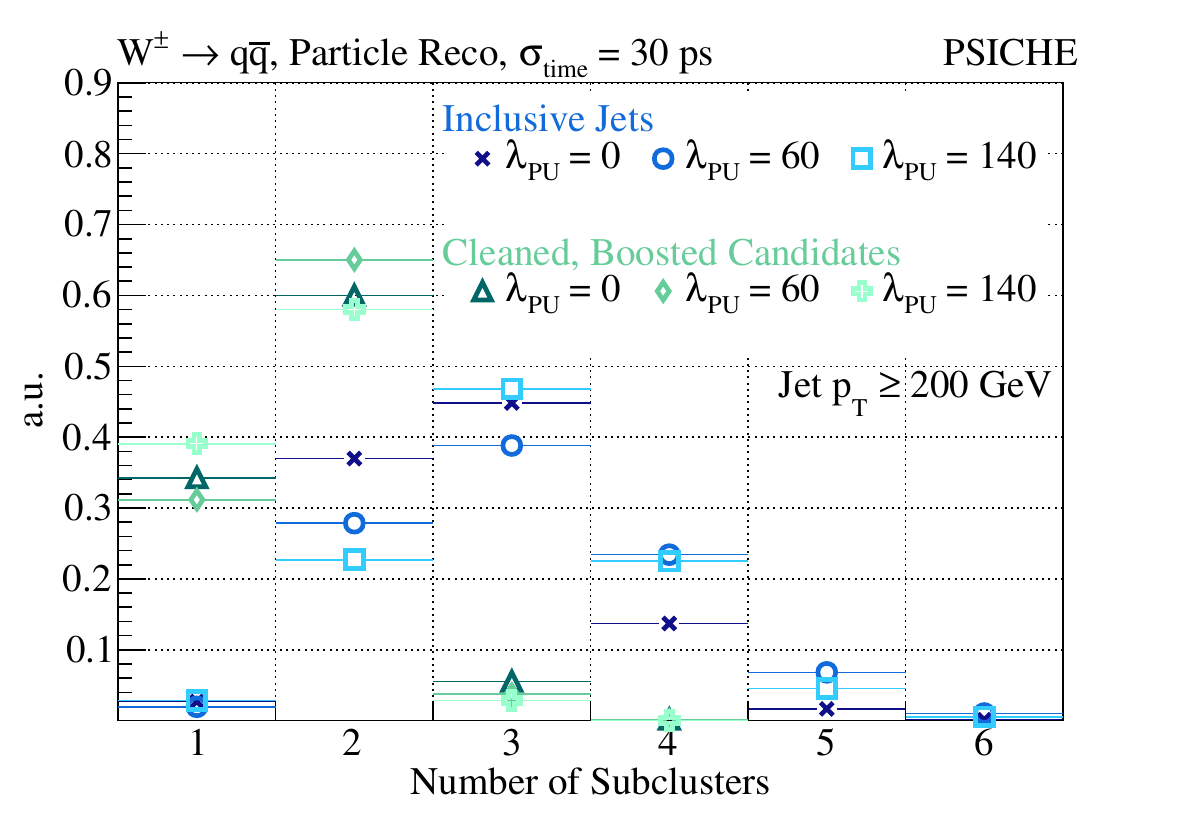} 
 \hspace{0.3cm} 	\includegraphics[width=0.48\linewidth]{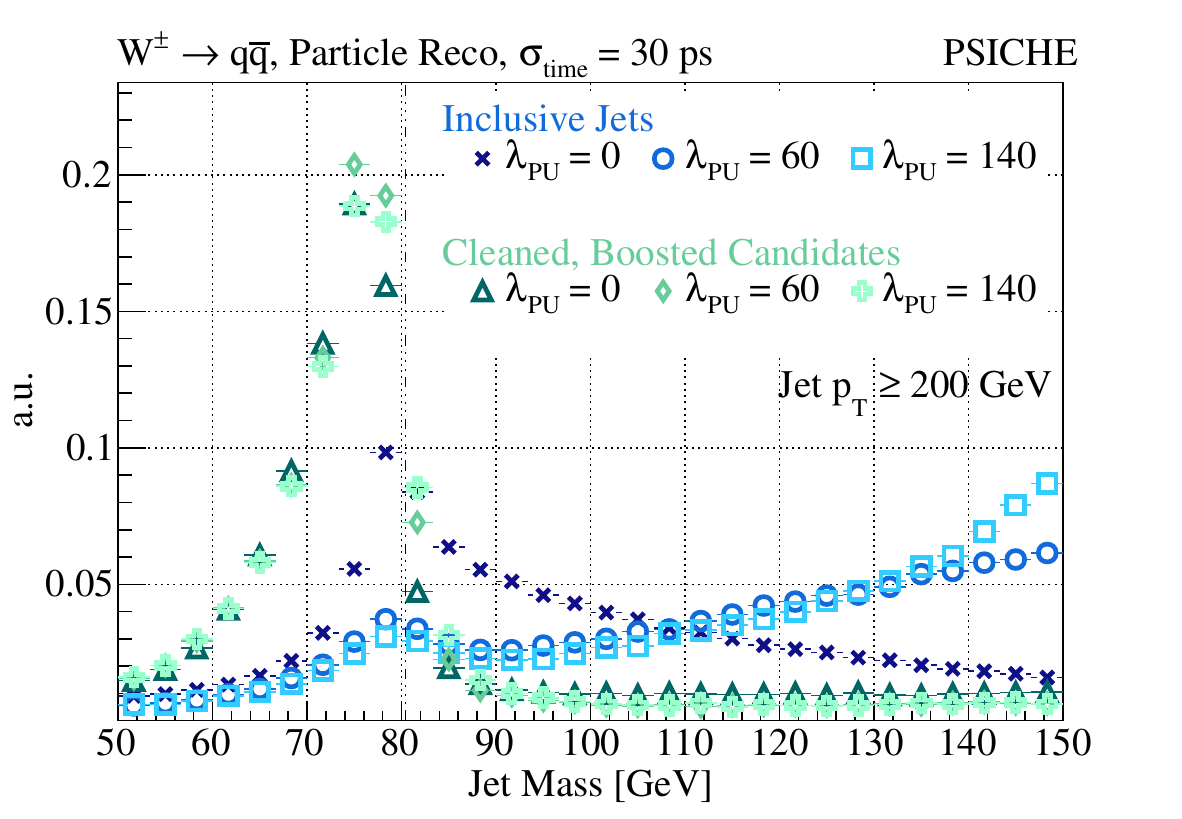} 
\caption{\small Distributions of subcluster multiplicity (left, within a window of the $W$ mass) and mass (right) for PSICHE jets with (green curves) and without (blue curves) the pileup mitigation subcluster selection applied. Various pileup levels ranging from none (dark blue x's, dark green triangles), to moderate (blue circles, green diamonds), to extreme (light blue squares, light blue crosses) are simulated for the single $W$ sample used here, and are denoted by their associated shapes in the legend.}\label{fig:mass_pu_cleaned} 
\end{figure}

\twocolumngrid

Figure \ref{fig:mass_pu_cleaned} demonstrates the effectiveness of this time-and-kinematic-based PU mitigation strategy by comparing the subcluster multiplicity and jet mass spectra before and after the cleaning is applied, with the blue curves corresponding to those from Figure \ref{fig:puscenarios_nsubclusters_mass}. Across all PU scenarios, the expected subcluster multiplicity and the central $W$ mass value are recovered. The most dramatic change can be seen in the highest PU scenario. Before the application of this PU cleaning method, the $W$ resonance is hardly recognizable. Now, a sharp, resolved peak at the $W$-boson mass is apparent in the cleaned, boosted spectrum. While the mode of the cleaned $W$-mass distribution is slightly shifted below the pole mass due to removed energy, analyzers can (and do, in practice) correct for such jet energy-scale effects through calibration. Irrespective of PU scenario, the mass and subcluster multiplicity peak positions and resolutions are similar among these cleaned jets, illustrating PSICHE's ability to resolve a latent structure of interest in a largely unsupervised manner, despite significant confounding effects.

\vspace{-0.3cm}
\section{Further Developments and Applications} \label{sec:future}

\subsection{Algorithmic Improvements and Embellishments}
With its highly modular form, there are many possibilities for improvement and refinement throughout PSICHE. This could range from changes to the treatment of observation elements at the mixture model level to embellishments and additions to the overarching probabilistic hierarchy. 

A natural, albeit challenging, extension would be the possibility of incorporating more than two clustering scales, beyond a vGMM embedded into a BHC tree structure. One could, for example, imagine each cluster itself becoming a mixture of mixtures, and even schemes with dynamic evolution in the number/depth of additional (sub)scales. The danger when adding further clustering scales is the introduction of non-identifiability and singularities into the model, as different configurations can become probabilistically indistinguishable. In the application to jet clustering described in this work, such degeneracies are mitigated through the transformation of observables into local jet/cluster reference frames (described in Section~\ref{sec:telescoping}). As the location coordinates of elements depends on their cluster assignment, hypothetical subclusters describing the same measurements will be distinct between different clusters. Adopting an analogous subcluster-dependence of observations could allow for additional hierarchies to be probabilistically incorporated within.

Perhaps a more fruitful, but increasingly domain-dependent, approach would be to replace the normal distributions appearing in the vGMM with alternative, less generic, likelihood models for observations. The only practical restriction for inclusion into the PSICHE framework  is that a tractable, sufficiently accurate, approximation of the posterior distribution of any latent random variables can be found. Such a probabilistic mixture component could, for example, incorporate even more information about low-level detector elements and particle showers into the jet clustering application. We have seen that the incorporation of observation precisions/uncertainties is crucial when combining heterogenous coordinates such as time and spatial location in the same clustering domain, and richer, domain-specific models beyond normal distributions could be of great benefit.

As discussed in Section~\ref{sec:voronoi}, the adoption of the BHC, tree-consistent partitioning in PSICHE introduces an order $N^{2}$ dependence in algorithmic complexity, if not for additional restrictions to merges. In the current implementation, only nearest-neighbors in 2D space are considered for combination, with the asymptotic bottleneck then appearing in the order $N \mathrm{log} N$ Voronoi tessellation construction needed for this evaluation. The addition of more dimensions, like time, in the nearest-neighbor calculation is possible, although this will cause the computational burden of triangulation to exceed the unrestricted BHC, obviating the use of this heuristic beyond its utility as an inductive prior~\cite{Delauny_divide}. There is also the possibility of improving how observations are incorporated into nearest-neighbor evaluations, by using a distance metric that more carefully captures the local jet transformations of measurements. Distance, and nearest-neighbor, evaluations based on Bregman divergences could be a natural adoption for mixtures of exponential family distributions, where the resulting information geometry can be informed by the model itself~\cite{info_geo}.

\subsection{Applications and Phenomenology of \\ Jet Clustering}
In its current design, PSICHE is already well-poised to be deployed in other particle physics applications involving jets. For example, the wealth of information learned from the posterior sufficient statistics of the vGMM could be leveraged for jet tagging, a task that relies on jet substructure observables. As demonstrated in this paper, prong multiplicity is one indicator of jet type. However, methods of jet tagging can also rely on the shape and kinematic characterization of jet substructure. While PSICHE's subjet information may be compatible with, for example, the calculation of n-subjettiness \cite{Thaler_2011}, the sufficient statistics describing this structure may be used to define new sets of jet observables, or be used in a higher-level and/or machine learning-based substructure analysis.

Jet tagging is especially important in searches for Beyond Standard Model (BSM) physics. These searches can be guided by a variety of theoretical frameworks, each with distinct detector signatures. Of particular relevance is a class of frameworks known as Hidden Valley scenarios \cite{strassler_echoes_2007}, which feature an extended symmetry of the SM and unconventional jet signatures, like jets containing a non-trivial amount of missing energy, or jets with a diversity of sizes and multiplicities \cite{carloni_visible_2010}. Because PSICHE is timing-aware and capable of co-learning multiple scales at once, elucidating this non-trivial substructure may be feasible.

Combined with dynamic jet sizes, PSICHE's simultaneously-learned substructure can be leveraged in BSM searches for jet classification, both identifying expected patterns through labelled optimization or even generically allowing for unexpected characteristics through unsupervised learning and class discovery. A common prediction in most BSM scenarios is the existence of new particles with unknown, variable masses. Phenomenologically, these could range in appearance in high energy interactions from heavy, large aperture jets to light, highly collimated ones. Generally, search efforts exclusively target distinct ranges or extremes among these possibilities, with strict jet sizes chosen appropriately \cite{cms_collaboration_search_2020, Aaboud_2019, Aad_2024, Sirunyan_2019}. PSICHE's dynamically learned jet size could expand the potential of these searches beyond a narrow kinematic regime, accommodating variable jet types within the same analysis, or even same interaction. 

The inclusion of timing within the PSICHE framework not only offers a handle on PU mitigation (without the need for a mass veto \cite{cmscollaboration2026searchresonanceeventsquarks}), but also provides a new dimension of information for BSM searches. Clustering in time is especially important when the hypothesized BSM physics predicts new particles with non-negligible lifetimes, due to either small couplings, decays with massive mediators, and/or reduced kinematic phase space of the decaying particle \cite{Giudice_1999,Evans_2016,Meade_2009,nagata_cornering_2017,davoudiasl_long-lived_2026,chan_lhc_2012}. These long-lived particles can manifest both as jet initiators, leading to displaced or ``emerging" jets, or they can appear within jets as displaced vertices. In either case, the introduction of time into jet clustering can further resolve these features. Furthermore, timing within jets can be used to calculate kinematic information of long-lived particles, such as velocities and masses \cite{flowers_timing_2020}, which could be powerful discriminators in BSM searches and provide crucial interpretative information in the case of a discovery.

\section{Conclusion} \label{sec:conclusion}
We have presented PSICHE, a probabilistic clustering algorithm with jet reconstruction as an example use-case. While principles of the algorithm itself are field-agnostic, there are domain-specific features in this iteration that make its design suitable for this application. The general features of PSICHE are variational Gaussian mixture models embedded in a tree-consistent Bayesian hierarchical partitioning. This nested, probabilistic model recursively leverages historical information to guide the initialization of the current mixture model, avoiding an heuristic initialization scheme. Gaussian subcluster multiplicity is further controlled with a probabilistic model comparison step, akin to a likelihood ratio. Furthermore, experimental uncertainties and observational weights are incorporated into the model to provide a gauge of absolute and relative importance among measurements and between clustering-domain dimensions. The result is a wholly-unsupervised, fully-probabilistic clustering algorithm that allows for flexible cluster sizes, simultaneously-learned substructure and clustering across dimensions with non-uniform resolutions.

 Additionally, domain-specific features, such as a transformation of observations to mimic the projective geometry of particle detectors, tune PSICHE specifically towards the task of jet reconstruction in the domain of high energy physics. Jet scale is not determined by a fixed, external size parameter, but rather is learned dynamically by the algorithm. Moreover, jet substructure is symbiotically co-learned with the jet scale in space and time. Finally, the presence and intensity of pileup is studied and a pileup mitigation technique leveraging PSICHE's dynamic, multi-scale, spatiotemporal clustering is presented. This mitigation method is able to recover expected resonances, even in pileup conditions as extreme as those expected at the High Luminosity LHC. 

Development of the PSICHE algorithm can be extended with modeling of additional, nested scales or clustering dimensions, potentially with more informative metrics. Due to its generic, underlying framework, PSICHE may be adapted for clustering in other domains, especially in multidimensional, hierarchical data with associated observational uncertainties. In its current iteration, PSICHE's learned information may be leveraged in other particle physics jet-related tasks, such as jet tagging. Due to its timing-inherent, multi-scale, structure-intrinsic design PSICHE also may be a promising avenue for reconstructing jet signatures of Beyond Standard Model physics, such as Hidden Valley models and/or frameworks predicting yet-unseen, long-lived particles.

%Appendices require a title and some information.

\begin{acknowledgments}
This work was made possible by U. S. National Science Foundation awards
1945038  and 2310030. ML wishes to thank the Madison and Lila Self
Graduate Fellowship for their generous support. The authors would also
like to thank KC Kong, Justin Anguiano, and collaborators from the University of Kansas High Energy Physics group for helpful discussions and feedback on this manuscript.
\end{acknowledgments}

%\bibliography{apstemplate}
\bibliography{main_biblio}{}

@article{Meade_2009,
   title={General Gauge Mediation},
   volume={177},
   ISSN={0375-9687},
   url={http://dx.doi.org/10.1143/PTPS.177.143},
   DOI={10.1143/ptps.177.143},
   journal={Progress of Theoretical Physics Supplement},
   publisher={Oxford University Press (OUP)},
   author={Meade, Patrick and Seiberg, Nathan and Shih, David},
   year={2009},
   pages={143–158} }

@article{flowers_timing_2020,
	title = {Timing information at {HL}-{LHC}: complete determination of masses of dark matter and long lived particle},
	volume = {2020},
	copyright = {2021 The Author(s)},
	issn = {1029-8479},
	shorttitle = {Timing information at {HL}-{LHC}},
	url = {https://link.springer.com/article/10.1007/JHEP03(2020)132},
	doi = {10.1007/JHEP03(2020)132},
	language = {En},
	number = {3},
	urldate = {2025-06-10},
	journal = {Journal of High Energy Physics},
	publisher = {Springer},
	author = {Flowers, Zachary and Kang, Dong Woo and Meier, Quinn and Park, Seong Chan and Rogan, Christopher},
	month = mar,
	year = {2020},
	note = {Number: 3},
	pages = {1--20},
}

@article{chan_lhc_2012,
	title = {{LHC} {Signatures} of a {Minimal} {Supersymmetric} {Hidden} {Valley}},
	volume = {2012},
	issn = {1029-8479},
	url = {http://arxiv.org/abs/1112.2705},
	doi = {10.1007/JHEP05(2012)155},
	number = {5},
	urldate = {2026-08-01},
	journal = {Journal of High Energy Physics},
	author = {Chan, Yuk Fung and Low, Matthew and Morrissey, David E. and Spray, Andrew P.},
	month = may,
	year = {2012},
	note = {arXiv:1112.2705 [hep-ph]},
	pages = {155},
}

@misc{davoudiasl_long-lived_2026,
	title = {Long-{Lived} {Dark} {Hadrons} at the {Electron}-{Ion} {Collider}},
	url = {http://arxiv.org/abs/2607.16400},
	doi = {10.48550/arXiv.2607.16400},
	urldate = {2026-07-22},
	publisher = {arXiv},
	author = {Davoudiasl, Hooman and Liu, Hongkai and Neil, Ethan T.},
	month = jul,
	year = {2026},
	note = {arXiv:2607.16400 [hep-ph]},
}

@article{nagata_cornering_2017,
	title = {Cornering {Compressed} {Gluino} at the {LHC}},
	volume = {2017},
	issn = {1029-8479},
	url = {http://arxiv.org/abs/1701.07664},
	doi = {10.1007/JHEP03(2017)025},
	number = {3},
	urldate = {2026-09-04},
	journal = {Journal of High Energy Physics},
	author = {Nagata, Natsumi and Otono, Hidetoshi and Shirai, Satoshi},
	month = mar,
	year = {2017},
	note = {arXiv:1701.07664 [hep-ph]},
	pages = {25},
}

@article{Evans_2016,
   title={Long-lived staus and displaced leptons at the LHC},
   volume={2016},
   ISSN={1029-8479},
   url={http://dx.doi.org/10.1007/JHEP04(2016)056},
   DOI={10.1007/jhep04(2016)056},
   number={4},
   journal={Journal of High Energy Physics},
   publisher={Springer Science and Business Media LLC},
   author={Evans, Jared A. and Shelton, Jessie},
   year={2016},
   month=Apr, pages={1–39} }

@article{Giudice_1999,
   title={Theories with gauge-mediated supersymmetry breaking},
   volume={322},
   ISSN={0370-1573},
   url={http://dx.doi.org/10.1016/S0370-1573(99)00042-3},
   DOI={10.1016/s0370-1573(99)00042-3},
   number={6},
   journal={Physics Reports},
   publisher={Elsevier BV},
   author={Giudice, G.F. and Rattazzi, R.},
   year={1999},
   month=Dec, pages={419–499} }

@inproceedings{attias1999inferring,
  title={Inferring parameters and structure of latent variable models by variational Bayes},
  author={Attias, Hagai},
  booktitle={Proceedings of the Fifteenth Conference on Uncertainty in Artificial Intelligence (UAI'99)},
  pages={21--30},
  year={1999},
  publisher={Morgan Kaufmann Publishers Inc.},
  address={San Francisco, CA, USA}
}

@article{lloyd1982least,
  title={Least squares quantization in PCM},
  author={Lloyd, Stuart P.},
  journal={IEEE Transactions on Information Theory},
  volume={28},
  number={2},
  pages={129--137},
  year={1982},
  publisher={IEEE}
}

@incollection{goos_comparing_2003,
	location = {Berlin, Heidelberg},
	title = {Comparing Clusterings by the Variation of Information},
	volume = {2777},
	isbn = {978-3-540-40720-1 978-3-540-45167-9},
	url = {http://link.springer.com/10.1007/978-3-540-45167-9_14},
	doi = {10.1007/978-3-540-45167-9_14},
	pages = {173--187},
	booktitle = {Learning Theory and Kernel Machines},
	publisher = {Springer Berlin Heidelberg},
	author = {Meilă, Marina},
	editor = {Schölkopf, Bernhard and Warmuth, Manfred K.},
	editorb = {Goos, Gerhard and Hartmanis, Juris and Van Leeuwen, Jan},
	editorbtype = {redactor},
	urldate = {2024-11-14},
	year = {2003},
	note = {Series Title: Lecture Notes in Computer Science},
}

@article{dempster_maximum_1977,
	title = {Maximum {Likelihood} from {Incomplete} {Data} {Via} the \textit{{EM}} {Algorithm}},
	volume = {39},
	copyright = {https://academic.oup.com/journals/pages/open\_access/funder\_policies/chorus/standard\_publication\_model},
	issn = {1369-7412, 1467-9868},
	url = {https://academic.oup.com/jrsssb/article/39/1/1/7027539},
	doi = {10.1111/j.2517-6161.1977.tb01600.x},
	language = {en},
	number = {1},
	urldate = {2026-04-09},
	journal = {Journal of the Royal Statistical Society Series B: Statistical Methodology},
	author = {Dempster, A. P. and Laird, N. M. and Rubin, D. B.},
	month = sep,
	year = {1977},
	pages = {1--22},
}

@book{mclachlan_finite_2019,
	title = {Finite Mixture Models},
	author = {{McLachlan}, Geoffrey J and Lee, Sharon X and Rathnayake, Suren I},
	year = {2019},
	langid = {english},
}

@article{heller_bayesian_2005,
	title = {Bayesian Hierarchical Clustering},
	url = {https://mlg.eng.cam.ac.uk/zoubin/papers/icml05heller.pdf},
	journal = {Proceedings of the 22nd International Conference on Machine Learning},
	author = {Heller, Katherine A. and Ghahramani, Zoubin},
	urldate = {2020-07-31},
	year = {2005},
	langid = {english},
}

@inbook{Stolcke_1994,
   title={Inducing probabilistic grammars by Bayesian model merging},
   ISBN={9783540489856},
   ISSN={1611-3349},
   url={http://dx.doi.org/10.1007/3-540-58473-0_141},
   DOI={10.1007/3-540-58473-0_141},
   booktitle={Grammatical Inference and Applications},
   publisher={Springer Berlin Heidelberg},
   author={Stolcke, Andreas and Omohundro, Stephen},
   year={1994},
   pages={106–118} }

@article{Banfield1993ModelbasedGA,
  title={Model-based Gaussian and non-Gaussian clustering},
  author={Jeffrey D. Banfield and Adrian E. Raftery},
  journal={Biometrics},
  year={1993},
  volume={49},
  pages={803-821},
  url={https://api.semanticscholar.org/CorpusID:17507406}
}

@inproceedings{vaithyanathan2013modelbasedhierarchicalclustering,
  title={Model-Based Hierarchical Clustering},
  author={Shivakumar Vaithyanathan and Byron Dom},
  booktitle={Conference on Uncertainty in Artificial Intelligence},
  year={2000},
  url={https://api.semanticscholar.org/CorpusID:2155333}
}

@article{blei_variational_2017,
	title = {Variational {Inference}: {A} {Review} for {Statisticians}},
	volume = {112},
	issn = {0162-1459, 1537-274X},
	shorttitle = {Variational {Inference}},
	url = {http://arxiv.org/abs/1601.00670},
	doi = {10.1080/01621459.2017.1285773},
	number = {518},
	urldate = {2023-06-14},
	journal = {Journal of the American Statistical Association},
	author = {Blei, David M. and Kucukelbir, Alp and McAuliffe, Jon D.},
	month = apr,
	year = {2017},
	note = {arXiv:1601.00670 [cs, stat]},
	pages = {859--877},
}

@inproceedings{blei_variational_2004,
	location = {Banff, Alberta, Canada},
	title = {Variational methods for the Dirichlet process},
	rights = {https://www.acm.org/publications/policies/copyright\_policy\#Background},
	url = {http://portal.acm.org/citation.cfm?doid=1015330.1015439},
	doi = {10.1145/1015330.1015439},
	eventtitle = {Twenty-first international conference},
	pages = {12},
	booktitle = {Twenty-first international conference on Machine learning  - {ICML} '04},
	publisher = {{ACM} Press},
	author = {Blei, David M. and Jordan, Michael I.},
	urldate = {2026-04-12},
	year = {2004},
	langid = {english},
}

@inproceedings{rasmussen_nite_nodate,
 author = {Rasmussen, Carl},
 booktitle = {Advances in Neural Information Processing Systems},
 editor = {S. Solla and T. Leen and K. M\"{u}ller},
 pages = {},
 publisher = {MIT Press},
 title = {The Infinite Gaussian Mixture Model},
 url = {https://proceedings.neurips.cc/paper_files/paper/1999/file/97d98119037c5b8a9663cb21fb8ebf47-Paper.pdf},
 volume = {12},
 year = {1999}
}

@inproceedings{chen_variational_2016,
	location = {Singapore},
	title = {A Variational Bayesian Approach for Unsupervised Clustering},
	isbn = {978-981-10-0539-8},
	pages = {651--660},
	booktitle = {Frontier Computing},
	publisher = {Springer Singapore},
	author = {Chen, Mu-Song and Wang, Hsuan-Fu and Hwang, Chi-Pan and Ho, Tze-Yee and Hung, Chan-Hsiang},
	editor = {Hung, Jason C and Yen, Neil Y. and Li, Kuan-Ching},
	year = {2016},
}

@article{guan_variational_nodate,
	title = {Variational Inference for Nonparametric Multiple Clustering},
	author = {Guan, Yue and Dy, Jennifer G and Niu, Donglin and Ghahramani, Zoubin},
	langid = {english},
}

@article{grigoriev_optimal_2003,
	title = {Optimal Jet Finder},
	volume = {155},
	issn = {00104655},
	url = {http://arxiv.org/abs/hep-ph/0301226},
	doi = {10.1016/S0010-4655(03)00291-1},
	pages = {42--64},
	number = {1},
	journal = {Computer Physics Communications},
	shortjournal = {Computer Physics Communications},
	author = {Grigoriev, D. Yu and Jankowski, E. and Tkachov, F. V.},
	urldate = {2026-04-12},
	year = {2003},
	month={09},
	eprinttype = {arxiv},
	eprint = {hep-ph/0301226},
}

@article{Ellis_2010,
   title={Recombination algorithms and jet substructure: Pruning as a tool for heavy particle searches},
   volume={81},
   ISSN={1550-2368},
   url={http://dx.doi.org/10.1103/PhysRevD.81.094023},
   DOI={10.1103/physrevd.81.094023},
   number={9},
   journal={Physical Review D},
   publisher={American Physical Society (APS)},
   author={Ellis, Stephen D. and Vermilion, Christopher K. and Walsh, Jonathan R.},
   year={2010},
   month=may }

@article{Li_2025,
   title={Jet reconstruction with Mamba networks in collider events},
   volume={112},
   ISSN={2470-0029},
   url={http://dx.doi.org/10.1103/pwp8-ffqk},
   DOI={10.1103/pwp8-ffqk},
   number={5},
   journal={Physical Review D},
   publisher={American Physical Society (APS)},
   author={Li, Jinmian and Li, Peng and Long, Bingwei and Zhang, Rao},
   year={2025},
   month=Sept }

@article{Chekanov_2006,
   title={A new jet algorithm based on the k-means clustering for the reconstruction of heavy states from jets},
   volume={47},
   ISSN={1434-6052},
   url={http://dx.doi.org/10.1140/epjc/s2006-02618-3},
   DOI={10.1140/epjc/s2006-02618-3},
   number={3},
   journal={The European Physical Journal C},
   publisher={Springer Science and Business Media LLC},
   author={Chekanov, S.},
   year={2006},
   month=July, pages={611–616} }

@article{De_Simone_2019,
   title={Guiding new physics searches with unsupervised learning},
   volume={79},
   ISSN={1434-6052},
   url={http://dx.doi.org/10.1140/epjc/s10052-019-6787-3},
   DOI={10.1140/epjc/s10052-019-6787-3},
   number={4},
   journal={The European Physical Journal C},
   publisher={Springer Science and Business Media LLC},
   author={De Simone, Andrea and Jacques, Thomas},
   year={2019},
   month=Mar }

@inproceedings{brehmer2020hierarchicalclusteringparticlephysics,
    author = "Brehmer, Johann and Macaluso, Sebastian and Pappadopulo, Duccio and Cranmer, Kyle",
    title = "{Hierarchical clustering in particle physics through reinforcement learning}",
    booktitle = "{34th Conference on Neural Information Processing Systems}",
    eprint = "2011.08191",
    archivePrefix = "arXiv",
    primaryClass = "cs.AI",
    month = "11",
    year = "2020"
}

@inproceedings{Huth:1990pfa,
    author        = "Huth, J. and others",
    title         = "{Toward a Standardization of Jet Definitions}",
    booktitle     = "{Research Directions for the Decade: Proceedings of the 1990 DPF Summer Study on High Energy Physics (Snowmass '90)}",
    editor        = "Berger, E. L.",
    address       = "Singapore",
    publisher     = "World Scientific",
    year          = "1992",
    pages         = "134--136",
    eprint        = "FERMILAB-CONF-90-249-E",
    archivePrefix = "arXiv",
    note          = "Presented at Snowmass 1990"
}

@techreport{cms_collaboration_cambridge-aachen_2009,
      collaboration = "CMS",
      title         = "{A Cambridge-Aachen (C-A) based Jet Algorithm for boosted
                       top-jet tagging}",
      institution   = "CERN",
      reportNumber  = "CMS-PAS-JME-09-001",
      address       = "Geneva",
      year          = "2009",
      url           = "https://cds.cern.ch/record/1194489",
}

@article{ju_supervised_2020,
	title = {Supervised Jet Clustering with Graph Neural Networks for Lorentz Boosted Bosons},
	volume = {102},
	issn = {2470-0010, 2470-0029},
	url = {http://arxiv.org/abs/2008.06064},
	doi = {10.1103/PhysRevD.102.075014},
	pages = {075014},
	number = {7},
	journal = {Physical Review D},
	shortjournal = {Phys. Rev. D},
	author = {Ju, Xiangyang and Nachman, Benjamin},
	urldate = {2026-04-12},
	year = {2020},
	month={10},
	day={13},
	eprinttype = {arxiv},
	eprint = {2008.06064 [hep-ph]},
}

@article{komiske_energy_2018,
	title = {Energy flow polynomials: A complete linear basis for jet substructure},
	volume = {2018},
	issn = {1029-8479},
	url = {http://arxiv.org/abs/1712.07124},
	doi = {10.1007/JHEP04(2018)013},
	shorttitle = {Energy flow polynomials},
	pages = {13},
	number = {4},
	journal = {Journal of High Energy Physics},
	shortjournal = {J. High Energ. Phys.},
	author = {Komiske, Patrick T. and Metodiev, Eric M. and Thaler, Jesse},
	urldate = {2026-08-10},
	year = {2018},
	month={04},
	eprinttype = {arxiv},
	eprint = {1712.07124 [hep-ph]},
}

@article{larkoski_jet_2023,
	title = {Jet {SIFT}-ing: a new scale-invariant jet clustering algorithm for the substructure era},
	volume = {108},
	issn = {2470-0010, 2470-0029},
	url = {http://arxiv.org/abs/2302.08609},
	doi = {10.1103/PhysRevD.108.016005},
	shorttitle = {Jet {SIFT}-ing},
	pages = {016005},
	number = {1},
	journal = {Physical Review D},
	shortjournal = {Phys. Rev. D},
	author = {Larkoski, Andrew J. and Rathjens, Denis and Veatch, Jason and Walker, Joel W.},
	urldate = {2026-04-12},
	year = {2023},
	month={07},
	day={11},
	eprinttype = {arxiv},
	eprint = {2302.08609 [hep-ph]},
}

@article{ellis_qjets_2012,
	title = {Qjets: A Non-Deterministic Approach to Tree-Based Jet Substructure},
	volume = {108},
	issn = {0031-9007, 1079-7114},
	url = {http://arxiv.org/abs/1201.1914},
	doi = {10.1103/PhysRevLett.108.182003},
	shorttitle = {Qjets},
	pages = {182003},
	number = {18},
	journal = {Physical Review Letters},
	shortjournal = {Phys. Rev. Lett.},
	author = {Ellis, Stephen D. and Hornig, Andrew and Krohn, David and Roy, Tuhin S. and Schwartz, Matthew D.},
	urldate = {2025-08-29},
	year = {2012},
	month={05},
	day={04},
	eprinttype = {arxiv},
	eprint = {1201.1914 [hep-ph]},
}

@article{mackey_fuzzy_2016,
	title = {Fuzzy Jets},
	volume = {2016},
	issn = {1029-8479},
	url = {http://arxiv.org/abs/1509.02216},
	doi = {10.1007/JHEP06(2016)010},
	pages = {10},
	number = {6},
	journal = {Journal of High Energy Physics},
	shortjournal = {J. High Energ. Phys.},
	author = {Mackey, Lester and Nachman, Benjamin and Schwartzman, Ariel and Stansbury, Conrad},
	urldate = {2023-04-10},
	year = {2016},
	month={06},
	eprinttype = {arxiv},
	eprint = {1509.02216 [hep-ph, stat]},
}

@article{larkoski_soft_2014,
	title = {Soft Drop},
	volume = {2014},
	issn = {1029-8479},
	url = {http://arxiv.org/abs/1402.2657},
	doi = {10.1007/JHEP05(2014)146},
	pages = {146},
	number = {5},
	journal = {Journal of High Energy Physics},
	shortjournal = {J. High Energ. Phys.},
	author = {Larkoski, Andrew J. and Marzani, Simone and Soyez, Gregory and Thaler, Jesse},
	urldate = {2025-03-19},
	year = {2014},
	month={05},
	eprinttype = {arxiv},
	eprint = {1402.2657 [hep-ph]},
}

@article{Cacciari_2008,
   title={Pileup subtraction using jet areas},
   volume={659},
   ISSN={0370-2693},
   url={http://dx.doi.org/10.1016/j.physletb.2007.09.077},
   DOI={10.1016/j.physletb.2007.09.077},
   number={1–2},
   journal={Physics Letters B},
   publisher={Elsevier BV},
   author={Cacciari, Matteo and Salam, Gavin P.},
   year={2008},
   month=jan, pages={119–126} }

@article{Krohn_2010,
   title={Jet trimming},
   volume={2010},
   ISSN={1029-8479},
   url={http://dx.doi.org/10.1007/JHEP02(2010)084},
   DOI={10.1007/jhep02(2010)084},
   number={2},
   journal={Journal of High Energy Physics},
   publisher={Springer Science and Business Media LLC},
   author={Krohn, David and Thaler, Jesse and Wang, Lian-Tao},
   year={2010},
   month=feb }

@article{Berta_2014,
   title={Particle-level pileup subtraction for jets and jet shapes},
   volume={2014},
   ISSN={1029-8479},
   url={http://dx.doi.org/10.1007/JHEP06(2014)092},
   DOI={10.1007/jhep06(2014)092},
   number={6},
   journal={Journal of High Energy Physics},
   publisher={Springer Science and Business Media LLC},
   author={Berta, Peter and Spousta, Martin and Miller, David W. and Leitner, Rupert},
   year={2014},
   month=jun }

@article{cacciari_softkiller_2015,
	title = {{SoftKiller}, a particle-level pileup removal method},
	volume = {75},
	issn = {1434-6044, 1434-6052},
	url = {http://arxiv.org/abs/1407.0408},
	doi = {10.1140/epjc/s10052-015-3267-2},
	pages = {59},
	number = {2},
	journal = {The European Physical Journal C},
	shortjournal = {Eur. Phys. J. C},
	author = {Cacciari, Matteo and Salam, Gavin P. and Soyez, Gregory},
	urldate = {2025-03-19},
	year = {2015},
	month={02},
	eprinttype = {arxiv},
	eprint = {1407.0408 [hep-ph]},
}

@techreport{CERN_HLLHC,
      title         = "{The High-Luminosity LHC Project}",
      reportNumber  = "CERN/SPC/1068, CERN/FC/6014, CERN/3255",
      year          = "2016",
      url           = "https://cds.cern.ch/record/2199189",
}

@article{Krohn_2014,
   title={Jet cleansing: Separating data from secondary collision induced radiation at high luminosity},
   volume={90},
   ISSN={1550-2368},
   url={http://dx.doi.org/10.1103/PhysRevD.90.065020},
   DOI={10.1103/physrevd.90.065020},
   number={6},
   journal={Physical Review D},
   publisher={American Physical Society (APS)},
   author={Krohn, David and Schwartz, Matthew D. and Low, Matthew and Wang, Lian-Tao},
   year={2014},
   month=sep }

@article{bertolini_pileup_2014,
	title = {Pileup Per Particle Identification},
	volume = {2014},
	issn = {1029-8479},
	url = {http://arxiv.org/abs/1407.6013},
	doi = {10.1007/JHEP10(2014)059},
	pages = {59},
	number = {10},
	journal = {Journal of High Energy Physics},
	shortjournal = {J. High Energ. Phys.},
	author = {Bertolini, Daniele and Harris, Philip and Low, Matthew and Tran, Nhan},
	urldate = {2023-03-24},
	year = {2014},
	month={10},
	eprinttype = {arxiv},
	eprint = {1407.6013 [hep-ex, physics:hep-ph]},
}

@article{Scott:2024txs,
    author = "Scott, Jacob L. and Dong, Zhongtian and Kim, Taejoon and Kong, Kyoungchul and Park, Myeonghun",
    title = "{Hybrid quantum-classical approach for combinatorial problems at hadron colliders}",
    eprint = "2410.22417",
    archivePrefix = "arXiv",
    primaryClass = "hep-ph",
    month = "10",
    year = "2024"
}

@article{Jackson_2017a,
   title={Recursive jigsaw reconstruction: HEP event analysis in the presence of kinematic and combinatoric ambiguities},
   volume={96},
   ISSN={2470-0029},
   url={http://dx.doi.org/10.1103/PhysRevD.96.112007},
   DOI={10.1103/physrevd.96.112007},
   number={11},
   journal={Physical Review D},
   publisher={American Physical Society (APS)},
   author={Jackson, Paul and Rogan, Christopher},
   year={2017},
   month=Dec }

@article{Jackson_2017b,
   title={Sparticles in motion: Analyzing compressed SUSY scenarios with a new method of event reconstruction},
   volume={95},
   ISSN={2470-0029},
   url={http://dx.doi.org/10.1103/PhysRevD.95.035031},
   DOI={10.1103/physrevd.95.035031},
   number={3},
   journal={Physical Review D},
   publisher={American Physical Society (APS)},
   author={Jackson, Paul and Rogan, Christopher and Santoni, Marco},
   year={2017},
   month=Feb }

@article{Lee:2023tfx,
    author = "Lee, Lawrence and Bell, Charles and Lawless, John and Nash, Cordney and Nibigira, Emery",
    title = "{Experimental impact of jet fragmentation reference frames at particle colliders}",
    eprint = "2308.10951",
    archivePrefix = "arXiv",
    primaryClass = "hep-ph",
    doi = "10.1016/j.physletb.2025.139561",
    journal = "Phys. Lett. B",
    volume = "866",
    pages = "139561",
    year = "2025"
}

@article{bogatskiy_explainable_2024,
	title = {Explainable Equivariant Neural Networks for Particle Physics: {PELICAN}},
	volume = {2024},
	issn = {1029-8479},
	url = {http://arxiv.org/abs/2307.16506},
	doi = {10.1007/JHEP03(2024)113},
	shorttitle = {Explainable Equivariant Neural Networks for Particle Physics},
	pages = {113},
	number = {3},
	journal = {Journal of High Energy Physics},
	shortjournal = {J. High Energ. Phys.},
	author = {Bogatskiy, Alexander and Hoffman, Timothy and Miller, David W. and Offermann, Jan T. and Liu, Xiaoyang},
	urldate = {2025-01-30},
	year = {2024},
	month={03},
	day={20},
	eprinttype = {arxiv},
	eprint = {2307.16506 [hep-ph]},
}

@article{Thaler_2011,
   title={Identifying boosted objects with N-subjettiness},
   volume={2011},
   ISSN={1029-8479},
   url={http://dx.doi.org/10.1007/JHEP03(2011)015},
   DOI={10.1007/jhep03(2011)015},
   number={3},
   journal={Journal of High Energy Physics},
   publisher={Springer Science and Business Media LLC},
   author={Thaler, Jesse and Van Tilburg, Ken},
   year={2011},
   month=mar }

@article{chien_telescoping_2020,
	title = {Telescoping jet substructure},
	volume = {101},
	issn = {2470-0010, 2470-0029},
	url = {http://arxiv.org/abs/1711.11041},
	doi = {10.1103/PhysRevD.101.114006},
	pages = {114006},
	number = {11},
	journal = {Physical Review D},
	shortjournal = {Phys. Rev. D},
	author = {Chien, Yang-Ting and Emerman, Alex and Hsu, Shih-Chieh and Meehan, Samuel and Montague, Zachery},
	urldate = {2026-04-08},
	year = {2020},
	month={06},
	day={09},
	eprinttype = {arxiv},
	eprint = {1711.11041 [hep-ph]},
}

@article{marzani_looking_2019,
	title = {Looking inside jets: an introduction to jet substructure and boosted-object phenomenology},
	volume = {958},
	url = {http://arxiv.org/abs/1901.10342},
	doi = {10.1007/978-3-030-15709-8},
	shorttitle = {Looking inside jets},
	journal = {{arXiv}:1901.10342 [hep-ex, physics:hep-ph]},
	author = {Marzani, Simone and Soyez, Gregory and Spannowsky, Michael},
	urldate = {2021-11-08},
	year = {2019},
	eprinttype = {arxiv},
	eprint = {1901.10342},
}

@article{dillon_uncovering_2019,
	title = {Uncovering latent jet substructure},
	volume = {100},
	issn = {2470-0010, 2470-0029},
	url = {http://arxiv.org/abs/1904.04200},
	doi = {10.1103/PhysRevD.100.056002},
	pages = {056002},
	number = {5},
	journal = {Physical Review D},
	shortjournal = {Phys. Rev. D},
	author = {Dillon, Barry M. and Faroughy, Darius A. and Kamenik, Jernej F.},
	urldate = {2021-09-24},
	year = {2019},
	month={09},
	day={03},
	eprinttype = {arxiv},
	eprint = {1904.04200},
}

@article{klimek_time_2021,
	title = {The Time Substructure of Jets and Boosted Object Tagging},
	issn = {0954-3899, 1361-6471},
	url = {http://arxiv.org/abs/1911.11235},
	doi = {10.1088/1361-6471/ac446a},
	journal = {Journal of Physics G: Nuclear and Particle Physics},
	shortjournal = {J. Phys. G: Nucl. Part. Phys.},
	author = {Klimek, Matthew D.},
	urldate = {2022-01-19},
	year = {2021},
	month={12},
	day={17},
}

@article{dasgupta_towards_2013,
	title = {Towards an understanding of jet substructure},
	volume = {2013},
	issn = {1029-8479},
	url = {http://arxiv.org/abs/1307.0007},
	doi = {10.1007/JHEP09(2013)029},
	pages = {29},
	number = {9},
	journal = {Journal of High Energy Physics},
	shortjournal = {J. High Energ. Phys.},
	author = {Dasgupta, Mrinal and Fregoso, Alessandro and Marzani, Simone and Salam, Gavin P.},
	urldate = {2026-01-31},
	year = {2013},
	month={09},
	eprinttype = {arxiv},
	eprint = {1307.0007 [hep-ph]},
}

@article{feige_precision_2012,
	title = {Precision Jet Substructure from Boosted Event Shapes},
	volume = {109},
	issn = {0031-9007, 1079-7114},
	url = {http://arxiv.org/abs/1204.3898},
	doi = {10.1103/PhysRevLett.109.092001},
	pages = {092001},
	number = {9},
	journal = {Physical Review Letters},
	shortjournal = {Phys. Rev. Lett.},
	author = {Feige, Ilya and Schwartz, Matthew D. and Stewart, Iain W. and Thaler, Jesse},
	urldate = {2026-01-31},
	year = {2012},
	month={08},
	day={30},
	eprinttype = {arxiv},
	eprint = {1204.3898 [hep-ph]},
}

@article{butterworth_jet_2008,
	title = {Jet substructure as a new Higgs search channel at the {LHC}},
	volume = {100},
	issn = {0031-9007, 1079-7114},
	url = {http://arxiv.org/abs/0802.2470},
	doi = {10.1103/PhysRevLett.100.242001},
	pages = {242001},
	number = {24},
	journal = {Physical Review Letters},
	shortjournal = {Phys. Rev. Lett.},
	author = {Butterworth, Jonathan M. and Davison, Adam R. and Rubin, Mathieu and Salam, Gavin P.},
	urldate = {2025-09-09},
	year = {2008},
	month={06},
	day={18},
	eprinttype = {arxiv},
	eprint = {0802.2470 [hep-ph]},
}

@misc{delafuente2026simplexdemixingdisentanglingmultiple,
      title={Simplex Demixing: Disentangling Multiple Light-Flavor Jets at Colliders}, 
      author={Gregorio de la Fuente and Jesse Thaler},
      year={2026},
      eprint={2607.24921},
      archivePrefix={arXiv},
      primaryClass={hep-ph},
      url={https://arxiv.org/abs/2607.24921}, 
}

@unpublished{patrick_breheny_wishart_2013,
	location = {University of Kentucky},
	title = {Wishart Priors},
	url = {https://myweb.uiowa.edu/pbreheny/uk/teaching/701/notes/3-28.pdf},
	note = {{BST} 701: Bayesian Modeling in Biostatics},
	author = {{Patrick Breheny}},
	urldate = {2023-09-07},
	year = {2013},
	month={03},
	day={28},
}

@article{cms_collaboration_performance_2024,
	title = {Performance of the {CMS} electromagnetic calorimeter in pp collisions at $\sqrt{s}$ = 13 {TeV}},
	volume = {19},
	issn = {1748-0221},
	url = {http://arxiv.org/abs/2403.15518},
	doi = {10.48550/arXiv.2403.15518},
	number = {9},
	journal = {Journal of Instrumentation},
	shortjournal = {{JINST}},
	author = {{CMS Collaboration}},
	urldate = {2025-09-17},
	year = {2024},
	month={03},
	day={22},
	eprinttype = {arxiv},
	eprint = {2403.15518 [physics]},
	note = {version: 1},
}

@article{cms_collaboration_time_2010,
	title = {Time Reconstruction and Performance of the {CMS} Electromagnetic Calorimeter},
	volume = {5},
	issn = {1748-0221},
	url = {http://arxiv.org/abs/0911.4044},
	doi = {10.1088/1748-0221/5/03/T03011},
	pages = {T03011--T03011},
	number = {3},
	journal = {Journal of Instrumentation},
	shortjournal = {J. Inst.},
	author = {{CMS Collaboration}},
	urldate = {2025-09-17},
	year = {2010},
	month={03},
	day={19},
	eprinttype = {arxiv},
	eprint = {0911.4044 [physics]},
}

@misc{chekanov2002jetalgorithmsminireview,
      title={Jet algorithms: a minireview}, 
      author={S. V. Chekanov},
      year={2002},
      eprint={hep-ph/0211298},
      archivePrefix={arXiv},
      primaryClass={hep-ph},
      url={https://arxiv.org/abs/hep-ph/0211298}, 
}

@article{gell-mann_schematic_1964,
	title = {A Schematic Model of Baryons and Mesons},
	language = {en},
	journal = {Physics Letters},
	author = {Gell-Mann, M.},
	year = {1964},
}

@article{GellMann:1962xb,
    author = "Gell-Mann, Murray",
    title = "{Symmetries of Baryons and Mesons}",
    journal = "Phys. Rev.",
    volume = "125",
    pages = "1067--1084",
    year = "1962",
    doi = "10.1103/PhysRev.125.1067"
}

@techreport{Zweig:570209,
      author        = "Zweig, G",
      title         = "{An {SU}$_3$ model for strong interaction symmetry and its
                       breaking}",
      institution   = "CERN",
      reportNumber  = "CERN-TH-401",
      address       = "Geneva",
      year          = "1964",
      url           = "https://cds.cern.ch/record/352337",
      note          = "An updated version of the paper is available as \href{https://cds.cern.ch/record/570209}{CERN-TH-412},
                       Feb. 1964.",
      doi           = "10.17181/CERN-TH-401",
}

@article{Andersson:1983jt,
    author = "Andersson, B. and Gustafson, G. and S{\"o}derberg, B.",
    title = "{A general model for jet fragmentation}",
    journal = "Z. Phys. C",
    volume = "20",
    pages = "317--329",
    year = "1983",
    doi = "10.1007/BF01407824"
}

@article{SJOSTRAND1984469,
title = {Jet fragmentation of multiparton configurations in a string framework},
journal = {Nuclear Physics B},
volume = {248},
number = {2},
pages = {469-502},
year = {1984},
issn = {0550-3213},
doi = {https://doi.org/10.1016/0550-3213(84)90607-2},
url = {https://www.sciencedirect.com/science/article/pii/0550321384906072},
author = {Torbjörn Sjöstrand}
}

@article{dreyer_lund_2018,
	title = {The Lund Jet Plane},
	volume = {2018},
	issn = {1029-8479},
	url = {http://arxiv.org/abs/1807.04758},
	doi = {10.1007/JHEP12(2018)064},
	pages = {64},
	number = {12},
	journal = {Journal of High Energy Physics},
	shortjournal = {J. High Energ. Phys.},
	author = {Dreyer, Frederic A. and Salam, Gavin P. and Soyez, Gregory},
	urldate = {2024-08-07},
	year = {2018},
	month={12},
	eprinttype = {arxiv},
	eprint = {1807.04758 [hep-ex, physics:hep-ph]},
}

@article{2091129,
    collaboration = "ATLAS",
    title = "{A High-Granularity Timing Detector for the ATLAS Phase-II Upgrade: Technical Design Report}",
    reportNumber = "CERN-LHCC-2020-007, ATLAS-TDR-031"
}

@techreport{cms_collaboration_mip_2019,
      author = {{CMS Collaboration}},
      title         = "{A MIP Timing Detector for the CMS Phase-2 Upgrade}",
      institution   = "CERN",
      reportNumber  = "CERN-LHCC-2019-003, CMS-TDR-020",
      address       = "Geneva",
      year          = "2019",
      url           = "https://cds.cern.ch/record/2667167",
}

@article{collaboration_measurement_2024,
	title = {Measurement of boosted Higgs bosons produced via vector boson fusion or gluon fusion in the H $\to$ $\mathrm{b\bar{b}}$ decay mode using {LHC} proton-proton collision data at $\sqrt{s}$ = 13 {TeV}},
	volume = {2024},
	issn = {1029-8479},
	url = {http://arxiv.org/abs/2407.08012},
	doi = {10.1007/JHEP12(2024)035},
	pages = {35},
	number = {12},
	journal = {Journal of High Energy Physics},
	shortjournal = {J. High Energ. Phys.},
	author = {{CMS Collaboration}},
	urldate = {2025-09-11},
	year = {2024},
	month={12},
	day={04},
	eprinttype = {arxiv},
	eprint = {2407.08012 [hep-ex]},
}

@article{Aaboud_2019,
   title={Search for light resonances decaying to boosted quark pairs and produced in association with a photon or a jet in proton–proton collisions at $\sqrt{s} = 13$ TeV with the ATLAS detector},
   volume={788},
   ISSN={0370-2693},
   url={http://dx.doi.org/10.1016/j.physletb.2018.09.062},
   DOI={10.1016/j.physletb.2018.09.062},
   journal={Physics Letters B},
   publisher={Elsevier BV},
   author={{ATLAS Collaboration}},
      year={2019},
   month=Jan, pages={316–335} 
   }

@article{Sirunyan_2019,
   title={Search for low mass vector resonances decaying into quark-antiquark pairs in proton-proton collisions at $\sqrt{s} = 13$ TeV},
   volume={100},
   ISSN={2470-0029},
   url={http://dx.doi.org/10.1103/PhysRevD.100.112007},
   DOI={10.1103/physrevd.100.112007},
   number={11},
   journal={Physical Review D},
   publisher={American Physical Society (APS)},
   author={{CMS Collaboration}},
      year={2019},
   month=Dec 
   }

@article{Aad_2024,
   title={Search for low-mass resonances decaying into two jets and produced in association with a photon or a jet at $\sqrt{s} = 13$ TeV with the ATLAS detector},
   volume={110},
   ISSN={2470-0029},
   url={http://dx.doi.org/10.1103/PhysRevD.110.032002},
   DOI={10.1103/physrevd.110.032002},
   number={3},
   journal={Physical Review D},
   publisher={American Physical Society (APS)},
   author={{ATLAS Collaboration}},
      year={2024},
   month=Aug 
   }

@article{Sirunyan_2020,
doi = {10.1088/1748-0221/15/06/P06005},
url = {https://doi.org/10.1088/1748-0221/15/06/P06005},
year = {2020},
month = {jun},
publisher = {},
volume = {15},
number = {06},
pages = {P06005},
author = {{CMS Collaboration}},
title = {Identification of heavy, energetic, hadronically decaying particles using  machine-learning techniques},
journal = {Journal of Instrumentation}
}

@article{Lapsien_2016,
   title={A new tagger for hadronically decaying heavy particles at the LHC},
   volume={76},
   ISSN={1434-6052},
   url={http://dx.doi.org/10.1140/epjc/s10052-016-4443-8},
   DOI={10.1140/epjc/s10052-016-4443-8},
   number={11},
   journal={The European Physical Journal C},
   publisher={Springer Science and Business Media LLC},
   author={Lapsien, T. and Kogler, R. and Haller, J.},
   year={2016},
   month=Nov }

@article{Krohn_2009,
   title={Jets with variable R},
   volume={2009},
   ISSN={1029-8479},
   url={http://dx.doi.org/10.1088/1126-6708/2009/06/059},
   DOI={10.1088/1126-6708/2009/06/059},
   number={06},
   journal={Journal of High Energy Physics},
   publisher={Springer Science and Business Media LLC},
   author={Krohn, David and Thaler, Jesse and Wang, Lian-Tao},
   year={2009},
   month=June, pages={059–059} }

@misc{cmscollaboration2026searchresonanceeventsquarks,
      title={Search for a resonance in events with four top quarks decaying into two leptons and jets in proton-proton collisions}, 
      author={{CMS Collaboration}},
      year={2026},
      eprint={2608.10148},
      archivePrefix={arXiv},
      primaryClass={hep-ex},
      url={https://arxiv.org/abs/2608.10148}, 
}

@article{cms_collaboration_search_2020,
	title = {Search for electroweak production of a vector-like T quark using fully hadronic final states},
	volume = {2020},
	issn = {1029-8479},
	url = {http://arxiv.org/abs/1909.04721},
	doi = {10.1007/JHEP01(2020)036},
	pages = {36},
	number = {1},
	journal = {Journal of High Energy Physics},
	shortjournal = {J. High Energ. Phys.},
	author = {{CMS Collaboration}},
	urldate = {2025-06-16},
	year = {2020},
	month={01},
	day={08},
	eprinttype = {arxiv},
	eprint = {1909.04721 [hep-ex]},
}

@article{ParticleFlow_2017,
   title={Particle-flow reconstruction and global event description with the CMS detector},
   volume={12},
   ISSN={1748-0221},
   url={http://dx.doi.org/10.1088/1748-0221/12/10/P10003},
   DOI={10.1088/1748-0221/12/10/p10003},
   number={10},
   journal={Journal of Instrumentation},
   publisher={IOP Publishing},
   author={CMS Collaboration},
   year={2017},
   month=oct, pages={P10003–P10003} }

@book{raiffa_applied_1961,
	address = {Boston, MA},
	title = {Applied {Statistical} {Decision} {Theory}},
	publisher = {Division of Research, Graduate School of Business Administration, Harvard University},
	author = {Raiffa, Howard and Schlaifer, Robert},
	year = {1961},
}

@article{80cb41ee-7d47-3d90-ae24-2998213a9236,
 ISSN = {00063444, 14643510},
 URL = {http://www.jstor.org/stable/2332750},
 author = {G. B. Wetherill},
 journal = {Biometrika},
 number = {3/4},
 pages = {281--292},
 publisher = {[Oxford University Press, Biometrika Trust]},
 title = {Bayesian Sequential Analysis},
 urldate = {2025-08-29},
 volume = {48},
 year = {1961}
}

@book{bishop_pattern_2006,
	address = {New York},
	series = {Information science and statistics},
	title = {Pattern recognition and machine learning},
	isbn = {978-0-387-31073-2},
	language = {en},
	publisher = {Springer},
	author = {Bishop, Christopher M.},
	year = {2006},
}

@article{cacciari_anti-k_t_2008,
	title = {The anti-k\_t jet clustering algorithm},
	volume = {2008},
	issn = {1029-8479},
	url = {http://arxiv.org/abs/0802.1189},
	doi = {10.1088/1126-6708/2008/04/063},
	number = {04},
	urldate = {2023-09-29},
	journal = {Journal of High Energy Physics},
	author = {Cacciari, Matteo and Salam, Gavin P. and Soyez, Gregory},
	month = apr,
	year = {2008},
	note = {arXiv:0802.1189 [hep-ph]},
	pages = {063--063},
}

@article{bierlich_comprehensive_2022,
	title = {A comprehensive guide to the physics and usage of {PYTHIA} 8.3},
	copyright = {https://creativecommons.org/licenses/by/4.0},
	url = {https://scipost.org/10.21468/SciPostPhysCodeb.8},
	doi = {10.21468/scipostphyscodeb.8},
	language = {en},
	urldate = {2025-07-18},
	journal = {SciPost Physics Codebases},
	author = {Bierlich, Christian and Chakraborty, Smita and Desai, Nishita and Gellersen, Leif and Helenius, Ilkka and Ilten, Philip and Lönnblad, Leif and Mrenna, Stephen and Prestel, Stefan and Preuss, Christian Tobias and Sjöstrand, Torbjörn and Skands, Peter and Utheim, Marius and Verheyen, Rob},
	month = nov,
	year = {2022},
}

@article{cacciari_fastjet_2012,
	title = {{FastJet} user manual (for version 3.4.2)},
	volume = {72},
	issn = {1434-6044, 1434-6052},
	shorttitle = {{FastJet} user manual},
	url = {http://link.springer.com/10.1140/epjc/s10052-012-1896-2},
	doi = {10.1140/epjc/s10052-012-1896-2},
	language = {en},
	number = {3},
	urldate = {2023-10-02},
	journal = {The European Physical Journal C},
	author = {Cacciari, Matteo and Salam, Gavin P. and Soyez, Gregory},
	month = mar,
	year = {2012},
	pages = {1896},
}

@article{ellis_successive_1993,
	title = {Successive combination jet algorithm for hadron collisions},
	volume = {48},
	doi = {10.1103/PhysRevD.48.3160},
	number = {7},
	journal = {Physical Review D},
	author = {Ellis, Stephen D. and Soper, Davison E.},
	year = {1993},
	pages = {3160--3166},
}

@article{cacciari_dispelling_2006,
	title = {Dispelling the {N}{\textasciicircum}3 myth for the {Kt} jet-finder},
	volume = {641},
	issn = {03702693},
	url = {http://arxiv.org/abs/hep-ph/0512210},
	doi = {10.1016/j.physletb.2006.08.037},
	number = {1},
	urldate = {2023-09-29},
	journal = {Physics Letters B},
	author = {Cacciari, Matteo and Salam, Gavin P.},
	month = sep,
	year = {2006},
	note = {arXiv:hep-ph/0512210},
	pages = {57--61},
}

@book{acosta_cms_2006,
	address = {Genève},
	title = {{CMS} {Physics} technical design report. {Volume} {I}, {Volume} {I},},
	isbn = {978-92-9083-268-3},
	language = {English},
	publisher = {CERN},
	author = {Acosta, Darin and {Centre européen pour la recherche nucléaire}},
	year = {2006},
	note = {OCLC: 819303207},
}

@book{cms_collaboration_cms_1997,
	series = {Technical design report. {CMS}},
	title = {The {CMS} electromagnetic calorimeter project: {Technical} {Design} {Report}},
	shorttitle = {The {CMS} electromagnetic calorimeter project},
	url = {https://cds.cern.ch/record/349375},
	urldate = {2024-01-18},
	publisher = {CERN},
	author = {{CMS Collaboration}},
	year = {1997},
	note = {Place: Geneva},
}

@article{Del34,
  author  = {Boris Delaunay},
  title   = {Sur la sph{\`e}re vide. {\`A} la m{\'e}moire de Georges Vorono{\"i}},
  journal = {Bulletin de l'Acad{\'e}mie des Sciences de l'URSS. Classe des sciences math{\'e}matiques et naturelles},
  year    = {1934},
  number  = {6},
  pages   = {793--800},
  url     = {http://mi.mathnet.ru/eng/im4937},
  note    = {Zbl 0010.41101},
}

@article{voronoi1908nouvelles,
  title={Nouvelles applications des param{\`e}tres continus {\`a} la th{\'e}orie des formes quadratiques. Premier m{\'e}moire. Sur quelques propri{\'e}t{\'e}s des formes quadratiques positives parfaites},
  author={Vorono{\"\i}, Georges},
  journal={Journal f{\"u}r die reine und angewandte Mathematik (Crelles Journal)},
  volume={1908},
  number={133},
  pages={97--178},
  year={1908},
  publisher={De Gruyter},
  doi={10.1515/crll.1908.133.97},
  url={http://eudml.org/doc/149276}
}

@article{strassler_echoes_2007,
	title = {Echoes of a {Hidden} {Valley} at {Hadron} {Colliders}},
	volume = {651},
	issn = {03702693},
	url = {http://arxiv.org/abs/hep-ph/0604261},
	doi = {10.1016/j.physletb.2007.06.055},
	number = {5-6},
	urldate = {2026-05-14},
	journal = {Physics Letters B},
	author = {Strassler, Matthew J. and Zurek, Kathryn M.},
	month = aug,
	year = {2007},
	note = {arXiv:hep-ph/0604261},
	pages = {374--379},
}

@article{carloni_visible_2010,
	title = {Visible {Effects} of {Invisible} {Hidden} {Valley} {Radiation}},
	volume = {2010},
	issn = {1029-8479},
	url = {http://arxiv.org/abs/1006.2911},
	doi = {10.1007/JHEP09(2010)105},
	number = {9},
	urldate = {2026-06-24},
	journal = {Journal of High Energy Physics},
	author = {Carloni, Lisa and Sjostrand, Torbjorn},
	month = sep,
	year = {2010},
	note = {arXiv:1006.2911 [hep-ph]},
	pages = {105},
}

@article{info_geo,
	author = {Boissonnat, Jean-Daniel and Nielsen, Frank and Nock, Richard},
	da = {2010/09/01},
	doi = {10.1007/s00454-010-9256-1},
	id = {Boissonnat2010},
	isbn = {1432-0444},
	journal = {Discrete \& Computational Geometry},
	number = {2},
	pages = {281--307},
	title = {Bregman Voronoi Diagrams},
	ty = {JOUR},
	url = {https://doi.org/10.1007/s00454-010-9256-1},
	volume = {44},
	year = {2010}}

@article{Delauny_divide,
author = {Cignoni, Paolo and Montani, Claudio and Scopigno, Roberto},
year = {1998},
month = {04},
pages = {333-341},
title = {DeWall: A fast divide and conquer Delaunay triangulation algorithm in E d},
volume = {30},
journal = {Computer-Aided Design},
doi = {10.1016/S0010-4485(97)00082-1}
}
\bibliographystyle{unsrt}

\end{document}